\documentclass[12pt]{report}%
\usepackage{ amsmath, graphics, rotating}%

\begin{document}
\large
\begin{center}{\large\bf QUANTUM THEORY ON A GALOIS
FIELD}\end{center}
\vskip 1em \begin{center} {\large Felix M. Lev} \end{center}
\vskip 1em \begin{center} {\it Artwork Conversion Software Inc.,
1201 Morningside Drive, Manhattan Beach, CA 90266, USA
(E-mail:  felixlev@hotmail.com)} \end{center}
\vskip 1em

{\it Abstract:}
\vskip 0.5em

Systems of free particles in a quantum theory based 
on a Galois field (GFQT) are discussed in detail. 
In this approach infinities cannot exist, the
cosmological constant problem does not arise and one
irreducible representation of the
symmetry algebra necessarily describes a particle
and its antiparticle simultaneously. As a consequence,
well known results of the standard
theory (spin-statistics theorem; a particle and
its antiparticle have the same masses and spins but
opposite charges etc.) can be proved without 
involving local covariant equations. The
spin-statistics theorem is simply a requirement that 
quantum theory should be based on
complex numbers. Some new features of GFQT are as
follows: i) elementary particles cannot be neutral;
ii) the Dirac vacuum energy problem has a natural 
solution and the vacuum energy
(which in the standard theory is infinite and negative)
equals zero as it should be; iii) the charge operator has 
correct properties only for massless particles with the
spins 0 and 1/2.  
In the AdS version of the theory there exists a
dilemma that either the notion of particles and
antiparticles is absolute and then only particles
with a half-integer spin can be elementary or
the notion is valid only when energies are not
asymptotically large and then supersymmetry is 
possible.

\begin{flushleft} PACS: 02.10.De, 03.65.Ta, 11.30.Fs, 11.30.Ly\end{flushleft}

\begin{flushleft} Keywords: quantum theory, Galois fields, elementary particles\end{flushleft}

\vfill\eject

\tableofcontents

\vfill\eject

\chapter{Introduction}
\label{C1}

The phenomenon of local quantum field theory (LQFT) has no
analogs in the history of science. There is no branch of
science where so impressive agreements between theory and
experiment have been achieved. The theory has successfully
predicted the existence of many new particles
and even new interactions.
It is hard to believe that all these achievements
are only coincidences. At the same time, the level of
mathematical rigor in LQFT is very poor and, as a
result, LQFT has several well known difficulties and
inconsistencies. The absolute majority of physicists
believes that agreement with experiment is much more
important than the lack of mathematical rigor, but not
all of them think so. For example, Dirac's opinion is
as follows \cite{DirMath}:
$'$The agreement
with observation is presumably by coincidence, just like the
original calculation of the hydrogen spectrum with Bohr orbits.
Such coincidences are no reason for turning a blind eye to the
faults of the theory. Quantum electrodynamics is rather like
Klein-Gordon equation. It was built up from physical ideas
that were not correctly incorporated into the theory and it
has no sound mathematical foundation.$'$ In addition, 
LQFT fails in quantizing gravity since
quantum gravity is not renormalizable.

Usually there is no
need to require that the level of mathematical rigor
in physics should be the same as in mathematics.
However, physicists should
have a feeling that, at least in principle, mathematical
statements used in the theory can be substantiated.
The absence of a well-substantiated LQFT by no means
can be treated as a
pure academic problem. This becomes immediately
clear when one wishes to work beyond perturbation
theory. The problem arises to
what extent the difficulties of LQFT can be overcome
in the framework of LQFT itself or LQFT can only be a
special case of a more general theory based on essentially new
ideas. Weinberg's opinion \cite{Wein} is that LQFT can
be treated $'$in the way it is$'$, but at the same
time it is a $'$low energy approximation to a
deeper theory that may not even be a field theory,
but something different like a string theory$'$.

The main problem is the choice of strategy
for constructing a new quantum theory. Since
nobody knows for sure which strategy is the best one,
different approaches should be investigated. In the
present paper we are trying to follow Dirac's
advice given in Ref. \cite{DirMath}: $'$I learned
to distrust all
physical concepts as a basis for a theory. Instead
one should
put one's trust in a mathematical scheme, even if
the scheme does not appear at first sight to be
connected with physics. One should concentrate on
getting an interesting mathematics.$'$

The mostly known consequence of the poor definition of
local operators in LQFT is the existence of
divergencies. The fact that in renormalizable theories
one can get rid off divergencies in the perturbation
expansion for the S-matrix does not imply that the
other operators (e.g. the Hamiltonian) automatically
become well-defined. In Dirac's opinion \cite{DirMath},
$'$the renormalization idea would be sensible
only if it was applied with finite renormalization
factors, not infinite ones.$'$ Weinberg \cite{Wein}
describes the problem of infinities as follows:
$'$Disappointingly this problem appeared with even
greater severity in the early days of quantum theory,
and although greatly ameliorated by
subsequent improvements in the theory, it remains
with us to the present day.$'$ A reader might also
be interested in the discussion of infinities given
in $'$t Hooft's Nobel Lecture \cite{tHooft}. A desire
to have a theory without divergencies was probably the
main motivation for developing modern theories
extending LFQT (loop quantum gravity,
noncommutative quantum theory, string theory etc.).

There exists a wide literature aiming to solve
the difficulties of LQFT by replacing the field
of complex numbers by quaternions, p-adic numbers
or other constructions. In the present paper we 
accept the 
philosophy that the ultimate quantum theory
will not contain the actual infinity at all. Then
quantum theory can be based only on Galois fields
(GFQT). Since any Galois field is finite, 
the problem of
infinities in GFQT does not exist in principle
and all operators are well defined. In other words,
GFQT solves the problem of infinities once and
for all. Some implications
of Galois field in quantum physics have been
investigated by several authors (see e.g. Ref.
\cite{Galois} and references
therein) but they usually considered a replacement of
the conventional spacetime by a Galois field and the
theory involved other nonfinite fields. In contrast
to those approaches, our one does not contain space-time
coordinates at all (in the spirit of Heisenberg ideas)
but (instead of the S-matrix) the main ingredient of
the theory is linear spaces and operators
over Galois fields.

A well known historical fact is that originally
quantum theory has been proposed in two formalisms
which seemed to be essentially different: the
Schroedinger wave formalism and the Heisenberg
operator (matrix) formalism. It has been shown later
by Born, von Neumann and others that the both
formalisms are equivalent and, in addition, the
path integral formalism has been developed. From time
to time physicists change their preferences and at
present the wave approach prevails. GFQT is a
direct generalization of the operator formalism
when the field of complex numbers is replaced by
a Galois field. However, as it will be clear in Chap.
\ref{C4}, such a generalization is meaningful only
if the symmetry algebra is of the de Sitter type
(e.g. so(2,3) or so(1,4)) but not Poincare one.

A standard requirement to any new theory is that there
should be a correspondence principle with the
existing theory, i.e. there should exist conditions
when the existing theory and the new one give close
predictions. For this reason our first goal is to 
describe a formulation of the standard
quantum theory which can be directly
extended to GFQT. Such a formulation is
discussed in Chap. \ref{C2}. In Chap. \ref{C3}
{\it the standard} anti de Sitter theory of free 
particles is described in detail. We deliberately 
discuss the standard and GFQT theories not in
parallel but separately. We hope that after the
reader gets used to the the standard theory
in our approach, he or she will realize that the
transition to GFQT contains nothing mysterious
and is very natural. As already noted, GFQT can
be treated as the Heisenberg version
to quantum theory but complex numbers are replaced
by a Galois field. In that case the infinite
matrices representing selfadjoint operators in
Hilbert spaces become automatically truncated
such that they represent operators in finite 
dimensional spaces over a Galois field.

The notion of Galois fields appears first in
Chap. \ref{C4} and for reading the present paper, 
only very elementary
knowledge (if any) of such fields is needed. 
Although this notion is extremely simple and 
elegant, the 
majority of physicists is not familiar
with it. For this reason an attempt is made to 
explain the basic facts about
Galois fields in a simplest possible way and using
arguments which, hopefully, can be accepted by
physicists. The readers who are not familiar with
Galois fields can also obtain basic knowledge from
standard textbooks (see e.g. Refs. \cite{VDW}).

A description of elementary particles 
in the anti de Sitter version of GFQT is given
in Chap. \ref{C5}. This chapter is preparatory
for Chap. \ref{C6} where most important new
results of GFQT are described in detail.  

\vfill\eject

\begin{sloppypar}
\chapter{Motivation}
\label{C2}
\end{sloppypar}

\section{Space-time and operator formalism}
\label{S2.1}

The physical meaning of spacetime is one of the main problems
in modern physics. In the standard approach to elementary particle
theory it is assumed from the beginning that there exists
a background spacetime (e.g. Minkowski or de Sitter spacetime),
and the system under consideration is described by local
quantum fields defined on that spacetime. Then by using
Lagrangian formalism and Noether theorem, one can
(at least in principle) construct global quantized operators
(e.g. the four-momentum operator) for the system as a whole.
It is interesting to note that after this stage has been
implemented, one can safely forget about spacetime and
concentrate his or her efforts on calculating S-matrix and
other physical observables.

If we accept the standard operator approach
then, to be consistent, we should assume that {\it any}
physical quantity is described by a selfadjoint operator
in the Hilbert space of states for our system
(we will not discuss the difference between
selfadjoint and Hermitian operators).
Then the first question which immediately arises is that,
even in nonrelativistic quantum mechanics, there is no
operator corresponding to time \cite{time}. It is also
well known that, when quantum mechanics is combined with
relativity, there is no operator satisfying all the
properties of the spatial position operator
(see e.g. Ref. \cite{NW}). For these reasons the
quantity $x$ in the Lagrangian density $L(x)$ is only
a parameter which becomes the coordinate in the
classical limit.

These facts were well known already in 30th of the last
century and became very popular in 60th (recall the
famous Heisenberg S-matrix program). In the first section
of the well-known textbook \cite{BLP} it is claimed that
spacetime and local quantum fields are rudimentary
notions which will disappear in the ultimate quantum theory.
Since that time, no arguments questioning those ideas have
been given, but in view of the great success of gauge
theories in 70th and 80th,
such ideas became almost forgotten.

The problem of whether the empty classical spacetime has a
physical meaning, has been discussed for a long time. In
particular, according to the famous Mach's principle,
the properties of space at a given point depend on the
distribution of masses in the whole Universe. As described
in a wide literature (see e.g. Refs. 
\cite{Mach1,Mach2,Wein2,Mach}
and references therein), Mach's principle was a guiding
one for Einstein in developing general relativity (GR), but
when it has been constructed, it has been realized that it 
does not contain Mach's principle at all! As noted in Refs. 
\cite{Mach1,Mach2,Wein2}, this problem is not closed.

Consider now how one should define the notion of
elementary particles.
Although particles are observable and fields are not,
in the spirit of LQFT, fields are more fundamental
than particles, and a possible definition is as
follows \cite{Wein1}: 'It is simply a particle whose
field appears in the Lagrangian. It does not matter if
it's stable, unstable, heavy, light --- if its field
appears in the Lagrangian then it's elementary,
otherwise it's composite'.

Another approach has been developed by Wigner in his
investigations of unitary irreducible representations
(IRs) of the Poincare group \cite{Wigner}. In view
of this approach, one might postulate that a particle
is elementary if the set
of its wave functions is the space of a unitary IR
of the symmetry group in the given theory
(see also Ref. \cite{JMLL}).

Although in standard well-known theories (QED, electroweak
theory and QCD) the above approaches are equivalent,
the following problem arises. The symmetry
group is usually chosen as a group of motions of
some classical
manifold. How does this agree with the above discussion
that quantum theory in the operator formulation should
not contain spacetime? A possible answer is as follows.
One can notice that for calculating observables (e.g. the
spectrum of the Hamiltonian) we need in fact not a
representation of the group but a representation of its
Lie algebra by Hermitian operators. After such a
representation has been constructed, we have only
operators acting in the Hilbert space and this is all
we need in the operator approach. The representation
operators of the group are needed only if it is
necessary to calculate some macroscopic
transformation, e.g. time evolution. In the approximation
when classical time is a good approximate parameter,
one can calculate evolution, but nothing guarantees
that this is always the case (e.g. at the very early
stage of the Universe). An interesting discussion of
this problem can be found in Ref. \cite{JMLL1}.
Let us also note that in
the stationary formulation of scattering theory, the
S-matrix can be defined without any mentioning of
time (see e.g. Ref. \cite{Kato}). For these reasons
we can assume that on quantum level the symmetry
algebra is more fundamental than the symmetry group.

In other words, instead of saying that some operators
satisfy commutation relations of a Lie algebra
$A$ because spacetime $X$ has a group of motions $G$ such
that $A$ is the Lie algebra of $G$, we say that there
exist operators satisfying  commutation
relations of the Lie algebra $A$ such that: for some
operator functions $\{O\}$ of them the classical
limit is a good approximation, a set $X$ of the eigenvalues
of the operators $\{O\}$ represents a classical manifold with
the group of motions $G$ and its Lie algebra is $A$.
This is not of course in the spirit of famous Klein's Erlangen
program or LQFT.

Consider for illustration the well-known example of
nonrelativistic quantum mechanics. Usually
the existence of the Galilei spacetime is assumed from
the beginning. Let $({\bf r},t)$ be the spacetime coordinates
of a particle in that spacetime. Then
the particle momentum operator is $-i\partial/\partial {\bf r}$
and the Hamiltonian describes evolution by the
Schroedinger equation.
In our approach one starts from an IR of the Galilei algebra.
The momentum operator and the Hamiltonian represent four of ten
generators of such a representation. If it is implemented in
a space of functions
$\psi({\bf p})$ then the momentum operator is simply the
operator of multiplication by ${\bf p}$. Then the
position operator can be {\it defined} as
$i\partial/\partial {\bf p}$
and time can be {\it defined} as an evolution
parameter such that
evolution is described by the Schroedinger equation with
the given Hamiltonian. Mathematically the both approaches are
equivalent since they are related to each other by the Fourier
transform. However, the philosophies behind them are
essentially different. In the second approach there is
no empty spacetime (in the spirit of Mach's principle) and
the spacetime coordinates have a physical meaning only if
there are particles for which the coordinates
can be measured.

Summarizing our discussion, we assume that,
{\it by definition}, on quantum level a Lie algebra is
the symmetry algebra if there exist physical
observables such that their operators
satisfy the commutation relations characterizing the
algebra. Then, a particle is called elementary if the
set of its wave functions is a space of IR of this 
algebra by Hermitian operators.
Such an approach is in the spirit of that considered
by Dirac in Ref. \cite{Dir}. By using the abbreviation
'IR' we will always mean an irreducible representation
by Hermitian operators.

\section{PCT, spin-statistics and all that}
\label{S2.2}

The title of this section is borrowed from that in the
well known book \cite{SW} where the famous results of
the particle theory are derived from LQFT. In this
theory each elementary particle is
described in two ways: i) by using an IR of the Poincare
algebra; ii) by using a local Poincare covariant
equation. For each values of the mass and spin, there exist
two IRs - with positive and negative energies, respectively.
However, the negative energy IRs are not used since
particles and its antiparticles are described only by
positive energy IRs. It is usually assumed that a particle
and its antiparticle are described by their own equivalent
positive energy IRs. In Poincare invariant theories
there exists a possibility that a massless particle and
its antiparticle are described by different IRs. For
example, the massles neutrino can be described by an IR
with the left-handed helicity while the antineutrino ---
by an IR with the right-handed helicity. At the same
time, in the AdS case there exists a nonzero probability
for transitions between the left-handed and right-handed
massless states (see e.g. Ref. \cite{tmf} and Sect. \ref{S3.2}).

A description in terms of a covariant equation seems to be
more fundamental
since one equation describes a particle and its antiparticle
simultaneously. Namely, the negative energy solutions of
the covariant equation are associated with antiparticles
by means of quantization such that the
creation and annihilation operators for the antiparticle
have the usual meaning but they enter the quantum
Lagrangian with the coefficients representing the
negative energy solutions. On the other hand, covariant
equations describe functions defined on a classical
spacetime and for this reason their applicability is
limited to local theories only.

The necessity to have negative energy solutions is
related to the implementation of the idea that the
creation or annihilation of an antiparticle can be treated,
respectively, as the annihilation or creation of the
corresponding particle with the negative energy. However,
since negative energies have no direct physical meaning in
the standard theory, this idea is implemented implicitly
rather than explicitly.

The fact that on the level of Hilbert spaces a particle
and its antiparticle are treated independently,
poses a problem
why they have equal masses, spins and lifetimes. The usual
explanation (see e.g. \cite{AB,BLP,Wein,SW})
is that this is a consequence of CPT invariance.
Therefore if it appears that the masses of a particle and
its antiparticle were not equal, this would indicate the
violation of CPT invariance. In turn, as shown in
well-known works \cite{CPT,SW}, any local Poincare invariant
quantum theory is automatically CPT invariant.

Such an explanation seems to be not quite convincing.
Although at present there are no theories which explain
the existing data better than the standard model based on
LQFT, there is no guarantee that
the ultimate quantum theory will be necessarily local.
The modern theories aiming to unify all the known
interactions (loop quantum gravity, noncommutative
quantum theory, string theory etc.) do not
adopt the exact locality. Therefore each of those
theories should give its own explanation.

Consider a model example when isotopic invariance
is exact (i.e. electromagnetic and weak interactions are
absent). Then the proton and the neutron have equal masses
and spins as a consequence of the fact that they belong
to the same IR of the isotopic algebra. In this example
the proton and the neutron are simply different states
of the same object - the nucleon, and the problem of why
they have equal masses and spins has a natural explanation.

It is clear from this example that in theories where
a particle and its antiparticle are described by the
same IR of the symmetry algebra, the fact that they have
equal masses and spins has a natural explanation. In
such theories the very existence of antiparticles is
inevitable (i.e. the existence of a particle without
its antiparticle is impossible) and a particle and its
antiparticle are different states of the same object.
We will see in Chap. \ref{C5} that in GFQT such
a situation indeed takes place.

Another famous result of LQFT is the Pauli spin-statistics
theorem \cite{Pauli}. After the original Pauli proof,
many authors investigated
more general approaches to the theorem (see e.g. Ref.
\cite{Kuckert} and references therein) but in all the approaches
the locality was used by assuming that particles are described by
covariant equations. Meanwhile, if, for simplicity, we consider
only free particles then all the information about them is known
from the corresponding IR while local covariant wave functions
are needed only for constructing interaction Lagrangian in 
LQFT. From this point view the problem arises whether, at least
in the case of free particles, the spin-statistics theorem can be
proved by using only the properties of the particle IRs. An analogous
problem can be posed about the well known result that the spatial
parity is real for bosons and imaginary for fermions
\cite{AB,BLP,Wein-susy}). We will see in Chap. \ref{C6} that 
since in GFQT one IR
describes a particle and its antiparticle simultaneously, the
spin-statistics theorem and the property of the spatial parity can
be proved only on the level of IRs.

\section{Poincare invariance vs. de Sitter invariance}
\label{S2.3}

As follows from our definition of symmetry on quantum
level, the theory is Poincare invariant if the
representation operators for the system under
consideration satisfy the well-known commutation relations
$$[P^{\mu},P^{\nu}]=0, \quad [M^{\mu\nu}, P^{\rho}]=
-2i(g^{\mu\rho}P^{\nu}-g^{\nu\rho}P^{\mu}),$$
\begin{equation}
[M^{\mu\nu},M^{\rho\sigma}]=-2i (g^{\mu\rho}M^{\nu\sigma}+
g^{\nu \sigma}M^{\mu\rho}-g^{\mu\sigma}M^{\nu\rho}-g^{\nu\rho}
M^{\mu\sigma})
\label{2.1}
\end{equation}
where $\mu,\nu,\rho,\sigma=0,1,2,3$, $P^{\mu}$ are the
four-momentum operators, $M^{\mu\nu}$ are the angular momentum
operators, the metric tensor in Minkowski space has the nonzero
components $g^{00}=-g^{11}=-g^{22}=-g^{33}=1$, and for
further convenience we use the
system of units with $\hbar/2=c=1$. The operators
$M^{\mu\nu}$ are antisymmetric: $M^{\mu\nu}=-M^{\nu\mu}$ and
therefore there are
only six independent angular momentum operators.

The question arises whether Poincare invariant quantum
theory can be a starting point for its generalization
to GFQT. The answer is probably 'no'
and the reason is the following. GFQT is discrete
and finite because the only numbers it can contain are
elements of a Galois field. By analogy with integers,
those numbers can have no dimension and all operators in
GFQT cannot have the continuous spectrum. In the Poincare
invariant quantum theory the angular momentum operators are
dimensionless (if $\hbar/2=c=1$) but the momentum
operators have the dimension of the inverse length.
In addition, the momentum operators and the generators of
the Lorentz boosts $M^{0i}$
$(i=1,2,3)$ contain the continuous spectrum.

This observation might prompt a skeptical reader
to immediately conclude that no GFQT can
describe the nature. However, a
simple way out of this situation is as follows.

\begin{sloppypar}
First we recall the well-known fact that conventional
Poincare invariant theory is a special case of 
de Sitter invariant one. The symmetry algebra of the de Sitter
invariant quantum theory can be either so(2,3) or so(1,4).
The algebra so(2,3) is the Lie algebra of
symmetry group of the four-dimensional manifold in the
five-dimensional space, defined by the equation
\begin{equation}
x_5^2+x_0^2-x_1^2-x_2^2-x_3^2=R^2
\label{2.2}
\end{equation}
where a constant $R$ has the dimension of length.
We use $x_0$ to denote the conventional time coordinate and
$x_5$ to denote the fifth coordinate. The notation $x_5$
rather than $x_4$ is used since in the literature the
latter is sometimes used to denote $ix_0$.
Analogously, so(1,4) is the Lie algebra of the symmetry
group of the four-dimensional manifold in the
five-dimensional space, defined by the equation
\begin{equation}
x_0^2-x_1^2-x_2^2-x_3^2-x_5^2=-R^2
\label{2.3}
\end{equation}
\end{sloppypar}

The quantity $R^2$ is often written as $R^2=3/\Lambda$
where $\Lambda$ is the cosmological constant. The
existing astronomical data show that it is very
small. The nomenclature is such that $\Lambda < 0$
corresponds to the case of Eq. (\ref{2.2}) while
$\Lambda >0$ - to the case of Eq. (\ref{2.3}). In the
literature the latter is often called the
de Sitter (dS) space while the former is called the
anti de Sitter (AdS) one. Analogously, some authors
prefer to call only so(1,4)
as the dS algebra while so(2,3) is called the AdS
one. In view of recent cosmological investigations it
is now believed that $\Lambda > 0$ (see e.g. Ref.
\cite{Perlmutter}).

The both de Sitter algebras are ten-parametric,
as well as the Poincare algebra. However,
in contrast to the Poincare algebra, all
the representation
operators of the de Sitter algebras  are angular
momenta, and in the units $\hbar/2=c=1$ they are
dimensionless. The commutation relations now can be
written in the form of one tensor equation
\begin{equation}
[M^{ab},M^{cd}]=-2i (g^{ac}M^{bd}+g^{bd}M^{cd}-
g^{ad}M^{bc}-g^{bc}M^{ad})
\label{2.4}
\end{equation}
where $a,b,c,d$ take the values 0,1,2,3,5 and the operators
$M^{ab}$ are antisymmetric. The diagonal metric tensor
has the
components $g^{00}=-g^{11}=-g^{22}=-g^{33}=1$ as usual,
while $g^{55} =1$ for the algebra so(2,3) and
$g^{55}=-1$ for the algebra so(1,4).

When $R$ is very large, the transition from the de Sitter
symmetry to Poincare one (this procedure is called
contraction \cite{IW}) is performed as follows. We define
the operators
$P^{\mu} = M^{\mu 5}/2R$. Then, when
$M^{\mu 5}\rightarrow \infty$, $R\rightarrow \infty$, but
their ratio is finite,
Eq. (\ref{2.4}) splits into the set of expressions given by
Eq. (\ref{2.1}).

Note that our definition of the de Sitter symmetry on
quantum level does not involve the cosmological
constant at all. It appears only if
one is interested in interpreting results in terms of
the de Sitter spacetime or in the Poincare limit.
Since all the
operators $M^{ab}$ are dimensionless in units $\hbar/2=c=1$,
the de Sitter invariant quantum theories can be formulated
only in terms of dimensionless variables.
In particular one might expect that the gravitational
and cosmological constants are not fundamental in the
framework of such theories. Mirmovich has proposed a
hypothesis \cite{Mirmovich} that only
quantities having the dimension of the angular momentum
can be fundamental.

If one assumes that spacetime is fundamental then in the
spirit of GR it is natural to think that
the empty space is flat, i.e. that the cosmological
constant is equal to zero. This was the subject of the
well-known dispute between Einstein and de Sitter
described in a wide literature (see e.g. Refs.
\cite{Mach2,Einst-dS} and references therein). In the
LQFT the cosmological constant
is given by a contribution of vacuum diagrams,
and the problem is to explain why it is so small. On the
other hand, if we assume that symmetry on quantum level in
our formulation is more fundamental then 
the cosmological constant problem does not arise at all. 
Instead we
have a problem of why nowadays Poincare symmetry is so
good approximate symmetry. It seems natural to
involve the anthropic principle for the explanation of
this phenomenon (see e.g. Ref. \cite{Linde} and references
therein).

Let us note that there is no continuous transition from
de Sitter invariant to Poincare invariant theories.
If $R$ is finite then we have de Sitter invariance even
if $R$ is very large. Therefore if we accept the de Sitter
symmetry from the beginning, we should accept that energy
and momentum are no longer fundamental physical quantities.
We can use them at certain conditions but only as
approximations.

In local Poincare invariant theories, the energy and
momentum can be written as integrals from energy-momentum
tensor over a space-like hypersurface, while the angular
momentum operators can
be written as integrals from the angular momentum
tensor over the same hypersurface. In local de
Sitter invariant theories all the representation
generators are angular momenta. Therefore the
energy-momentum tensor in that case is not needed at
all and all the representation generators can be written
as integrals from a five-dimensional angular momentum
tensor over some hypersurface.
Meanwhile, in GR and quantum theories in curved spaces
the energy-momentum tensor is used in
situations where spacetime is much
more complicated than the dS or AdS space-time.
It is well-known that Einstein was not
satisfied by the presence of this tensor
in his theory, and this was the subject of numerous
discussions in the literature.

Finally we discuss the differences between
the so(2,3) and so(1,4) invariant theories. There exists
a wide literature devoted to the both of them.

The former has many features analogous to those in
Poincare invariant theories. There exist
IRs of the AdS algebra where the operator $M^{05}$,
which is the de Sitter analog of the energy operator,
is bounded below by some positive value which can be
treated as the AdS analog of the mass (see also
Chap. \ref{C3}). With such an interpretation, a
system of free particles with the AdS masses
$m_1,m_2...m_n$ has the minimum energy equal
to $m_1+m_2+...m_n$. There also exist IRs with negative
energies where the energy operator is bounded above.
The so(2,3) invariant theories can be easily generalized
if supersymmetry is required while supersymmetric
generalizations of the so(1,4) invariant theories is
problematic.

On the contrary, the latter has many unusual features. For
example, even in IRs, the operator $M_{05}$ (which is
the SO(1,4) analog of the energy operator) has the
spectrum in the interval $(-\infty,+\infty)$. The dS mass
operator of the system of free particles with the dS
masses  $m_1,m_2...m_n$  is not bounded below by
the value of  $m_1+m_2+...m_n$ and also has the spectrum
in the interval $(-\infty,+\infty)$ \cite{lev1,lev2}.

For these reasons there existed an opinion
that de Sitter invariant quantum theories can be based
only on the algebra so(2,3) and not so(1,4).
However, as noted above, in view of recent cosmological
observations
it is now believed that $\Lambda > 0$, and the opinion
has been changed. As noted in our works
\cite{lev1,lev2,lev3,jpa} (done before these developments),
so(1,4) invariant theories have many
interesting properties, and the fact that they are
unusual does not mean that they contradict experiment.
For example, a particle has only an
exponentially small probability to
have the energy less than its mass (see e.g. Ref.
\cite{lev2}). The fact that the mass operator is not
bounded below by  the value of  $m_1+m_2+...m_n$ poses an
interesting question whether some interactions
(even gravity) can be a direct consequence of
the dS symmetry \cite{lev2,lev3}. A very interesting
property of the so(1,4) invariant theories is that they
do not contain bound states at all and, as a consequence,
the free and interacting operators are unitarily
equivalent \cite{lev3}. This poses the problem
whether the notion of interactions is needed at all.
Finally, as shown in Ref. \cite{jpa}, each
IR of the so(1,4) algebra cannot be interpreted in a
standard way but describe a particle and its antiparticle
simultaneously.

As argued by Witten \cite{Witten}, the Hilbert space
for quantum gravity in the
dS space is finite-dimensional and therefore there are
problems in quantizing General Relativity in that
space (because any unitary
representation of the SO(1,4) group is necessarily
infinite-dimensional).
However this difficulty disappears in GFQT since
linear spaces in this theory are finite-dimensional.

The main goal of the present paper is to convince
the reader that GFQT is a more natural quantum
theory than the standard one based on complex
numbers. For this reason (in view of the above remarks)
we consider below only the so(2,3) version of GFQT.
The problem of the Galois field generalization of
so(1,4) invariant theories has been discussed in
Ref. \cite{lev2,gravity}.

\vfill\eject

\chapter{Elementary particles in standard so(2,3)
invariant theory}
\label{C3}

\section{IRs of the sp(2) algebra}
\label{S3.1}

The key role in constructing IRs of the so(2,3)
algebra is played by IRs of the sp(2) subalgebra.
They are described by a set of operators $(a',a",h)$
satisfying the commutation relations
\begin{equation}
[h,a']=-2a'\quad [h,a"]=2a"\quad [a',a"]=h
\label{3.1}
\end{equation}
and the Hermiticity conditions $a^{'*}=a^"$ and $h^*=h$.
As usual, if $A$ is an operator in the Hilbert space
under consideration then $A^*$ is used to denote the
operator adjoined to $A$. 

The  Casimir operator of the second order for the algebra
(\ref{3.1}) has the form
\begin{equation}
K=h^2-2h-4a"a'=h^2+2h-4a'a"
\label{3.2}
\end{equation}
We first consider representations with the vector $e_0$,
such that
\begin{equation}
a'e_0=0,\quad he_0=q_0e_0,\quad (e_0,e_0)=1,\quad q_0>0
\label{3.3}
\end{equation}
where we use $(...,...)$ to denote the scalar product in
the representation space. Denote $e_n =(a")^ne_0$.
Then it follows from Eqs. (\ref{3.2}) and (\ref{3.3}),
that for any $n=0,1,2,...$
\begin{equation}
he_n=(q_0+2n)e_n,\quad Ke_n=q_0(q_0-2)e_n,
\label{3.4}
\end{equation}
\begin{equation}
a'a"e_n=(n+1)(q_0+n)e_n
\label{3.5}
\end{equation}
\begin{equation}
(e_{n+1},e_{n+1})=(n+1)(q_0+n)(e_n,e_n)
\label{3.6}
\end{equation}

The elements $e_n$ form a basis of the IR where
$e_0$ is a vector with a minimum eigenvalue of the
operator $h$ (minimum weight) and there are no
vectors with the maximum weight.
This is in agreement with the well known fact that
IRs of noncompact algebras are infinite dimensional.
It is clear that the operator $h$ is positive definite
and bounded below by the quantity $q_0$. The above
UIRs are used for constructing positive energy
UIRs of the so(2,3) algebra (see the next section).

Analogously, one can construct UIRs starting from
the element $e_0'$ such that
\begin{equation}
a"e_0'=0,\quad he_0'=-q_0e_0',\quad
(e_0',e_0')=1,\quad q_0> 0
\label{3.7}
\end{equation}
and the elements $e_n'$ can be defined as
$e_n'=(a')^ne_0'$.
Then $e_0'$ is the vector with the maximum weight,
there are no vectors with the minimum weight, the
operator $h$ is negative definite and bounded above
by the quantity $-q_0$. Such IRs are used for
constructing negative energy IRs of the so(2,3)
algebra.

\section{IRs of the so(2,3) algebra}
\label{S3.2}

IRs of the so(2,3) algebra relevant for describing
elementary particles have been considered by many
authors. We need a description which can be directly
generalized to the case of Galois field. Our description
in this section is a combination of two elegant ones
given in Ref. \cite{Evans} for standard IRs and in Ref.
\cite{Braden} for IRs over a Galois field.

As noted in Sect. \ref{S2.3}, the representation
operators of the so(2,3) algebra in units
$\hbar/2=c=1$ are given by Eq. (\ref{2.4}).
In these units the spin of fermions is odd, and
the spin of bosons is even. If $s$ is the particle spin
then the corresponding IR of the su(2) algebra has
the dimension $s+1$. Note that if $s$ is interpreted
in such a way then it does not depend on the choice of
units (in contrast to the maximum eigenvalue of the
$z$ projection of the spin operator).

For analyzing IRs implementing Eq. (\ref{2.4}), it is
convenient to work with another set of ten operators.
Let $(a_j',a_j",h_j)$ $(j=1,2)$ be two independent sets
of operators satisfying the
commutation relations for the sp(2) algebra
\begin{equation}
[h_j,a_j']=-2a_j'\quad [h_j,a_j"]=2a_j"\quad [a_j',a_j"]=h_j
\label{3.8}
\end{equation}
The sets are independent in the sense that
for different $j$ they mutually commute with each other.
We denote additional four operators as $b', b",L_+,L_-$.
The meaning of $L_+,L_-$ is as follows. The operators
$L_3=h_1-h_2,L_+,L_-$ satisfy the commutation relations
of the su(2) algebra
\begin{equation}
[L_3,L_+]=2L_+\quad [L_3,L_-]=-2L_-\quad [L_+,L_-]=L_3
\label{3.9}
\end{equation}
while the other commutation relations are as follows
\begin{eqnarray}
&[a_1',b']=[a_2',b']=[a_1",b"]=[a_2",b"]=\nonumber\\
&[a_1',L_-]=[a_1",L_+]=[a_2',L_+]=[a_2",L_-]=0\nonumber\\
&[h_j,b']=-b'\quad [h_j,b"]=b"\quad
[h_1,L_{\pm}]=\pm L_{\pm},\nonumber\\
&[h_2,L_{\pm}]=\mp L_{\pm}\quad [b',b"]=h_1+h_2\nonumber\\
&[b',L_-]=2a_1'\quad [b',L_+]=2a_2'\quad [b",L_-]=-2a_2"\nonumber\\
&[b",L_+]=-2a_1",\quad [a_1',b"]=[b',a_2"]=L_-\nonumber\\
&[a_2',b"]=[b',a_1"]=L_+,\quad [a_1',L_+]=[a_2',L_-]=b'\nonumber\\
&[a_2",L_+]=[a_1",L_-]=-b"
\label{3.10}
\end{eqnarray}
At first glance these relations might seem rather
chaotic but in fact they are very natural in the Weyl basis
of the so(2,3) algebra.

The relation between the above sets of ten operators is
as follows
\begin{eqnarray}
&M_{10}=i(a_1"-a_1'-a_2"+a_2')\quad M_{15}=a_2"+a_2'-a_1"-a_1'\nonumber\\
&M_{20}=a_1"+a_2"+a_1'+a_2'\quad M_{25}=i(a_1"+a_2"-a_1'-a_2')\nonumber\\
&M_{12}=L_3\quad M_{23}=L_++L_-\quad M_{31}=-i(L_+-L_-)\nonumber\\
&M_{05}=h_1+h_2\quad M_{35}=b'+b"\quad M_{30}=-i(b"-b')
\label{3.11}
\end{eqnarray}
In addition, if
$L_+^*=L_-$, $a_j^{'*}=a_j"$, $b^{'*}=b"$ and $h_j^*=h_j$ then
the operators $M^{ab}$ are Hermitian.

We use the basis in which the operators
$(h_j,K_j)$ $(j=1,2)$ are diagonal. Here $K_j$ is the
Casimir operator (\ref{3.2}) for algebra $(a_j',a_j",h_j)$.
For constructing IRs we need operators relating different
representations of the sp(2)$\times$sp(2) algebra.
By analogy with Refs. \cite{Evans,Braden}, one of the
possible choices is as follows
\begin{eqnarray}
&A^{++}=b"(h_1-1)(h_2-1)-a_1"L_-(h_2-1)-a_2"L_+(h_1-1)
+\nonumber\\
&a_1"a_2"b' \quad A^{+-}=L_+(h_1-1)-a_1"b'\nonumber\\
&A^{-+}=L_-(h_2-1)-a_2"b'\quad A^{--}=b'
\label{3.12}
\end{eqnarray}
As noted in Ref. \cite{lev2}, such a choice has several
advantages and one of them is that
\begin{equation}
[A^{++},A^{+-}]=[A^{++},A^{-+}]=[A^{+-},A^{--}]=[A^{-+},A^{--}]=0
\label{3.13}
\end{equation}
On the other hand, such a choice
is not convenient in the massless case. The description of massles
particles in GFQT has been discussed
in Ref. \cite{tmf} and we will mention the main results a bit later.

We consider the action of these operators only on the
space of "minimal"
sp(2)$\times$sp(2) vectors, i.e. such vectors $x$ that
$a_j'x=0$ for $j=1,2$, and $x$ is the eigenvector of the
operators $h_j$. It is easy to see that if $x$ is a minimal
vector such that
$h_jx=\alpha_jx$ then $A^{++}x$ is the minimal
eigenvector of the
operators $h_j$ with the eigenvalues $\alpha_j+1$, $A^{+-}x$ -
with the eigenvalues $(\alpha_1+1,\alpha_2-1)$,
$A^{-+}x$ - with the eigenvalues $(\alpha_1-1,\alpha_2+1)$,
and $A^{--}x$ - with the eigenvalues $\alpha_j-1$.

By analogy with Refs. \cite{Evans,Braden}, we require
the existence of the vector $e_0$ satisfying the conditions
\begin{eqnarray}
&a_j'e_0=b'e_0=L_+e_0=0\quad h_je_0=q_je_0\nonumber\\
&(e_0,e_0)\neq 0\quad (j=1,2)
\label{3.14}
\end{eqnarray}

As noted in Sect. \ref{S2.3}, $M^{05}=h_1+h_2$ is the AdS
analog of the energy operator, since $M^{05}/2R$ becomes
the usual energy when the AdS algebra is contracted to
the Poincare one. As follows from
Eqs. (\ref{3.8}) and (\ref{3.12}), the operators
$(a_1',a_2',b')$ reduce the AdS energy by two units.
Therefore $e_0$ is the state with the minimum energy
which can be called the rest state, and the spin in our
units is equal to the maximum value
of the operator $L_3=h_1-h_2$ in that state. For these
reasons we use $s$ to denote $q_1-q_2$
and $m$ to denote $q_1+q_2$. Therefore the quantity
$q_1-q_2$ should always be a natural number (we will
always treat zero as a natural number).
In the standard classification
\cite{Evans}, the massive case is characterized by
the condition $q_2>1$ and massless one --- by
the condition $q_2=1$. There also exist two
exceptional IRs discovered by Dirac \cite{DiracS}
(Dirac singletons). They are characterized by the
values of $(q_1,q_2)$ equal to (1/2,1/2) and (3/2,1/2).

As follows from the above remarks, the elements
\begin{equation}
e_{nk}=(A^{++})^n(A^{-+})^ke_0
\label{3.15}
\end{equation}
represent the minimal sp(2)$\times$sp(2) vectors with the
eigenvalues of the operators $h_1$ and $h_2$ equal to
$Q_1(n,k)=q_1+n-k$ and $Q_2(n,k)=q_2+n+k$, respectively.
It can be shown by a direct calculation that
\begin{equation}
A^{--}A^{++}e_{nk}=(n+1)(m+n-2)(q_1+n)(q_2+n-1)e_{nk}
\label{3.16}
\end{equation}
\begin{eqnarray}
&(e_{n+1,k},e_{n+1,k})=(q_1+n-k-1)(q_2+n-k-1)(n+1)\nonumber\\
&\times (m+n-2)(q_1+n)(q_2+n-1)(e_{nk},e_{nk})
\label{3.17}
\end{eqnarray}
\begin{equation}
A^{+-}A^{-+}e_{nk}=(k+1)(s-k)(q_1-k-2)(q_2+k-1)e_{nk}
\label{3.18}
\end{equation}
\begin{eqnarray}
&(e_{n,k+1},e_{n,k+1})=(q_2+n+k-1)(q_1-k-2)(q_2+k-1)\nonumber\\
&\times (k+1)(s-k)(e_{nk},e_{nk})/(q_1+n-k-2)
\label{3.19}
\end{eqnarray}

In the massive case, as follows from Eqs. (\ref{3.18}) and
(\ref{3.19}), $k$ can assume only the values $0,1,...s$
while as follows from Eqs. (\ref{3.16}) and
(\ref{3.17}), $n=0,1,...\infty$. The same expressions
also show that in the singleton case the possible
combinations of $(n,k)$ are $(0,0)$ and $(1,0)$ for
$q_1=q_2=1/2$, and $(0,0)$ and $(0,1)$ for
$q_1=3/2\,\,q_2=1/2$.

In the massless case some matrix elements contain ambiguities
0/0 which should be resolved. The solution is described e.g. in
Refs. \cite{Evans,tmf}. The quantity $n$ can take the values
of $0,1,...\infty$ only if $k=0$ or $k=s$ while if $0<k<s$
then only $n=0$ is possible. This is an analog of the situation
in Poincare invariant theory where massless particles are
described by IRs characterizing by helicity which can be either
$s$ or $-s$. However, in the AdS case one IR describes the
particles with both helicities. Since in the AdS case the rest
states characterized by $n=0$ do not have measure zero (as in
Poincare invariant theory), there exists a nonzero probability
for transitions between the helicities $s$ and $-s$. Note that
$0<k<s$ is possible only if $s\geq 2$ (i.e. $s\geq 1$ in the
usual units) and some authors treat only this case as massless.
We will treat IR as massless only if $q_2=1$. In that case the
quantities $(q_1,q_2)$ take the values of $(1+s,1)$ and $m=2+s$.

The full basis of the representation space can be chosen
in the form
\begin{equation}
e(n_1n_2nk)=(a_1")^{n_1}(a_2")^{n_2}e_{nk}
\label{3.20}
\end{equation}
where, as follows from the
results of the preceding section,
$n_1$ and $n_2$ can be any natural numbers.

As follows from Eqs. (\ref{3.6}) and (\ref{3.20}),
the quantity
\begin{equation}
Norm(n_1n_2nk)=(e(n_1n_2nk),e(n_1n_2nk))
\label{3.21}
\end{equation}
can be represented as
\begin{equation}
Norm(n_1n_2nk)=F(n_1n_2nk)G(nk)
\label{3.22}
\end{equation}
where
\begin{eqnarray}
&F(n_1n_2nk)= n_1!(Q_1(n,k)+n_1-1)!n_2!(Q_2(n,k)+n_2-1)!\nonumber\\
&G(nk)=\{(q_2+k-2)!n!(m+n-3)!(q_1+n-1)!\nonumber\\
&(q_2+n-2)!k!s!\}\{(q_1-k-2)![(q_2-2)!]^3(q_1-1)!\nonumber\\
&(m-3)!(s-k)![Q_1(n,k)-1][Q_2(n,k)-1]\}^{-1}
\label{3.23}
\end{eqnarray}
Note that we deliberately do not normalize the basis
vectors to one since the above expressions can be
directly used in the GFQT.

In standard Poincare and AdS theories there also exist IRs with
negative energies (see the discussion in Sects. \ref{S2.2}
and \ref{S2.3}).
They can be constructed by analogy with positive energy IRs.
Instead of Eq. (\ref{3.14}) one can require the existence of the
vector $e_0'$ such that
\begin{eqnarray}
&a_j"e_0'=b"e_0'=L_-e_0'=0\quad h_je_0'=-q_je_0'\nonumber\\
&(e_0',e_0')\neq 0\quad (j=1,2)
\label{3.24}
\end{eqnarray}
where the quantities $q_1,q_2$ are the same as for positive
energy UIRs. It is obvious that positive and negative energy
IRs are fully independent since the spectrum of the operator
$M^{05}$ for such UIRs is positive and negative, respectively.
As noted in Sect. \ref{S2.2}, the negative energy IRs are
not used in the standard approach.

\section{Matrix elements of representation operators}
\label{S3.3}

The matrix elements of the operator $M^{ab}$ are defined as
\begin{equation}
M^{ab}e(n_1n_2nk)=\sum_{n_1'n_2'n'k'}
M^{ab}(n_1'n_2'n'k';n_1n_2nk)e(n_1'n_2'n'k')
\label{3.25}
\end{equation}
where the sum is taken over all possible values of
$(n_1',n_2',n',k')$.

Consider first the diagonal operators $h_1$ and $h_2$.
As follows from Eqs. (\ref{3.15}), (\ref{3.20}) and
(\ref{3.4})
\begin{eqnarray}
&h_1e(n_1n_2nk)=[Q_1(n,k)+2n_1]e(n_1n_2nk)\nonumber\\
&h_2e(n_1n_2nk)=[Q_2(n,k)+2n_2]e(n_1n_2nk)
\label{3.26}
\end{eqnarray}
It now follows from Eq. (\ref{3.11}) that
\begin{eqnarray}
&M_{05}e(n_1n_2nk)=[m+2(n+n_1+n_2)]e(n_1n_2nk)\nonumber\\
&L_3e(n_1n_2nk)=[s+2(n_1-n_2-k)]e(n_1n_2nk)
\label{3.27}
\end{eqnarray}
As follows from Eqs. (\ref{3.15}), (\ref{3.20}) and
(\ref{3.5})
\begin{eqnarray}
&a_1'e(n_1n_2nk)=n_1[Q_1(n,k)+n_1-1]e(n_1-1,n_2nk)\nonumber\\
&a_1"e(n_1n_2nk)=e(n_1+1,n_2nk)\nonumber\\
&a_2'e(n_1n_2nk)=n_2[Q_2(n,k)+n_2-1]e(n_1,n_2-1,nk)\nonumber\\
&a_2"e(n_1n_2nk)=e(n_1,n_2+1,nk)
\label{3.28}
\end{eqnarray}
We will always use a convention that $e(n_1n_2nk)$ is a null
vector if some of the numbers $(n_1n_2nk)$ are not in the 
range described above.  

Consider the operator $b"$. Since it commutes
with $a_1"$ and $a_2"$ (see Eq. (\ref{3.10})), it follows
from Eq. (\ref{3.20}) that
\begin{equation}
b"e(n_1n_2nk)=(a_1")^{n_1}(a_2")^{n_2}b"e_{nk}
\label{3.29}
\end{equation}
and, as follows from Eq. (\ref{3.12})
\begin{eqnarray}
&b"e_{nk}=\{[Q_1(n,k)-1][Q_2(n,k)-1]\}^{-1}(A^{++}+a_1"A^{-+}+\nonumber\\
&a_2"A^{+-}+a_1"a_2"A^{--})e_{nk}
\label{3.30}
\end{eqnarray}
By using this expression and Eqs. (\ref{3.13}),
(\ref{3.15}), (\ref{3.16}, (\ref{3.18}) and (\ref{3.30}),
one gets
\begin{eqnarray}
&b"e(n_1n_2nk)=\{[Q_1(n,k)-1][Q_2(n,k)-1]\}^{-1}\nonumber\\
&[k(s+1-k)(q_1-k-1)(q_2+k-2)e(n_1,n_2+1,n,k-1)+\nonumber\\
&n(m+n-3)(q_1+n-1)(q_2+n-2)e(n_1+1,n_2+1,n-1,k)+\nonumber\\
&e(n_1,n_2,n+1,k)+e(n_1+1,n_2,n,k+1)]
\label{3.31}
\end{eqnarray}

Consider now the operator $b'$. We first use the
commutation relations (\ref{3.10}) and obtain
\begin{eqnarray}
&b'e(n_1n_2nk)=b'(a_1")^{n_1}(a_2")^{n_2}e_{nk}=
(a_1")^{n_1}b'(a_2")^{n_2}e_{nk}+\nonumber\\
&n_1(a_1")^{n_1-1}L_+(a_2")^{n_2}e_{nk}=
(a_1")^{n_1}(a_2")^{n_2}b'e_{nk}+\nonumber\\
&n_2(a_1")^{n_1}(a_2")^{n_2-1}L_-e_{nk}+
n_1(a_1")^{n_1-1}(a_2")^{n_2}L_+e_{nk}+\nonumber\\
&n_1n_2(a_1")^{n_1-1}(a_2")^{n_2-1}b"e_{nk}
\label{3.32}
\end{eqnarray}
The rest of the calculations is analogous to that
for the operator $b"$ and the result is
\begin{eqnarray}
&b'e(n_1n_2nk)=\{[Q_1(n,k)-1][Q_2(n,k)-1]\}^{-1}
[n(m+n-3)\nonumber\\
&(q_1+n-1)(q_2+n-2)(q_1+n-k+n_1-1)(q_2+n+k+n_2-1)\nonumber\\
&e(n_1n_2,n-1,k)+n_2(q_1+n-k+n_1-1)e(n_1,n_2-1,n,k+1)+\nonumber\\
&n_1(q_2+n+k+n_2-1)k(s+1-k)(q_1-k-1)(q_2+k-2)\nonumber\\
&e(n_1-1,n_2,n,k-1)+n_1n_2e(n_1-1,n_2-1,n+1,k)]
\label{3.33}
\end{eqnarray}

By analogy with the above calculations one can obtain
\begin{eqnarray}
&L_+e(n_1n_2nk)=\{[Q_1(n,k)-1][Q_2(n,k)-1]\}^{-1}
\{(q_2+n+k+\nonumber\\
&n_2-1)[k(s+1-k)(q_1-k-1)(q_2+k-2)e(n_1n_2n,k-1)+\nonumber\\
&n(m+n-3)(q_1+n-1)(q_2+n-2)e(n_1+1,n_2,n-1,k)]+\nonumber\\
&n_2[e(n_1,n_2-1,n+1,k)+e(n_1+1,n_2-1,n,k+1)]\}
\label{3.34}
\end{eqnarray}
\begin{eqnarray}
&L_-e(n_1n_2nk)=\{[Q_1(n,k)-1][Q_2(n,k)-1]\}^{-1}
\{n_1[k(s+1-k)\nonumber\\
&(q_1-k-1)(q_2+k-2)e(n_1-1,n_2n,k-1)+e(n_1-1,n_2,\nonumber\\
&n+1,k)]+(q_1+n-k+n_1-1)[e(n_1n_2n,k+1)+n(m+n-3)\nonumber\\
&(q_1+n-1)(q_2+n-2)e(n_1,n_2+1,n-1,k)]\}
\label{3.35}
\end{eqnarray}

\section{Quantization and discrete symmetries}
\label{S3.4}

In the standard approach, the operators of
physical quantities act in the Fock space of the
given system. Suppose that the system consists of free
particles and their antiparticles. Then to define the
Fock space we have to define first the annihilation and
creation operators for them.

As already noted, in the
standard approach it is assumed that a particle
and its antiparticle
are described by equivalent UIRs of the symmetry algebra
with positive energies. Then, as follows from the above assumption,
$(n_1n_2nk)$ is the complete set of quantum numbers for
both the particle and its antiparticle.
Let $a(n_1n_2nk)$ be the
operator of particle annihilation in the state described
by the vector $e(n_1n_2nk)$. Then the adjoint operator
$a(n_1n_2nk)^*$ has the meaning of particle creation in
that state. Since we do not normalize the states
$e(n_1n_2nk)$ to one, we require that the operators
$a(n_1n_2nk)$ and $a(n_1n_2nk)^*$ should satisfy
either the anticommutation relations
\begin{eqnarray}
&\{a(n_1n_2nk),a(n_1'n_2'n'k')^*\}=\nonumber\\
&Norm(n_1n_2nk)
\delta_{n_1n_1'}\delta_{n_2n_2'}\delta_{nn'}\delta_{kk'}
\label{3.36}
\end{eqnarray}
or the commutation relations
\begin{eqnarray}
&[a(n_1n_2nk),a(n_1'n_2'n'k')^*]=\nonumber\\
&Norm(n_1n_2nk)
\delta_{n_1n_1'}\delta_{n_2n_2'}\delta_{nn'}\delta_{kk'}
\label{3.37}
\end{eqnarray}
Analogously, if $b(n_1n_2nk)$ and $b(n_1n_2nk)^*$ are
operators of the antiparticle annihilation and creation
in the state $e(n_1n_2nk)$ then
\begin{eqnarray}
&\{b(n_1n_2nk),b(n_1'n_2'n'k')^*\}=\nonumber\\
&Norm(n_1n_2nk)
\delta_{n_1n_1'}\delta_{n_2n_2'}\delta_{nn'}\delta_{kk'}
\label{3.38}
\end{eqnarray}
in the case of anticommutation relations and
\begin{eqnarray}
&[b(n_1n_2nk),b(n_1'n_2'n'k')^*]=\nonumber\\
&Norm(n_1n_2nk)
\delta_{n_1n_1'}\delta_{n_2n_2'}\delta_{nn'}\delta_{kk'}
\label{3.39}
\end{eqnarray}
in the case of commutation relations.
We also assume that in the case of anticommutation
relations all the operators $(a,a^*)$ anticommute with
all the operators $(b,b^*)$ while in the case of
commutation relations they commute with each other. It is
also assumed that the Fock space contains the vacuum vector
$\Phi_0$ such that
\begin{equation}
a(n_1n_2nk)\Phi_0=b(n_1n_2nk)\Phi_0=0\quad \forall n_1,n_2,n,k
\label{3.40}
\end{equation}

The Fock space can now be defined as a linear combination
of all elements obtained by the action of the operators
$(a^*,b^*)$ on the vacuum vector, and the problem of
second quantization of representation operators can
now be formulated as follows. Let $(A_1,A_2....A_n)$
be representation
operators describing IR of the AdS algebra. One should
replace them by operators acting in the Fock space
such that the commutation relations between their
images in the Fock space are the same as for original
operators (in other words, we should have a homomorphism
of Lie algebras of operators acting in the space of IR
and in the Fock space). We can also require that our
map should be compatible with the Hermitian
conjugation in both spaces. It is easy to verify that
a possible solution satisfying all the requirements is
as follows. Taking into account the fact that the
matrix elements satisfy the proper commutation relations,
the operators $M^{ab}$ in the quantized form
\begin{eqnarray}
&M^{ab}=\sum M^{ab}(n_1'n_2'n'k';n_1n_2nk)
[a(n_1'n_2'n'k')^*a(n_1n_2nk)+\nonumber\\
&b(n_1'n_2'n'k')^*b(n_1n_2nk)]/Norm(n_1n_2nk)
\label{3.41}
\end{eqnarray}
satisfy the commutation relations in the form
(\ref{2.4}) or (\ref{3.8}-\ref{3.10}). We will not use
special notations for operators in the Fock space since
in each case it will be clear whether the operator in
question acts in the space of IR or in the Fock space.

\begin{sloppypar}
A well known problem in the standard theory is that
the quantization procedure does not define the order
of the annihilation and creation operators uniquely.
For example, another possible solution is
\begin{eqnarray}
&M^{ab}=\mp \sum M^{ab}(n_1'n_2'n'k';n_1n_2nk)
[a(n_1n_2nk)a(n_1'n_2'n'k')^*+\nonumber\\
&b(n_1n_2nk)b(n_1'n_2'n'k')^*]/Norm(n_1n_2nk)
\label{3.42}
\end{eqnarray}
for the cases of anticommutation and commutation
relations, respectively. It is clear that the 
solutions (\ref{3.41}) and
(\ref{3.42}) are different since the energy operators
$M^{05}$ in these expressions differ by an infinite
constant. In the standard theory the solution 
(\ref{3.41}) is selected by imposing an additional
requirement that all operators should be written 
in the normal form where annihilation operators 
should always precede creation ones. Then the vacuum 
has zero energy
and Eq. (\ref{3.42}) should be rejected. Such a
requirement does not follow from the
theory. Ideally there should be a procedure which 
correctly defines the order of operators from first
principles. We will see
in Sect. \ref{S5.3} that in GFQT the requirement about
the normal form is no needed since the analogs of 
Eqs. (\ref{3.41}) and (\ref{3.42}) are the same.
\end{sloppypar}

In the standard theory the charge conjugation is
defined as a unitary transformation such that
\begin{equation}
a(n_1n_2nk)\rightarrow \eta_C b(n_1n_2nk)\quad
a(n_1n_2nk)^* \rightarrow \eta_C^* b(n_1n_2nk)^*
\label{3.43}
\end{equation}
where $\eta_C$ is the charge parity such that
$|\eta_C|$ =1. It is obvious from Eq. (\ref{3.41})
that all the representation operators 
are invariant under the
transformation (\ref{3.43}) (C invariant). In fact, Eq.
(\ref{3.43}) is simply
another form of requiring that a particle and its
antiparticle are described by equivalent IRs.

In the standard theory there also exist neutral
particles. In that case there is no need to have two
independent sets of operators $(a,a^*)$ and $(b,b^*)$,
and Eq. (\ref{3.41}) should be written without the
$(b,b^*)$ operators. The problem of neutral particles
in GFQT is discussed in Sect. \ref{S6.3}.

Consider now the space inversion. In terms of the
representation operators it is defined as a
transformation $M_{ab}\rightarrow M_{ab}^P$ such that
\begin{eqnarray}
&M_{ik}^P=M_{ik}\quad M_{i0}^P=-M_{i0}\quad
M_{i5}^P=-M_{i5}\nonumber\\
&M_{05}^P=M_{05} \quad (i,k=1,2,3)
\label{3.44}
\end{eqnarray}
We have to find the transformation of the operators
$(a,a^*)$ and $(b,b^*)$ resulting in Eq. (\ref{3.44}).
A possible solution is as follows
\begin{eqnarray}
&a(n_1n_2nk)^*\rightarrow (-1)^{n_1+n_2+n}\eta_P^*
a(n_1n_2nk)^*\nonumber\\
&a(n_1n_2nk)\rightarrow (-1)^{n_1+n_2+n}\eta_P
a(n_1n_2nk)
\label{3.45}
\end{eqnarray}
and analogously for the $(b,b^*)$ operators. Here
$\eta_P$ is the spatial parity such that $|\eta_P|=1$.
As follows from C invariance, the spacial parities of
a particle and its antiparticles are the same.

The fact that Eq. (\ref{3.44}) is a consequence of
Eq. (\ref{3.45}) follows from the results of the
preceding section. Indeed, the expressions for matrix
elements of the representation operators show that the
operators $(a_j',a_j",b',b")$ ($j=1,2$) have nonzero
matrix elements only for transitions where the values
of $n_1+n_2+n$ differ by $\pm 1$. Therefore, as follows
from Eq. (\ref{3.42}), the quantized version of
those operators change their
sign under the transformation (\ref{3.45}). At the
same time, the operators $L_{\pm}$ and $h_j$ have
nonzero matrix elements only for transitions where
the values of $n_1+n_2+n$ in the initial and
final states are the same. 

In the standard theory \cite{AB,BLP,Wein-susy}
the result that $\eta_P$ is real for bosons and imaginary for
fermions can be obtained from covariant equations while no
limitation on this quantity can be obtained from the consideration
on the level of IRs.

\begin{sloppypar}
As already noted, in the standard theory only the CPT
invariance is fundamental. A well known problem with
the definition of this invariance is that if time
inversion is defined by analogy with space inversion
then the Hamiltonian will change its sign. There exist
two well known solution for avoiding this undesirable
behavior \cite{AB,BLP,Wein}: the Wigner solution
involves antiunitary operators and the Schwinger
solution involves transposed operators. Consider,
for example the Schwinger solution. If $A$ is an
operator then $A^T$ will denote the operator tranposed
to $A$. Then the CPT transformation in the Schwinger
formulation implies
\begin{equation}
P_{\mu}\rightarrow P_{\mu}^T\quad
M_{\mu\nu}\rightarrow -M_{\mu\nu}^T\quad
(\mu,\nu = 0,1,2,3)
\label{3.46}
\end{equation}
The operators obtained in such a way obviously
satisfy Eq. (\ref{2.1}) if the original operators
satisfy these expressions.
\end{sloppypar}

In the so(2,3) case an analogous transformation is
as follows. We first replace each of the
$M_{a5}$ operators by $-M_{a5}$ and then
each $M_{ab}$ operator by $-M_{ab}^T$. It is obvious
that these transformations preserve the commutation
relations (\ref{2.4}) and in the Poincare approximation
their result coincides with Eq. (\ref{3.46}). At the same
time, in the so(2,3) case the $M_{ab}$ operators with
$a=5$ or $b=5$ are on the same footing as the
the $M_{ab}$ operators with $a=0$ or $b=0$. For this
reason such an analog of the CPT transformation in the
so(2,3) theory is not so fundamental as in the Poincare
invariant theory. We will see below that in the GFQT
there exists a transformation which resembles some
features of the CPT one but at the same time has no
analog in the standard theory.

\section{Quantum numbers in Poincare approximation}
\label{S3.5}

The above description of elementary particles in the
so(2,3) invariant theories confirms the statement in
Sect. \ref{S2.3} that such theories in the operator
formalism do not involve the quantity $R$ or the
cosmological constant at all. However, these
quantities are
needed to describe the transition from the so(2,3)
to Poincare invariant theory. As noted in Sect.
\ref{S2.3}, from the formal point of view the Poincare
invariant theory can be treated as a special case
of the so(2,3) invariant one in the limit
$R\rightarrow \infty$.

The following important observation is now in order.
If we assume that the de Sitter symmetry is more
fundamental than
the Poincare one then the limit $R\rightarrow\infty$
{\it should not} be actually taken since in this case
the de Sitter symmetry will be lost and the preceding
consideration
will become useless. However, in the framework of
de Sitter invariant theories we can consider Poincare
invariance as the approximate symmetry when $R$ is very
large but $R\neq\infty$. The situation is analogous to
that when nonrelativistic theory is formally
treated as a special case of relativistic one in
the limit $c\rightarrow\infty$ and classical theory
is treated as a special case of quantum one in
the limit $\hbar\rightarrow 0$ but the limits are
not actually taken. Summarizing these remarks, we prefer
the term 'Poincare approximation' rather than
'Poincare limit'. {\it The term 'Poincare approximation'
will always imply that $R$ is very large but finite.}

As noted in Sect. \ref{S2.3}, the Poincare approximation
is meaningful if there exists a subset of states for which
the operators $M_{\mu 5}$ are much greater than the
other operators. Then we can define the standard
four-momentum operators as $P_{\mu}=M_{\mu 5}/2R$, and
they will satisfy Eq. (\ref{2.1}) with a good accuracy.

As noted in Sect. \ref{S3.2} (see Eq. (\ref{3.14})), the
quantity $m=q_1+q_2$ can be treated as the AdS mass. We
use $m_P$ to denote the standard Poincare
mass and $l_C$ to denote the Compton wave length
$\hbar /(m_Pc)$. In our units $l_C=2/m_P$. Therefore,
if $m=2Rm_P$ then $m$ is roughly the ratio of the AdS
radius to the Compton wave length. Hence even the AdS
masses of elementary particles are very large. For example,
if $R$ is, say, 20 billion light years then
the AdS mass of the electron is of order
$10^{39}$. Physicists are usually surprised by this fact
and their first impression is that the AdS or dS masses
are something unrealistic. The origin of the masses of
elementary particles is a serious problem of quantum
theory, but in any case the dS and AdS masses
are dimensionless and therefore more fundamental
than the Poincare mass which depends on the
system of units. 

Since $m=q_1+q_2$ and $s=q_1-q_2$ (see Sect. \ref{3.2})
then in the Poincare approximation we have that $m\gg s$
and $q_1\approx q_2\approx m/2$. As follows from the
first expression in Eq. (\ref{3.27}), the quantity
$2(n+n_1+n_2)$ is the AdS kinetic energy and therefore
the standard
kinetic energy is $(n+n_1+n_2)/R$. This is an indication
that in the Poincare approximation the quantum numbers
$(n_1n_2n)$ are asymptotically large. At the same time,
it follows from the second expression in Eq. (\ref{3.27})
that $n_1-n_2$ is not asymptotically large in that
approximation.

Since in the Poincare invariant theory the operators
$P_{\mu}$ commute with each other, one can find a basis
where these operators are diagonal simultaneously. This
means that to find a transition from the so(2,3) to
Poincare invariant theory we should find a subset of
states on which the operators $M_{\mu 5}$ are large
and approximately diagonal, i.e. with a high accuracy
the elements of this subset should be eigenvectors
of all the operators $M_{\mu 5}$ simultaneously.

The elementary particle in the so(2,3) invariant theory
is described only by dimensionless operators (in units
$\hbar /2=c=1$) which have only discrete spectrum. Is this
compatible with the fact that the Poincare invariant theory
also involves momentum operators which have only continuous
spectrum? The answer is that if instead of the operators
$M_{\mu 5}$ one considers the momenta $P_{\mu}=M_{\mu 5}/2R$
then these quantities still have the discrete spectrum but
the quantum of the momentum is $\hbar/(4R)$. This is an
extremely small value and therefore the discrete
spectrum can be effectively described by a continuous one.

\begin{sloppypar}
Instead of the basis vectors $e(n_1n_2nk)$ we will
now work with the vectors
\begin{equation}
{\tilde e}(n_1n_2nk)=(-1)^{n_2}exp[i(n_1-n_2)\varphi]
e(n_1n_2nk)/Norm(n_1n_2nk)^{1/2}
\label{3.47}
\end{equation}
which are mutually orthogonal and normalized to one.
We will consider a subset of states
\begin{equation}
x = \sum c(n_1n_2nk){\tilde e}(n_1n_2nk)
\label{3.48}
\end{equation}
such that the coefficients $c(n_1n_2nk)$ do not
change significantly when one of the quantum numbers
$(n_1n_2n)$ change by $\pm 1$. Then explicit
calculations using the results of Sects. \ref{S3.2}
and \ref{S3.3} show that when $q_1,q_2,n,n_1,n_2$
are very large, $q_1\approx q_2$ and $n_1\approx n_2$
one can obtain
\begin{eqnarray}
&M^1_5x\approx 4[n_1(q_1+n+n_1)]^{1/2}cos\varphi\, x\nonumber\\
&M^2_5x\approx 4[n_1(q_1+n+n_1)]^{1/2}sin\varphi\, x\nonumber\\
&M^3_5x\approx -2[n(m+n)]^{1/2}x
\label{3.49}
\end{eqnarray}
and the operators $M_{\mu 5}$ are much greater than the
other operators. It also follows from Eqs. (\ref{3.27})
and (\ref{3.49}) that
\begin{equation}
\{\sum_{\mu}M_{\mu 5}M^{\mu 5}\}x\approx m^2x
\label{3.50}
\end{equation}
as it should be.
\end{sloppypar}

We conclude that $\varphi$ has the meaning of the polar
angle, the meaning of the quantum numbers $(nn_1)$ in
the Poincare approximation is such that
$[2n_1(m+2n+2n_1)]^{1/2}/R$ is the magnitude of the
momentum projection onto the plane $xy$ and
$[n(m+n)]^{1/2}/R$ is the magnitude of the momentum
projection on the $z$ axis. Eq. (\ref{3.47}) might be
an indication that {\it Poincare invariance is not a
natural special case of the AdS one}. This conclusion
seems also to be reasonable from the fact that in the
AdS case there are no special reasons to prefer the
operators with the index 5.

\section{Supersymmetry}
\label{S3.6}

The superalgebra osp(1,4) is a natural  supersymmetric
generalization of the so(2,3) algebra since the latter is
the even part of the former. We will describe the osp(1,4)
superalgebra a bit later but first we note that its
IRs have several interesting distinctions from IRs of
the Poincare superalgebra. For this reason we first briefly
mention some well known facts about the
latter IRs (see e.g Ref. \cite{Wein-susy} for details).

Representations of the Poincare superalgebra are described
by 14 operators. Ten of them are the well known
representation operators
of the Poincare algebra --- four momentum operators and six
representation operators of the Lorentz algebra, which satisfy
the commutation relations (\ref{2.1}). In addition, there also
exist four linearly independent fermionic operators. The
anticommutators of the fermionic operators are linear
combinations of the momentum operators, and the commutators
of the fermionic operators
with the Lorentz algebra operators are linear combinations of the
fermionic operators. In addition, the fermionic operators
commute with the momentum operators.

From the formal point of view, representations of the osp(1,4)
superalgebra are also described by 14 operators --- ten
representation operators of the so(2,3) algebra and four
fermionic operators. There are three types of
relations: the operators
of the so(2,3) algebra commute with each other as usual
(see Sect. \ref{S3.2}), anticommutators of the
fermionic operators are linear combinations of the so(2,3)
operators and commutators of the latter with
the fermionic operators are their linear combinations.
However, in fact representations of the osp(1,4)
superalgebra can be described exclusively
in terms of the fermionic operators. The matter is
as follows. In the general case the anticommutators of four
operators form ten independent linear combinations.
Therefore, ten linearly independent bosonic operators can
be expressed in terms of fermionic ones. This is not the case
for the Poincare superalgebra since the Poincare algebra operators
are obtained from the so(2,3) ones by contraction.
One could say that the representations of the
osp(1,4) superalgebra is an implementation of the idea that
the supersymmetry is the extraction of the square root from
the usual symmetry (by analogy with the well known treatment of
the Dirac equation as a square
root from the Klein-Gordon one).

We denote the independent fermionic operators of the
osp(1,4) superalgebra as $(d_1,d_2,d_1^*,d_2^*)$ where the
$^*$ means the Hermitian conjugation as
usual. They should satisfy the following relations.
If $(A,B,C)$ are any fermionic operators, [...,...] is used
to denote a commutator and
$\{...,...\}$ to denote an anticommutator then
\begin{equation}
[A,\{ B,C\} ]=F(A,B)C + F(A,C)B
\label{3.51}
\end{equation}
where the form $F(A,B)$ is skew symmetric, $F(d_j,d_j^*)=1$
$(j=1,2)$ and
the other independent values of $F(A,B)$ are equal to zero.

We can now {\bf define} the so(2,3) operators
as follows:
\begin{eqnarray}
&b'=\{d_1,d_2\}\quad b"=\{d_1^*,d_2^*\}\quad
L_+=\{d_2,d_1^*\}\quad L_-=\{d_1,d_2^*\}\nonumber\\
&a_j'=(d_j)^2\quad a_j"=(d_j^*)^2\quad
h_j=\{d_j,d_j^*\} \quad (j=1,2)
\label{3.52}
\end{eqnarray}
Then by using Eq. (\ref{3.51}) and the properties of the form
$F(.,.)$, one can show by a direct calculations that so
defined operators satisfy the correct commutation relations
of the so(2,3) algebra (see Sect. \ref{S3.2}).
This result can be treated as a fact
that the representation operators of the so(2,3) algebra
are not fundamental, only the fermionic generators are.

A full classification of UIRs of the osp(1,4) superalgebra
has been given for the first time by Heidenreich
\cite{Heidenreich}. By analogy with his construction,
we require the existence of the cyclic vector $e_0$ satisfying the
conditions (compare with Eq. (\ref{3.14})):
\begin{eqnarray}
d_je_0=L_+e_0=0 \quad h_je_0=q_je_0\quad (e_0,e_0)\neq 0\quad (j=1,2)
\label{3.53}
\end{eqnarray}
The full representation space can be obtained by successively
acting by the fermionic operators on $e_0$ and taking all
possible linear combinations of such vectors (see Refs.
\cite{Heidenreich,levsusy} for details). As a result, one
can obtain the following Heidenreich classification. There
exist for types of IRs:

\begin{itemize}

\item If $q_2>1$ and $s\neq 0$ (massive IRs), the osp(1,4)
supermultiplets contain four IRs of the so(2,3) algebra
characterized by the values of the mass and spin
$$(m,s),(m+1,s+1),(m+1,s-1),(m+2,s).$$

\item If $q_2>1$ and $s=0$ (collapsed massive IRs), the osp(1,4)
supermultiplets contain three IRs of the so(2,3) algebra
characterized by the values of the mass and spin
$$(m,0),(m+1,1),(m+2,0).$$

\item If $q_2=1$ (massless IRs) the osp(1,4)
supermultiplets contains two IRs of the so(2,3) algebra
characterized by the spins $s$ and $s+1$.

\item Dirac supermultiplet containing two Dirac
singletons \cite{DiracS}.
\end{itemize}
The first three cases have well known analogs of IRs of the
super-Poincare algebra (see e.g. Ref. \cite{Wein-susy}) while
there is no super-Poincare analog of the Dirac supermultiplet.

The above discussion shows that in the AdS case supersymmetry
is even more attractive than in the Poincare one. As already
noted, there is no natural supersymmetric generalization of
the so(1,4) symmetry, and for this reason the criterion that
the theory should be supersymmetric could be a good criterion
for solving a long standing problem of which de Sitter
algebra is preferable. On the other hand, we will see below
that in GFQT a serious problem with supersymmetry arises.

\vfill\eject

\chapter{Why is GFQT more natural than the standard theory?}
\label{C4}

\section{What mathematics is most suitable for quantum
physics?}
\label{SG1}

Since the absolute majority of physicists are not familiar
with Galois fields, our first goal in this chapter is
to convince the reader that the notion of Galois
fields is not only very simple and elegant, but
also is a natural basis for quantum physics. If a reader
wishes to learn Galois fields on a more fundamental level,
he or she might start with standard textbooks (see e.g.
Ref. \cite{VDW}).

The existing quantum theory is based on
standard mathematics containing the notions of
infinitely small and infinitely large.

The notion of infinitely small is based on our
everyday experience that any macroscopic object can be
divided by two, three and even a million
parts. But is it possible to divide by two or three
the electron or neutrino? It seems obvious that
the very existence of elementary particles
indicates that the standard division has only a
limited sense. Indeed, let us consider, for example,
the gram-molecule of water having the mass 18 grams.
It contains the Avogadro number of molecules
$6\cdot 10^{23}$. We can divide this
gram-molecule by ten, million, billion, but when we
begin to divide by numbers greater than the Avogadro
one, the division operation loses its sense.

The notion of infinitely large is based on our belief
that {\it in principle} we can operate with any
large numbers. Suppose we wish to verify experimentally
whether addition is commutative, i.e. whether (a + b) =
(b + a) is always satisfied. If our Universe is finite
and contains not more than N elementary particles then
we shall not be able to do this if $a + b > N$.
In particular, if the Universe is finite then it is
impossible in principle to build a computer operating
with any large number of bits.

Let us look at mathematics from the point of
view of the famous Kronecker expression: 'God made
the natural numbers, all else is the work of man'.
Indeed, the natural numbers 0, 1, 2... have a
clear physical meaning. However only two operations
are always possible in the set of natural
numbers: addition and multiplication.

In order to
make addition reversible, we introduce negative
integers -1, -2 etc. Then, instead of the set of
natural numbers we can work
with the ring of integers where three operations
are always possible:
addition, subtraction and multiplication. However,
the negative numbers do not have a direct physical
meaning (we cannot say, for example, 'I have
minus two apples'). Their only role is to make
addition reversible.

The next step is the transition to the field of
rational numbers in which all
four operations excepting division by zero are possible.
However, as noted above, division has only a limited
sense.

In mathematics the notion of linear space is
widely used, and such important
notions as the basis and dimension are meaningful only
if the space is considered over a field or body.
Therefore if we start from natural numbers and wish
to have a field, then we have to introduce negative
and rational numbers. However, if, instead of all
natural numbers, we consider
only $p$ numbers 0, 1, 2, ... $p-1$ where $p$ is
prime, then we can easily construct a field without
adding any new elements. This construction, called
Galois field, contains nothing that could prevent
its understanding even by pupils of elementary
schools.

Let us denote the set of numbers 0, 1, 2,...$p-1$
as $F_p$. Define addition and multiplication as usual
but take the final result modulo $p$. For simplicity,
let us consider the case $p=5$. Then $F_5$ is the set 0,
1, 2, 3, 4. Then
$1+2=3$ and $1+3=4$ as usual, but $2+3=0$, $3+4=2$ etc.
Analogously, $1\cdot 2=2$,
$2\cdot 2=4$, but $2\cdot 3 =1$, $3\cdot 4=2$ etc.
By definition, the element
$y\in F_p$ is called opposite to $x\in F_p$ and is
denoted as $-x$ if $x+y=0$ in
$F_p$. For example, in $F_5$ we have -2=3, -4=1 etc.
Analogously $y\in F_p$ is
called inverse to $x\in F_p$ and is denoted as
$1/x$ if $xy=1$ in $F_p$.
For example, in $F_5$ we have 1/2=3, 1/4=4 etc. It is
easy to see that
addition is reversible for any natural $p>0$ but for
making multiplication
reversible we should choose $p$ to be a prime.
Otherwise the product of two
nonzero elements may be zero modulo $p$. If $p$ is
chosen to be a prime then
indeed $F_p$ becomes a field without introducing any
new objects (like negative numbers or fractions). For
example, in this field each element can obviously be
treated as positive and negative {\bf simultaneously}!

One might say: well, this is beautiful but impractical
since in physics and
everyday life 2+3 is always 5 but not 0. Let us suppose,
however that fundamental
physics is described not by 'usual mathematics' but by
'mathematics modulo $p$'
where $p$ is a very large number.
Then, operating with numbers much smaller than $p$ we
shall not notice this $p$,
at least if we only add and multiply. We will feel a
difference between 'usual mathematics' and 'mathematics
modulo p' only while operating with numbers
comparable to $p$.

We can easily extend the correspondence between $F_p$ and
the ring of integers $Z$
in such a way that subtraction will also be included.
To make it clearer we note
the following. Since the field $F_p$ is cyclic (adding
1 successively, we will
obtain 0 eventually), it is convenient to visually depict
its elements by the
points of a circle of the radius $p/2\pi$ on the plane $(x,y)$.
In Fig. 4.1 only a
part of the circle near the origin is depicted.
\begin{figure}[!ht]
\centerline{\scalebox{1.1}{\includegraphics{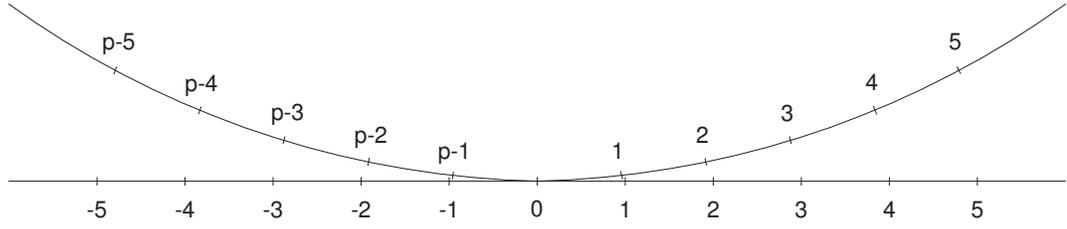}}}%
\caption{%
  Relation between $F_p$ and the ring of integers%
}%
\label{Fig.1}
\end{figure}
Then the distance between
neighboring elements of the field is equal to unity, and
the elements
0, 1, 2,... are situated on the circle counterclockwise. At
the same time we depict the elements of $Z$ as usual
such that each element $z\in Z$ is depicted by
a point with the
coordinates $(z,0)$. We can denote the elements of $F_p$
not only as 0, 1,... $p-1$
but also as 0, $\pm 1$, $\pm 2,$,...$\pm (p-1)/2$, and such
a set is called the
set of minimal residues. Let $f$ be a map from $F_p$ to Z,
such that the element
$f(a) \in Z$ corresponding to the minimal residue $a$ has
the same notation as
$a$ but is considered as the element of $Z$. Denote
$C(p) = p^{1/(lnp)^{1/2}}$ and
let $U_0$ be the set of elements $a\in F_p$ such that
$|f(a)|<C(p)$. Then if
$a_1,a_2,...a_n\in U_0$ and $n_1,n_2$ are such natural
numbers that
\begin{equation}
n_1<(p-1)/2C(p),\,\,n_2<ln((p-1)/2)/(lnp)^{1/2}
\end{equation}
\label{G1}
then
$$f(a_1\pm a_2\pm...a_n)=f(a_1)\pm f(a_2)\pm ...f(a_n)$$
if $n\leq n_1$ and
$$f(a_1 a_2...a_n)=f(a_1)f(a_2) ...f(a_n)$$ if $n\leq n_2$.
Thus though $f$ is not a homomorphism of rings $F_p$ and $Z$,
but if $p$ is sufficiently large, then for a sufficiently
large number of elements of $U_0$ the
addition, subtraction and multiplication are
performed according to the same rules
as for elements $z\in Z$ such that $|z|<C(p)$.
Therefore $f$ can be treated as a
local isomorphism of rings $F_p$ and $Z$.

The above discussion has a well known historical
analogy. For many years people believed
that our Earth was flat and infinite, and only
after a long period of time they realized that
it was finite and had a curvature. It is difficult
to notice the curvature when we deal only with
distances much less than the radius of the
curvature $R$. Analogously one might think that
the set of
numbers describing physics has a curvature defined
by a very large number $p$ but we do not notice
it when
we deal only with numbers much less than $p$.

Our arguments imply that the standard field of
real numbers will not be fundamental in future
quantum physics (although
there is no doubt that it is relevant for macroscopic
physics). Let us discuss this question in a greater detail
(see also Ref. \cite{FJD1}).
The notion of real numbers can be fundamental only if the
following property is valid: for any $\epsilon>0$, it is
possible (at least in principle) to build a computer
operating with a number of bits $N(\epsilon)$ such that
computations can be performed with the accuracy $\epsilon$. 
It is clear that this is not the case
if, for example, the Universe is finite.

Let us note that even for elements from $U_0$ the
result of division in the field
$F_p$ differs generally speaking, from the corresponding
result in the field of rational number $Q$. For example the
element 1/2 in $F_p$ is a very large number
$(p+1)/2$. For this reason one might think that physics
based on Galois
fields has nothing to with the reality. We will see in
the subsequent section that this is not so since the
spaces describing quantum systems are projective.

By analogy with the field of complex numbers, we can
consider a set $F_{p^2}$
of $p^2$ elements $a+bi$ where $a,b\in F_p$ and $i$ is
a formal element such
that $i^2=1$. The question arises whether $F_{p^2}$ is a
field, i.e.
we can define all the four operations excepting
division by zero.
The definition of addition, subtraction and multiplication
in $F_{p^2}$
is obvious and, by analogy with the field of complex
numbers, one could define division as
$1/(a+bi)\,=a/(a^2+b^2)\,-ib/(a^2+b^2)$.
This definition can be meaningful only if
$a^2+b^2\neq 0$ in $F_p$
for any $a,b\in F_p$ i.e. $a^2+b^2$ is not divisible by $p$.
Therefore the definition is meaningful only if
$p$ {\it cannot}
be represented as a sum of two squares and is
meaningless otherwise.
We will not consider the case $p=2$ and therefore $p$
is necessarily odd.
Then we have two possibilities: the value of $p\,(mod \,4)$ 
is either 1 or 3. The well known result of number theory
(see e.g. the
textbooks \cite{VDW}) is that a prime number $p$ can be
represented as a sum of two squares only in the
former case
and cannot in the latter one. Therefore the above
construction of
the field $F_{p^2}$ is correct only if
$p\,(mod \,4)\,=\,3$.
The first impression is that if Galois fields can somehow
replace the conventional field of complex numbers then this
can be done only for $p$ satisfying $p\,(mod \,4)\,=\,3$ and
therefore the case $p\,(mod \,4)\,=\,1$ is of no interest for
this purpose. We will see in the subsequent section that
correspondence between complex numbers and Galois fields
containing $p^2$ elements can also be established if
$p\,(mod \,4)\,=\,1$ but some results of GFQT discussed
below indicate that if quantum theory is based on a Galois
field then $p$ is probably such that $p\,(mod \,4)\,=\,3$
rather than $p\,(mod \,4)\,=\,1$. In general, it is well known
(see e.g. Ref. \cite{VDW}) that any Galois field consists of
$p^n$ elements where $p$ is prime and $n>0$ is natural.
The numbers
$p$ and $n$ define the field $F_{p^n}$ uniquely up to
isomorphism and $p$ is called the characteristic of the
Galois field.

\section{Modular representations of Lie algebras}
\label{SG2}

A well-known historical fact is that quantum mechanics
has been originally proposed by Heisenberg and
Schroedinger in two forms which seemed fully
incompatible with each other. While in the Heisenberg
operator (matrix)
formulation quantum states are described by infinite
columns and operators --- by infinite matrices, in the
Schroedinger wave formulations the states are described
by functions and operators --- by differential operators.
It has been shown later by Born, von Neumann and
others that
the both formulations are mathematically equivalent.

Quantum theory on a Galois field (GFQT) can be treated
as a version of the matrix formulation when complex numbers
are replaced by elements of a Galois field. We will see below
that in that case the columns and matrices are automatically
truncated in a certain way, and therefore the theory becomes
finite-dimensional (and even finite since any Galois field
is finite).

In conventional quantum theory the state of a system is
described by a vector $\tilde x$ from a separable Hilbert
space $H$. We will use a 'tilde' to denote elements
of Hilbert spaces and complex numbers while elements
of linear spaces over Galois fields and elements of the
fields will be denoted without a "tilde".

Let $(\tilde e_1,\tilde e_2,...)$ be a basis in $H$. This
means that $\tilde x$ can be represented as
\begin{equation}
\tilde x =\tilde c_1 \tilde e_1+\tilde c_2 \tilde e_2+...
\label{G2}
\end{equation}
where $(\tilde c_1,\tilde c_2,...)$ are complex numbers.
It is assumed
that there exists a complete set of commuting selfadjoint
operators $(\tilde A_1,\tilde A_2,...)$ in $H$ such that
each $\tilde e_i$ is the eigenvector of all these operators:
$\tilde A_j\tilde e_i =\tilde \lambda_{ji}\tilde e_i$. Then the
elements $(\tilde e_1,\tilde e_2,...)$ are mutually orthogonal:
$(\tilde e_i,\tilde e_j)=0$ if $i\neq j$ where (...,...) is
the scalar product in $H$. In that case the coefficients can
be calculated as
\begin{equation}
\tilde c_i = \frac{(\tilde e_i,\tilde x)}{(\tilde e_i,\tilde e_i)}
\label{G3}
\end{equation}
Their meaning is that
$|\tilde c_i|^2(\tilde e_i,\tilde e_i)/(\tilde x,\tilde x)$
represents the probability to find $\tilde x$ in the state
$\tilde e_i$. In particular, when $\tilde x$ and the basis
elements are normalized to one, the probability is exactly
equal to $|\tilde c_i|^2$.

Let us note that the Hilbert space contains a big
redundancy of elements, and we do not need to know all of
them. Indeed, with any desired accuracy we can approximate
each $\tilde x\in H$ by a finite linear combination
\begin{equation}
\tilde x =\tilde c_1 \tilde e_1+\tilde c_2 \tilde e_2+...
\tilde c_n\tilde e_n
\label{G4}
\end{equation}
where $(\tilde c_1,\tilde c_2,...\tilde c_n)$ are
rational complex numbers. In turn, the set Eq. (\ref{G4})
is redundant too. Indeed, we can use the fact that Hilbert
spaces in quantum theory are projective: $\psi$ and
$c\psi$ represent the same physical state. Then we can
multiply both parts of Eq. (\ref{G4}) by a common
denominator of the numbers
$(\tilde c_1,\tilde c_2,...\tilde c_n)$. As a result, we
can always assume that in Eq. (\ref{G4})
$\tilde c_j=\tilde a_j +i\tilde b_j$ where $\tilde a_j$
and $\tilde b_j$ are integers. If it is convenient, we can
approximate all physical states by another sets of elements.
For example, we can use Eq. (\ref{G4}), where
$\tilde c_j=\tilde a_j +\sqrt{2}i\tilde b_j$ and
$\tilde a_j$ and $\tilde b_j$ are integers.

The meaning of the fact that Hilbert
spaces in quantum theory are projective is very clear.
The matter is that not the probability itself but the
relative probabilities of different measurement outcomes
have a physical meaning.
We believe, the notion of probability is a
good illustration of the Kronecker expression about
natural numbers (see Sect. \ref{SG1}). Indeed, this notion
arises as follows. Suppose that conducting
experiment $n$ times we have seen the first event $n_1$ times,
the second event $n_2$ times etc. such that $n_1 + n_2 + ... = n$.
We introduce the quantities $w_i(n) = n_i/n$ (these quantities
depend on $n$) and $w_i = lim\, w_i(n)$ when $n \rightarrow
\infty$. Then $w_i$ is called the probability of the
$ith$ event. We see that all the information about the
experiment is given by a set of natural numbers.
However, in order to define
probabilities, people introduce additionally the notion of
rational numbers and the notion of limit. Of course,
the standard notion of probability can be used even
if quantum
theory is based entirely on natural numbers, but one should
realize that this is only a convention on how to describe
the measurement outcomes.

The Hilbert space is an example of a linear space over the
field of complex numbers. Roughly speaking this means that
one can multiply the elements of the space by the elements
of the field and use the properties
$\tilde a(\tilde b\tilde x)=(\tilde a\tilde b)\tilde x$
and $\tilde a(\tilde b\tilde x+\tilde c\tilde y)=
\tilde a\tilde b\tilde x +\tilde a\tilde c\tilde y$ where
$\tilde a,\tilde b,\tilde c$ are complex numbers and
$\tilde x,\tilde y$ are elements of the space. The fact that
complex numbers form a field is important for such notions
as linear dependence and the dimension of spaces over complex
numbers.

In general, it is possible to consider linear
spaces over any fields (see any textbook on modern
algebra, e.g. \cite{VDW}). It is natural to assume that
in GFQT the states of physical
systems should be described by elements of linear spaces over
a Galois field. Since we wish to have a correspondence
with the conventional theory, we assume that the Galois field
in question contains $p^2$ elements where $p$ is the
characteristic of the field. In Sect. \ref{SG1} we discussed
the correspondence between the ring of integers and
the field $F_p$. It has been also noted that if
$p=3\,(mod\,4)$ then the elements of $F_{p^2}$ can be
written as $a+bi$ where $a,b\in F_p$. We will now discuss
the construction of the field $F_{p^2}$ in more general
cases.

The field $F_{p^2}$ can be constructed by means of the
standard extension of the field $F_p$. Let the equation
$\kappa^2=-a_0$ $(a_0\in F_p)$ have no solutions in $F_p$.
Then $F_{p^2}$ can be formally described as a set of
elements $a+b\kappa$, where $a,b\in F_p$ and $\kappa$
satisfies the condition $\kappa^2=-a_0$. The actions in
$F_{p^2}$ are defined in the natural way. The condition
that the equation $\kappa^2=-a_0$ has no solutions in
$F_p$ is important in order to ensure that any nonzero
element from $F_{p^2}$ has an inverse.
Indeed, the definition
$(a+b\kappa)^{-1}=(a-b\kappa)/(a^2+b^2a_0)$ is correct
since the denominator can be equal to zero only if both,
$a$ and $b$ are equal to zero.

Consider three cases.

(1) $a_0=1$. In that case it is natural to write $i$
instead of $\kappa$. We must be sure that the element -1
in $F_p$ cannot be represented as a square of another
element from $F_p$. In number theory this is
expressed by saying that -1 is not a quadratic residue
in $F_p$. It is well known (see e.g. Ref. \cite{VDW})
that this is the case when $p=3\,(mod\,4)$ and is not
the case when $p=1\,(mod\,4)$. This fact is closely
related to one mentioned in Sect. \ref{SG1} that a
prime $p$ can be represented as a sum of two squares
only if $p=1\,(mod\,4)$.

(2) $a_0=2$. Assume that $p=1\,(mod\,4)$. Since -1 is
a quadratic residue in $F_p$ in this case, then the problem
is that 2 should not be a quadratic residue in this field.
This is the case (see e.g. Ref. \cite{VDW})
if $p=5\,(mod\,8)$.

(3) $a_0=3$. Again, assume that $p=1\,(mod\,4)$.
Then 3 should not be a quadratic residue in $F_p$.
This is, for example, the case (see e.g. Ref.
\cite{VDW}),
when $p$ is the prime of Fermat's type, i.e.
$p=2^n+1$.

In the general case, the field $F_p$ can be extended
to a field
containing $p^2$ elements as follows. First we note
a well known
fact \cite{VDW} that $F_p$ is a cyclic group with respect
to multiplication. There exists at least one element
$r\in F_p$ such that $(r,r^2,...r^{p-1}=1)$ represent
the set $(1,2,...p-1)$
in some order. The element $r$ is called the
primitive root of
the field $F_p$. It is clear from this observation
that $F_p$
contains the equal number $(p-1)/2$ of quadratic
residues and
nonresidues: the elements represented as odd powers
of $r$ are
nonresidues while those represented by even
powers of $r$ ---
residues. In particular, $r$ is not a quadratic
residue. Now
we can formally define $\kappa$ as an element
satisfying
the condition $\kappa^2=r$, and $F_{p^2}$ as a set of
elements $a+b\kappa$ $(a,b\in F_p)$. In this case the
definition
$(a+b\kappa)^{-1}=(a-b\kappa)/(a^2-b^2r)$ is correct since
$a^2-b^2r\neq 0$ if $a\neq 0$ or $b\neq 0$.

The above observation shows that if $F_p$ is extended to
$F_{p^2}$ then any element of $F_p$ has a square root
belonging to $F_{p^2}$.

It is well known (see e.g. Ref. \cite{VDW}) that the
field of $p^2$ elements has only one nontrivial
automorphism. In the above cases it can be defined as
$a+b\kappa\rightarrow a-b\kappa$ or as
$a+b\kappa\rightarrow (a+b\kappa)^p$. We will use a bar
to denote this automorphism. This means that if
$c=a+b\kappa$ $(a,b\in F_p)$ then $\bar{c}=a-b\kappa$.

By analogy with conventional quantum theory, we
require that linear spaces V over $F_{p^2}$, used for
describing physical states, are supplied by a
scalar product (...,...) such that for any $x,y\in V$
and $a\in F_{p^2}$, $(x,y)$ is an element of $F_{p^2}$
and the following properties are satisfied:
\begin{equation}
(x,y) =\overline{(y,x)},\quad (ax,y)=\bar{a}(x,y),\quad
(x,ay)=a(x,y)
\label{G5}
\end{equation}

In Sect. \ref{SG1} we discussed a map $f$ from $F_p$
to the ring of integers $Z$. We can extend the map in such
a way that it maps $F_{p^2}$ to a subset of complex numbers.
For example, if $p=3\,(mod\,4)$, $c\in F_{p^2}$, $c=a+bi$
$(a,b\in F_p)$, we can define $f(c)$ as $f(a)+f(b)i$.
If $a_0=-2$ and $c=a+b\kappa$, we define
$f(c)=f(a)+\sqrt{2}f(b)i$. Analogously, if $a_0=-3$ and
$c=a+b\kappa$, we define $f(c)=f(a)+\sqrt{3}f(b)i$.
Let $U$ be a subset of elements $c=a+b\kappa$ where
$a,b\in U_0$ (see Sect. \ref{SG1}). Then, for the elements
from $U$, addition, subtraction and multiplication look
the same as for the
elements $f(c)$ of the corresponding subset in the field of
complex numbers. This subset is of the form ${\tilde a} +
{\tilde b} i$ in the first case, ${\tilde a} + \sqrt{2}
{\tilde b} i$ in the second case and ${\tilde a} + \sqrt{3}
{\tilde b} i$ in the third one. Here ${\tilde a},
{\tilde b} \in Z$.

We will always consider only finite dimensional spaces
$V$ over
$F_{p^2}$. Let $(e_1,e_2,...e_N)$ be a basis in such a space.
Consider subsets in $V$ of the form $x=c_1e_1+c_2e_2+...c_ne_n$
where for any $i,j$
\begin{equation}
c_i\in U,\quad (e_i,e_j)\in U
\label{G6}
\end{equation}
On the other hand, as noted above, in conventional quantum
theory we can describe quantum states by subsets of the form
Eq. (\ref{G4}). If $n$ is much less than $p$,
\begin{equation}
f(c_i)=\tilde c_i,\quad f((e_i,e_j))=(\tilde e_i,\tilde e_j)
\label{G7}
\end{equation}
then we have the correspondence between the description of
physical states in projective spaces over $F_{p^2}$ on 
one hand and projective Hilbert spaces on the other.
This means
that if $p$ is very large then for a large number
of elements from $V$,
linear combinations with the coefficients belonging to
$U$ and
scalar products look in the same way as for the elements from
a corresponding subset in the Hilbert space.

In the general case a scalar product in $V$ does not define
any positive definite metric and thus there is no
probabilistic interpretation for all the elements from $V$.
In particular, $(e,e)=0$ does not necessarily imply
that $e=0$.
However, the probabilistic interpretation exists for such a
subset in $V$ that the conditions (\ref{G7}) are satisfied.
Roughly speaking this means that for elements
$c_1e_1+...c_ne_n$
such that $(e_i,e_i),c_i{\bar c}_i\ll p$, $f((e_i,e_i))>0$
and $c_i{\bar c}_i>0$ for all $i=1,...n$, the
probabilistic interpretation is valid.
It is also possible to explicitly construct
a basis $(e_1,...e_N)$ such that $(e_j,e_k)=0$ for
$j\neq k$ and
$(e_j,e_j)\neq 0$ for all $j$ (see the subsequent chapter).
Then $x=c_1e_1+...c_Ne_N$ $(c_j\in
F_{p^2})$ and the coefficients are uniquely defined by
$c_j=(e_j,x)/(e_j,e_j)$.

As usual, if $A_1$ and $A_2$ are linear operators
in $V$ such that
\begin{equation}
(A_1x,y)=(x,A_2y)\quad \forall x,y\in V
\label{G8}
\end{equation}
they are said to be conjugated: $A_2=A_1^*$.
It is easy to see
that $A_1^{**}=A_1$ and thus $A_2^*=A_1$.
If $A=A^*$ then the
operator $A$ is said to be Hermitian.

If $(e,e)\neq 0$, $Ae=ae$, $a\in F_{p^2}$,
and $A^*=A$, then it
is obvious that $a\in F_p$. At the same time, there exist
possibilities (see e.g. Ref. \cite{lev4}) when $e\neq 0$,
$(e,e)=0$ and the element $a$ is imaginary, i.e. $a=b\kappa$,
$b\in F_p$. Further, if
\begin{eqnarray}
&&Ae_1=a_1e_1,\quad Ae_2=a_2e_2,\quad (e_1,e_1)\neq 0,\nonumber\\
&& (e_2,e_2)\neq 0, \quad a_1\neq a_2
\label{G9}
\end{eqnarray}
then as in the usual case, one has $(e_1,e_2)=0$. At the same time,
the situation when
\begin{eqnarray}
&&(e_1,e_1)=(e_2,e_2)=0,\quad (e_1,e_2)\neq 0,\nonumber\\
&& a_1=b_1\kappa, \quad a_2=b_2\kappa\quad (b_1,b_2\in F_p)
\label{G10}
\end{eqnarray}
is also possible \cite{lev4}.

Let now $(A_1,...A_k)$ be a set of Hermitian
commuting operators
in $V$, and $(e_1,...e_N)$ be a basis in $V$
with the properties
described above, such that $A_je_i=\lambda_{ji}e_i$.
Further, let
$({\tilde A}_1,...{\tilde A}_k)$ be a set of
Hermitian  commuting
operators  in  some  Hilbert  space  $H$,   and
$(\tilde e_1,\tilde e_2,...)$ be some basis
in $H$ such that
$\tilde A_je_i={\tilde \lambda}_{ji}\tilde e_i$.
Consider a subset $c_1e_1+c_2e_2+...c_ne_n$
in $V$ such that,
in addition to the conditions (\ref{G7}), the
elements $e_i$
are the eigenvectors of the operators $A_j$ with
$\lambda_{ji}$
belonging to $U$ and such that
$f(\lambda_{ji})={\tilde \lambda}_{ji}$. Then the
action of the operators on such
elements have the same form as the action of corresponding
operators on the subsets of elements in Hilbert
spaces discussed above.

Summarizing this discussion, we conclude that if $p$
is large then there exists a correspondence between the
description of physical states on the language of
Hilbert spaces and selfadjoint operators
in them on one hand, and on the language of
linear spaces over
$F_{p^2}$ and Hermitian operators in them on the other.

The field of complex numbers is algebraically closed (see
standard textbooks on modern algebra, e.g. Ref. \cite{VDW}).
This implies that any equation of the $nth$ order
in this field
always has $n$ solutions. This is not, generally
speaking, the
case for the field $F_{p^2}$. As a consequence, not
every linear
operator in the finite-dimensional space over
$F_{p^2}$ has an
eigenvector (because the characteristic equation
may have no
solution in this field). One can define a field of
characteristic $p$ which is algebraically closed and
contains $F_{p^2}$. However such a field will
necessarily be
infinite and we will not use it. We will see in the
subsequent chapter
that uncloseness of the field $F_{p^2}$ does not
prevent one
from constructing physically meaningful representations
describing elementary particles in GFQT.

In physics one usually considers Lie  algebras over $R$ and
their representations by Hermitian  operators in
Hilbert spaces. It is clear that analogs of such
representations in our case are representations of Lie
algebras over $F_p$  by Hermitian operators
in spaces over $F_{p^2}$.
Representations in spaces over a field of nonzero
characteristics are called modular representations.
There exists a wide literature devoted to such
representations; detailed references can be found
for example in Ref.
\cite{FrPa} (see also Ref. \cite{lev4}).
In particular, it has been shown by Zassenhaus
\cite{Zass} that all modular IRs are finite-dimensional and
many papers have
dealt with the maximum dimension of such representations.
At the same time, it is worth noting that usually
mathematicians consider only representations
over an algebraically closed field.

From  the previous, it is natural to expect that the
correspondence  between
ordinary and modular representations of two Lie
algebras over $R$
and $F_p$, respectively, can be obtained if the structure
constants of the Lie algebra
over $F_p$ - $c_{kl}^j$, and the structure constants of the Lie
algebra over $R$ - ${\tilde c}_{kl}^j$, are such that
$f(c_{kl}^j)={\tilde c}_{kl}^j$
(the Chevalley basis \cite{Chev}), and
all the $c_{kl}^j$ belong to $U_0$. In Refs.
\cite{lev4,lev2, levsusy} modular analogs of IRs
of su(2), sp(2), so(2,3), so(1,4) algebras and the osp(1,4)
superalgebra have been considered. Also modular
representations describing strings
have been briefly mentioned. In all these cases the
quantities ${\tilde c}_{kl}^j$ take only the values
$0,\pm 1,\pm 2$ and the above correspondence
does have a place.

It is obvious that since all physical quantities in
GFQT are discrete, this theory cannot involve any
dimensional quantities and any operators having the
continuous spectrum. We have seen in the preceding chapter
than the so(2,3) invariant theory is dimensionless and
all the operators in UIRs have only discrete spectrum.
For this reason one might expect that this theory is a
natural candidate for its generalization to GFQT.
We will see in the subsequent chapter that there exists a
correspondendence in the above sense between modular IRs
of the finite field analog of the so(2,3) algebra and
UIRs of the standard so(2,3) algebra. At the same time,
there is no natural generalization of the Poincare
invariant theory to GFQT.

Since the main problems of LQFT originate from
the fact that local fields interact at the same point,
the idea of all modern theories aiming to improve
LQFT is to replace the interaction at a point by an
interaction in some small space-time region. From this
point of view, one could say that those theories involve
a fundamental length, explicitly or implicitly. Since
GFQT is a fully discrete theory, one might wonder
whether it could be treated as a version of quantum
theory with a fundamental length. Although in GFQT
all physical quantities are dimensionless and take
values in a Galois field, on a qualitative level
GFQT can be thought to be a theory with the fundamental
length in the following sense. The maximum value
of the angular momentum in GFQT cannot exceed
the characteristic of the Galois field $p$. Therefore
the Poincare momentum cannot exceed $p/R$. This
can be qualitatively interpreted in such a way that
the fundamental length in GFQT is of order $R/p$.

One might wonder how continuous transformations (e.g.
time evolution or rotations) can be described in the framework
of GFQT. A general remark is that if theory ${\cal B}$ is a
generalization of theory ${\cal A}$ then the relation between
them is not always straightforward. For example,
quantum mechanics is a generalization of classical
mechanics, but in quantum mechanics the experiment outcome
cannot be predicted unambiguously, a particle cannot be
always localized etc. As
noted in Sect. \ref{S2.1}, even in the framework of standard
quantum theory, the time evolution is well-defined only 
on macroscopic level. Suppose that this is the case and
the Hamiltonian $H_1$ in standard theory is a good approximation
for the Hamiltonian $H$ in GFQT. Then one might think that
$exp(-iH_1t)$ is a good approximation for $exp(-iHt)$. However,
such a straightforward conclusion is problematic for the
following reasons. First, there can be no continuous parameters
in GFQT. Second, even if $t$ is somehow discretized, 
it is not clear how the transformation $exp(-iHt)$ 
should be implemented in practice. On the macroscopic level the
quantity $Ht$ is very large and therefore the Taylor series
for $exp(-iHt)$ contains a large number of terms
which should be known with a good accuracy. On the other hand,
one can notice that for computing $exp(-iHt)$ it is sufficient
to know $Ht$ only modulo $2\pi$ but in this case the question
about the accuracy for $\pi$ arises.
We conclude that a direct correspondence between the
standard quantum theory and GFQT exists only on the level of
Lie algebras but on the level of Lie groups this problem requires 
further study.

\section{Why is quantum physics based on complex numbers?}
\label{SG3}

In the preceding section the choice of the field
$F_{p^2}$ in GFQT was motivated by the observation
that in that case there exists the correspondence with
the standard theory. However, if GFQT is treated as
a more general theory  than the standard one, the
philosophy should be changed: the former should explain
why the latter is based on complex numbers and not
the latter should explain why the former is based on
$F_{p^2}$.

Let us first discuss the arguments in favor of the
choice of complex numbers in the standard theory.
It involves the following assumptions:

\begin{itemize}

\item {\it Assumption 1: Quantum states are represented
by elements of a (projective) complex Hilbert space.}

\item {\it Assumption 2: Observable physical quantities
are represented by selfadjoint operators in this space.}

\end{itemize}

Since the field of complex numbers is algebraically closed,
any linear operator in a finite dimensional space has
at least one eigenvalue.
However, this is not necessarily the case if the
space is infinite-dimensional.

The usual motivation of Assumption 2 is that
since any physical quantity can take only real
values, the spectrum of the corresponding
operator should necessarily be real. According
to the spectral theorem for selfadjoint
operators in Hilbert spaces, this is indeed the
case. However, detailed arguments given in
Ref. \cite{JMLL3} and other works (see e.g.
Ref. \cite{Bender}) show that the real spectrum
and Assumption 2 are not necessary for
constructing meaningful quantum theory.
Note that (by definition) any complex
number is simply a pair of real numbers, and even
for this reason it is not clear why the case of
complex spectrum should be excluded. For example,
the complex spectrum can represent a pair of real
physical quantities.

In quantum theory it is also postulated that
the following requirement should be valid:

\begin{itemize}

\item {\it Requirement 1: Any linear operator
representing a physical quantity should have a
spectral decomposition.}

\end{itemize}

This implies that one can construct
a basis such that any its element is the eigenvector
of the given operator. As it is usual in quantum
physics, in the general case the
basis is understood in the sense of distributions,
i.e. points belonging to the continuous spectrum
are also treated as eigenvalues.

As follows from the spectral theorem, if one
accepts Assumption 2 then Requirement 1 is
satisfied automatically. However, the spectral
decomposition exists not only for selfadjoint
operators; for example, it also exists for unitary
operators. It is also clear that the spectral theorem
for selfadjoint operators is valid not only in complex
Hilbert spaces but in real Hilbert spaces as well.
Therefore if one accepts Assumption 2 then
Assumption 1 is not inevitable. The only  motivation of
Assumption 1 is that quantum theory based on complex
numbers successfully describes a wide range of physical
phenomena.

It is reasonable to believe that in future
quantum physics the choice of the number
field will be substantiated instead of saying that a
particular number field should be chosen because the
corresponding version of quantum theory is in
agreement with experimental data.
In the literature possibilities have been
considered that quantum theory is based not
on complex numbers but on quaternions, $p$-adic
numbers or other constructions. In each case the theory 
has its own interesting properties but the problem of the
motivation of the choice of the principal number field
remains.

If we accept that the ultimate theory should not
contain actual infinity at all then the only possible
choice of a number field is
the choice of a Galois field containing $p^n$
elements and the problem arises what the values
of $p$ and $n$ should be. If the Galois field in quantum
theory is characterized by some value of $p$ then we
still have to understand whether there exist
deep reasons for choosing this particular
value of $p$ or this is simply an accident
that our Universe has been created with this
value. In any case, if we accept that $p$ is
a universal constant then the problem arises
what the value of $n$ is and, in view of the
above discussion it is desirable not to postulate
that $n=2$ but to find a motivation for such a
choice.

By analogy with Assumption 1, we accept that
\begin{itemize}
\item {\it Assumption G1: Quantum states in
GFQT are represented by elements of a
linear projective space over a Galois field
$F_{p^n}$ containing $p^n$ elements and physical
quantities are represented by linear operators
in that space.}
\end{itemize}
Then we do not require any analog of Assumption 2.
Instead we accept Requirement 1 which in GFQT has
the same formulation. In the case
of finite-dimensional spaces, the existence of
the spectral decomposition for some operator $A$
means precisely that one can construct a basis
in the usual sense such that all its elements are
the eigenvectors of $A$.

As already noted, mathematicians usually consider
representations over algebraically closed (infinite)
fields while our approach is different. We
consider only (finite) Galois fields and investigate
what is the minimal extension of $F_p$ such that
modular IRs of the
symmetry algebra are fully decomposable.

Consider, for example, a modular analog of IRs of the
su(2) algebra.
Let ${\bf J}=(J_1, J_2, J_3)$ be the
operator of ordinary rotations in
the standard theory. In units $\hbar /2 =1$ the
commutation relations for the
components of ${\bf J}$ have the form
\begin{equation}
[J_1,J_2]=2iJ_3,\quad [J_3,J_1]=2iJ_2,\quad
[J_2,J_3]=2iJ_1
\label{G11}
\end{equation}
One can define the operators $J_{\pm}$ such that
\begin{equation}
J_1 = J_++J_- \quad J_2 = -i(J_+-J_-)
\label{G12}
\end{equation}
Then Eq. (\ref{G11}) can be rewritten as
\begin{equation}
[J_3,J_+]=2J_+\quad [J_3,J_-]=-2J_-\quad [J_+,J_-]=J_3
\label{G13}
\end{equation}
Since Eq. (\ref{G13}) does not contain the quantity $i$,
we now can require that in the modular case
the operators $(J_+J_-J_3)$ act in a space over a
Galois field and satisfy the same relations.

As follows from Eq. (\ref{G13}), the operator
\begin{equation}
K=J_3^2-2J_3+4J_+J_-=J_3^2+2J_3+4J_-J_+
\label{G14}
\end{equation}
is the Casimir operator for algebra $(J_+J_-J_3)$.
Consider the representations containing a vector
$e_0$ such that
\begin{equation}
J_+e_0=0,\quad J_3e_0=se_0
\label{G15}
\end{equation}
where $s\in F_p$. Then, as follows from
Eq. (\ref{G14}), $e_0$ is the eigenvector of the operator
$K$ with the eigenvalue $s(s+2)$. Denote
\begin{equation}
e_n=(J_-)^ne_0,\quad n=0,1,2...
\label{G16}
\end{equation}
Since $K$ is the Casimir operator, all the $e_n$ are its
eigenvectors with the same eigenvalue $s(s+2)$, and, as
follows from Eq. (\ref{G13}), $J_3e_n=(s-2n)e_n$.
Hence it follows from Eq. (\ref{G14}) that
\begin{equation}
J_+J_-e_n=(n+1)(s-n)e_n
\label{G17}
\end{equation}

The maximum value of $n$, $n_{max}$ is defined by
the condition that $J_-e_n=0$ if $n=n_{max}$. This
condition should be compatible with Eq. (\ref{G17})
and therefore $n_{max}=s$. It is easy to see that
the elements $e_n$ for $n=0,1,...s$ form a basis
of modular IR and therefore the dimension of
modular IR with a given $s$ is equal to $s+1$, as
in the standard case. The only difference is that
in the ordinary case $s$ can be any natural number
while in the modular case $s$ can take only the values
of $0,1,...p-1$.

In the standard case the operator $J_3$ is Hermitian
while $J_+^*=J_-$. One can assume that the modular IR
is considered in a space over $F_{p^2}$ and the same
Hermiticity conditions are satisfied. Then it follows
from Eq. (\ref{G17})
that \begin{equation}
(e_{n+1},e_{n+1})=(n+1)(s-n)(e_n,e_n)
\label{G18}
\end{equation}
while the elements $e_n$ with the different values
of $n$ are orthogonal to each other.
Therefore, if $(e_0,e_0)\neq 0$ then all the basis
elements have the nonzero
norm and they are orthogonal to each other. However,
we will not assume in advance that
modular IRs are considered in a space over
$F_{p^2}$. As explained above, our goal is to
investigate what is the minimal extension of $F_p$
such that modular IRs of the su(2) algebra have three
linearly independent observable operators.

The operator $J_3$ has the spectral decomposition
by construction. Consider now the operator
$J_1=J_++J_-$ which in the standard theory is the
$x$ component of the angular momentum. We use the
Pochhammer symbol $(a)_l$ to denote
$a(a+1)\cdots (a+l-1)$ and the standard notation
\begin{equation}
F(a,b;c;z)=\sum_l \frac{(a)_l(b)_lz^l}{l!(c)_l}
\label{G19}
\end{equation}
for the hypergeometric series. Let us define
\begin{equation}
e_j^{(x)}=\sum_{k=0}^s \frac{1}{k!}F(-j,-k;-s;2)e_k
\label{G20}
\end{equation}
As follows from Eqs. (\ref{G16}) and (\ref{G17}),
\begin{eqnarray}
&&J_1e_j^{(x)}=\sum_{k=0}^s \frac{1}{k!}
\{F(-j,-k-1;-s;2)(s-k)+\nonumber\\
&&F(-j,-k+1;-s;2)k\}e_k
\label{G21}
\end{eqnarray}
Now we use the following relation between the
hypergeometric functions (see e.g. Ref. \cite{BE}):
\begin{eqnarray}
&&[c-2a-(b-a)z]F(a,b;c;z)+a(1-z)F(a+1,b;c;z)-\nonumber\\
&&(c-a)F(a-1,b,c;z)=0
\label{G22}
\end{eqnarray}
Then it follows from Eq. (\ref{G21}) that
$J_1e_j^{(x)}=(s-2j)e_j^{(x)}$.

A possible way to prove that the elements $e_j^{(x)}$
$(j=0,1,..s)$ form a basis is to find a transformation
inverse to Eq. (\ref{G20}), i.e. to express the
elements $e_k$ $(k=0,1,...s)$ in terms of $e_j^{(x)}$.
This transformation has the form
\begin{equation}
e_k=\frac{s!}{2^s(s-k)!}\sum_{j=0}^s C_s^j
F(-j,-k;-s;2)e_j^{(x)}
\label{G23}
\end{equation}
($C_s^j=s!/[j!(s-j)!]$) and the proof is as follows.
First, as follows from Eq.
(\ref{G20}), the r.h.s. of Eq. (\ref{G23}) contains
$$\sum_{j=0}^s C_s^j F(-j,-k;-s;2)F(-j,-k;-s;2)$$
We represent this sum as a limit of
$$\sum_{j=0}^s C_s^j F(-j,-k;-s;2)F(-j,-k;-s;2)x^j$$
when $x\rightarrow 1$ and use the formula \cite{BE}
\begin{eqnarray}
&&\sum_{j=0}^s C_s^j F(-j,-k;-s;2)F(-j,-k';-s;2)x^j=\nonumber\\
&&(1+x)^{s-k-k'}(1-x)^{k+k'}F(-k,-k';-s;-\frac{4x}{(1-x)^2})
\label{G24}
\end{eqnarray}
As follows from Eq. (\ref{G19}), the series for the
hypergeometric function in Eq. (\ref{G24}) has the last
term corresponding to $l=min(k,k')$ and this term is
the most singular when $x\rightarrow 1$. Then it is
clear that if $k\neq k'$, the r.h.s. of Eq. (\ref{G24}) is
equal to zero in the limit $x\rightarrow 1$ while if
$k=k'$ then the limit is equal to $2^s(s-k)!/s!$.
This completes the proof of Eq. (\ref{G23})
and we conclude that the operator $J_1$ has the
spectral decomposition even without extending the
field $F_p$.

Consider now the operator $J_+-J_-$. Since in the
standard theory (see Eq. (\ref{G12})) it is equal to
$-iJ_2$ where $J_2$ is the $y$ projection of the
angular momentum, one might expect that in the
modular case $J_+-J_-$ can have a spectral
decomposition only if $F_p$ is extended.

Consider first the simplest nontrivial case when
$s=1$ ($s=1/2$ in the standard units). Then, as
follows from Eqs. (\ref{G16}) and
(\ref{G17}), the characteristic equation for the
operator $J_+-J_-$ is $\lambda^2 = -1$. As explained
above, in the case $p=3\,\,(mod\,\, 4)$ this
equation can be solved only by extending $F_p$.
However, if $p=1\,\,(mod\,\, 4)$, the equation
has solutions in $F_p$ and hence no extension
of $F_p$ is needed to ensure the spectral
decomposition of the operator $J_+-J_-$.
Nevertheless, if $p$ is very large and
$p=1\,\,(mod\,\, 4)$ then
the quantities $\lambda$ satisfying
$\lambda^2 = -1=p-1$ in $F_p$ are very large
(at least of
order $\sqrt{p}$). This obviously contradicts
experiment since in the representation of the
su(2) algebra with $s=1$ no quantities with
such large eigenvalues have been observed.

We conclude that the case
$p=1\,\,(mod\,\, 4)$ is probably incompatible
with the existing data. For this reason we
will consider only quadratic extensions of
$F_p$ in the case $p=3\,\,(mod\,\, 4)$. Then
by analogy with the above discussion one can
prove that the elements
\begin{equation}
e_j^{(y)}=\sum_{k=0}^s \frac{1}{k!}F(-j,-k;-s;2)i^ke_k
\label{G25}
\end{equation}
are the eigenvectors of the operator $J_+-J_-$
with the eigenvalues $i(s-2j)$ and they form the
basis in the representation space.

Our final conclusions in this section are as
follows. If quantum theory is based on a Galois
field then the number $p$ representing the
characteristic of the field is such that
$p=3\,\,(mod\,\, 4)$ rather than
$p=1\,\,(mod\,\, 4)$. Then the
complex extension of $F_p$ guarantees that
modular IRs of the su(2) algebra are fully
decomposable.

As shown in Ref. \cite{complex}, the quadratic
extension of $F_p$ is sufficient to ensure the
spectral decomposition of all the ten basis
representation operators in the case of spinless
modular IRs of the so(1,4) algebra. A generalization
of this result to IRs with the spin would
simultaneously be a proof in the so(2,3) case since
the both algebras are different real forms of the same
complex algebra. However, we will see in Sect.
\ref{S6.3} that in GFQT there also exists another
very interesting possibility for explaining why
this theory is based on $F_{p^2}$.

\vfill\eject

\chapter{Elementary particles in GFQT}
\label{C5}

\section{Modular IRs of the sp(2) algebra}
\label{S5.1}

In the preceding chapter we have described basic facts about
modular representations and their correspondence with the
standard ones. In Sect. \ref{S2.1} we argued that
in the standard theory, symmetry on
quantum level implies that the system under consideration is
described by a representation of some Lie algebra over the field
of real numbers by selfadjoint operators in some Hilbert space.
By analogy, in the present paper we assume that in GFQT
the symmetry algebra
is the Galois field analog of the AdS algebra so(2,3). For this
reason we will describe elementary particles by modular
IRs of this algebra. In turn, for this purpose we
should investigate modular IRs of the sp(2) algebra.

We will use the same notations for the representation
operators as
in Sect. \ref{S3.1} but now we assume that all the
operators are
considered in a space over a Galois field. In particular, Eqs.
(\ref{3.1}) and (\ref{3.2}) are now considered in a linear space
over a Galois field, and we can consider IRs containing
a vector
$e_0$ satisfying Eq. (\ref{3.3}). However, the quantity
$q_0$ should
now be understood as an element from $F_{p^2}$.
As noted in Sect. \ref{SG2}, one can prove that in fact $q_0$ is
"real" i.e. belongs to $F_p$. The case $q_0=0$ describes a
trivial IRs, so we will assume that $q_0\neq 0$.

We can derive Eqs. (\ref{3.4}-\ref{3.6}) in full
analogy with the
derivation in Sect. \ref{S3.1}. However, Eqs. (\ref{3.5}) and
(\ref{3.6}) now indicate that ordinary and modular IRs
differ each other in the following sense. The ordinary IRs
are infinite dimensional and $n=0,1,2...\infty$.
On the other hand,
in the modular case $q_0$ is one of the
numbers $1,...p-1$, and
the set $(e_0,e_1,...e_N)$ will be a basis of
IR if $a"e_i\neq 0$
for $i<N$ and $a"e_N=0$. These conditions
must be compatible with $a'a"e_N=0$.
Therefore, as follows from
Eq. (\ref{3.5}), $N$ is defined by the condition $q_0+N=0$ in
$F_p$. As a result, $N=p-q_0$ and the dimension of IR is equal
to $p-q_0+1$ (in agreement with the Zassenhaus
theorem \cite{Zass}).

One might say that $e_0$ is the vector with the
minimum weight while $e_N$ is the vector with
the maximum weight. However, the notions of
"less than" or "greater than" have only
a limited sense in $F_p$, as well as the notion of positive
and negative numbers in $F_p$ (see the preceding chapter). If
$q_0$ is positive in this sense then Eqs. (\ref{3.5}) and
(\ref{3.6}) indicate that the modular
IR under consideration can be
treated as a modular analog of IR with the minimum weight.
However, it is easy to see that $e_N$ is the
eigenvector of the operator $h$ with the
eigenvalue $-q_0$ in
$F_p$, and the same IR can be treated as a modular analog
of an IR with the maximum weight if $e_N$ is identified
with $e_0'$ in Eq. (\ref{3.7}).
Therefore one modular IR is an analog of two standard IRs
with the minimum and maximum weights, respectively.

Suppose that $q_0$ is positive and $q_0\ll p$ in the sense
of Sect. \ref{SG1}, i.e. $f(q_0)>0$ and $f(q_0)\ll p$.
Then, if $p$ is very large, there exists a large number
$N_0$ such that for
all $j,k<N_0$ the matrix elements of the operators
($a',a",h$) for all the transitions betweens the vectors
$e_j$ and $e_k$ belong to $U_0$. It is obvious that such
a quantity $N_0$ can be found since
all the expressions in Sect. \ref{S3.1} do not contain
division in $F_p$. In that case there exists the
correspondence between modular and standard IRs in the
sense of Sect. \ref{SG2}. This means that if one considers
the action of representation operators only on linear
combinations of elements $e_j$ with $j\ll N$ in the
projective Hilbert space and projective space over
$F_{p^2}$, respectively, then modular IRs are
practically indistinguishable from the standard ones
with the minimum weight. Analogously, if one considers
the action of representation operators only on linear
combinations of elements $e_j'$ with $j\ll N$ then
modular IRs are practically indistinguishable from the
standard ones with the maximum weight.

Let us forget for a moment that the eigenvalues of the
operator $h$ belong to $F_p$ and will treat them as
integers. Then, as follows from Eq. (\ref{3.4}), the
eigenvalues are
$$q_0,q_0+2,...,2p-2-q_0, 2p-q_0.$$
Therefore, if $f(q_0)>0$ and $f(q_0)\ll p$,
the maximum value of $q_0$ is equal to
$2p-q_0$, i.e. it is of order $2p$. In this case the
correspondence with standard IRs with the minimum weight
exists in the region where the eigenvalues of the
operator $h$ are positive and much less than $p$ while the
correspondence with standard IRs with the maximum weight
exists in the region where the eigenvalues of the
operator $h$ are close to $2p$ but less than
$2p$. This observation will be used below.

Finally we note that if $p=3\,\, (mod\, 4)$ then the
same arguments as in Sect. \ref{SG3} show that 
for modular IRs of the sp(2) algebra all
representation operators are fully decomposable.
This is also clear from the fact that su(2) and
sp(2) are different real forms of the same complex
Lie algebra.

\section{Modular IRs of the so(2,3) algebra}
\label{S5.2}

\begin{sloppypar}
We will decribe modular IRs of the $so(2,3)$ algebra
(more precisely its Galois field analog) by the set
of operators $(b',b',L_+,L_-)$ and $(a_j',a_j",h_j)$ ($j=1,2$)
satisfying the same commutation relations as in Sect.
\ref{S3.2}, but the operators act now in a space over
$F_{p^2}$. Analogously we can define the
$A$ operators as in Eq. (\ref{3.12}) and the vector
$e_0$ as in Eq. (\ref{3.14}), but the quantities $q_j$,
$m$ and $s$ now belong to $F_p$. Then if the $e_{nk}$ are
defined as in Eq. (\ref{3.15}), they will satisfy Eqs.
(\ref{3.16}-\ref{3.19}).
\end{sloppypar}

As follows from Eqs. (\ref{3.18}) and (\ref{3.19}), the
quantity $k$ takes the values $0,1,...s$ as in the
standard case. At the same time, it follows from Eqs.
(\ref{3.16}) and (\ref{3.17}), that in the modular case
$n=0,1,...n_{max}$ where $n_{max}$ is the first number
for which the numerator in Eqs. (\ref{3.16}) becomes
zero in $F_p$. In the massive case $n_{max}=p+2-m$
while in the massless one $n_{max}=p-1-s$ if $k=0$ or
$k=s$ and $n_{max}=0$ if $0<k<s$.

The full basis of the representation space can again be chosen
in the form (\ref{3.20}) but now, as follows from the
results of Sects. \ref{S3.2} and \ref{S5.1},
\begin{eqnarray}
&n_1=0,1,...N_1(n,k)\quad n_2=0,1,...N_2(n,k)\nonumber\\
&N_1(n,k)=p-q_1-n+k\quad N_2(n,k)=p-q_2-n-k
\label{5.1}
\end{eqnarray}
As a consequence, the dimension of modular IR
characterized by the mass $m$ and spin $s$ is given by
\begin{equation}
Dim(m,s)=\sum_{nk} (p-q_1-n+k+1)(p-q_2-n-k+1)
\label{dim}
\end{equation}
where the sum should be taken over all possible values of $n$
and $k$. A direct calculation shows that in the massive case
\begin{eqnarray}
&Dim(m,s)=\frac{1}{12}(p+3-m)(s+1)[4p^2-2p(m-3)+\nonumber\\
&(m-2)(m-4)-s(s-2)]
\label{massive}
\end{eqnarray}
in the massless case with $s<2$
\begin{equation}
Dim(2,0)==\frac{1}{6}p(2p+1)(p+1)\quad
Dim(3,1)=\frac{2}{3}p(p-1)(p+1)
\label{massless}
\end{equation}
in the massless case with $s\geq 2$
\begin{equation}
Dim(2+s,s)=\frac{1}{6}[4p^3+2p^2s-2p(3s^2-1)+s(s-1)(s+1)]
\label{massless2}
\end{equation}
and in the singleton case
\begin{equation}
Dim(1,0)=(p^2+1)/2\quad Dim(2,1)=(p^2-1)/2
\label{singleton}
\end{equation}
It is clear that Eqs. (\ref{dim}-\ref{singleton})
should be understood in
the usual sense (not in the sense of $F_p$). Therefore we
conclude that if the spin is not large, the dimension of IR
massive and massless IRs is or order $p^3$ while the
dimension of singleton IRs is of order $p^2/2$. Since any
Galois field is finite, each modular IR turns out
to be not only finite dimensional but even finite.

In the preceding section we discussed the correspondence
between modular and standard IRs of the sp(2) algebra.
In the case of the so(2,3) algebra the correspondence
exists if the following conditions are satisfied.
Suppose that $p$ is very large and each of the
quantities $(q_1,q_2,m,s)$ is
positive and much less than $p$ in the sense of Sect.
\ref{SG1}, e.g. $f(q_1)>0$ and $f(q_1)\ll p$
(this assumption
excludes the singleton cases since 1/2 and 3/2 are
very large numbers in $F_p$). Then
we can find a large number $N_0$
such that the matrix elements of the operators
$(a_j',a_j",h_j)$ ($j=1,2$) for the transitions
between the states $e(n_1n_2nk)$ and $e(n_1'n_2'n'k')$
belong to $U_0$ when all the quantum numbers in question
are less than $N_0$. In that case there exist many
states $\sum c(n_1n_2nk)e(n_1n_2nk)$
for which the action of the operators $(a_j',a_j",h_j)$
in the projective Hilbert
space and the projective space over $F_{p^2}$
are practically indistinguishable.

The problem arises whether it is possible to have the
correspondence for all the ten basis elements of our
IR. Indeed, the matrix elements of the operators
$(b',b",L_+,L_-)$ in the modular case can again be
written in the form of Eqs. (\ref{3.31}),
(\ref{3.33}-\ref{3.35}) but now division should be
understood in the sense of $F_p$. As shown in Ref.
\cite{lev4}, one can construct a nonorthogonal basis
of IR such that the correspondence exists for
all the ten operators. It is clear that the
correspondence should necessarily exist for the
diagonal operators since their eigenvalues represent
physical properties. It is not clear yet whether it
is important that the correspondence should also
exist for the remaining operators.

As noted in Sect. \ref{S3.2}, in the standard case there
also exist IRs with negative energies, which can be
constructed by using Eq. (\ref{3.24}).
At the same time, {\it the modular analog of a
positive energy IR characterized by $q_1,q_2$ in
Eq. (\ref{3.14}), and the modular
analog of a negative energy IR characterized by the same
values of $q_1,q_2$ in Eq. (\ref{3.24}) represent the same
modular IR.} This is the crucial difference between the
standard quantum theory and GFQT, and a proof is given
below.

\begin{sloppypar}
Let $e_0$ be a vector satisfying Eq. (\ref{3.14}). Denote
$N_1=p-q_1$ and $N_2=p-q_2$. Our goal is to prove
that the vector $x=(a_1")^{N_1}(a_2")^{N_2}e_0$ satisfies
the conditions
(\ref{3.24}), i.e. $x$ can be identified with $e_0'$.
\end{sloppypar}

As follows from the definition of $N_1,N_2$
and the results of Sects. \ref{S3.1} and \ref{S5.1},
the vector $x$ is the eigenvector of the operators $h_1$
and $h_2$ with the eigenvalues $-q_1$ and $-q_2$,
respectively, and, in
addition, it satisfies the conditions $a_1"x=a_2"x=0$.

Let us now prove that $b"x=0$. Since $b"$ commutes with the
$a_j"$, we can write $b"x$ in the form
\begin{equation}
b"x = (a_1")^{N_1}(a_2")^{N_2}b"e_0
\label{5.2}
\end{equation}
As follows from Eqs. (\ref{3.10}) and (\ref{3.14}),
$a_2'b"e_0=L_+e_0=0$ and $b"e_0$ is the eigenvector
of the operator $h_2$ with the eigenvalue $q_2+1$.
Therefore, $b"e_0$ is the minimal vector of the sp(2)
IR which has the dimension $p-q_2=N_2$.
Therefore $(a_2")^{N_2}b"e_0=0$ and $b"x=0$.

The next stage of the proof is to show that $L_-x=0$.
As follows from Eq. (\ref{3.10}) and the definition of
$x$,
\begin{equation}
L_-x = (a_1")^{N_1}(a_2")^{N_2}L_-e_0-
N_1(a_1")^{N_1-1}(a_2")^{N_2}b"e_0
\label{5.3}
\end{equation}
We have already shown that $(a_2")^{N_2}b"e_0=0$,
and therefore it is sufficient to prove that the first term
in the r.h.s. of Eq. (\ref{5.3}) is equal to zero. As follows
from Eqs. (\ref{3.10}) and (\ref{3.14}), $a_2'L_-e_0=b'e_0=0$,
and $L_-e_0$ is the eigenvector of the operator $h_2$ with the
eigenvalue $q_2+1$. Therefore $(a_2")^{N_2}L_-e_0=0$ and we
have proved that $L_-x=0$.

The fact that $(x,x)\neq 0$ immediately follows from the
definition of the vector $x$ and the results of the preceding
section. Therefore the vector $x$ can be indeed
identified with $e_0'$ and the above statement is proved.

By analogy with the consideration in the preceding section,
we forget for a moment that the eigenvalues of the operators
$h_1$ and $h_2$ should be treated as elements of $F_p$.
If they are treated as integers than one modular IR
corresponds to a standard IR with the positive energy
in the region where the eigenvalues of these operators
are positive and much less than $p$. The same modular
IR corresponds to an IR with the negative energy in the
region where the eigenvalues are each close to $2p$ but
less than $2p$. Recall that the AdS energy is $h_1+h_2$.
Therefore one modular IR corresponds to a standard
positive energy IR in the region where the energy is
positive and much less than $p$. At the same time, it
corresponds to an IR with the negative energy in the
region where the AdS energy is close to $4p$ but less
than $4p$. This observation will be used in the
discussion of the vacuum condition in Sect. \ref{S5.5}.

The matrix elements of the operator $M^{ab}$ can be again
defined by Eq. (\ref{3.25}). The important difference
between the standard and modular IRs is that in the
latter the trace of each operator $M^{ab}$ is
equal to zero while in the former this is obviously not the
case (for example, the energy operator is positive definite
for IRs defined by Eq. (\ref{3.14}) and negative definite
for IRs defined by Eq. (\ref{3.24})).
For the operators $(a_j',a_j",L_{\pm},b',b")$ the validity
of this statement is clear immediately: since they
necessarily change one of the quantum numbers $(n_1n_2nk)$
(see Sect. \ref{S3.3}), they do not contain
nonzero diagonal elements at all. The proof for the
diagonal operators $h_1$ and $h_2$ is as follows. For each
IR of the sp(2) algebra with the "minimal weight" $q_0$ and
the dimension $N+1$, the eigenvalues of the operator $h$ are
$(q_0,q_0+2,...q_0+2N)$. The sum of these eigenvalues is
equal to zero in $F_p$ since $q_0+N=0$ in $F_p$ (see the
preceding section). Therefore we conclude that
\begin{equation}
\sum_{n_1n_2nk} M^{ab}(n_1n_2nk;n_1n_2nk)=0
\label{5.4}
\end{equation}
This property is very important for investigating a new
symmetry between particles and antiparticles in GFQT,
which is discussed in the subsequent section.

\section{AB symmetry}
\label{S5.3}

In Sect. \ref{S3.4} we discussed the quantization
problem in the standard case and represented the
quantized representation operators in the form of
Eq. (\ref{3.41}) where the $(a,a^*)$ operators
refer to a particle and the $(b,b^*)$ operators ---
to its antiparticle. Such a form is a consequence of
the assumption that in the standard theory a
particle and its antiparticle are described by
equivalent but different positive energy IRs.
However, the results of the preceding section
show that one modular IR corresponds to two standard
IRs with the positive and negative energies,
respectively. This indicates to a possibility that
one modular IR describes a particle and its
antiparticle simultaneously. However, we don't
know yet what should be treated as a particle and
its antiparticle in the modular case. We have a
description of an object such that $(n_1n_2nk)$ is
the full set of its quantum numbers, and these
numbers take the values described in the preceding
section. For this reason we now assume that
$a(n_1n_2nk)$ is the operator describing
annihilation of the object with the quantum
numbers $(n_1n_2nk)$. Analogously $a(n_1n_2nk)^*$
describes creation of the object with the quantum
numbers $(n_1n_2nk)$. If these operators anticommute
then they satisfy Eq. (\ref{3.36}) while if they
commute then they satisfy Eq. (\ref{3.37}).
Then, by analogy with the consideration in Sect.
\ref{S3.4}, we conclude that the operators
\begin{eqnarray}
&M^{ab}=\sum M^{ab}(n_1'n_2'n'k';n_1n_2nk)\nonumber\\
&a(n_1'n_2'n'k')^*a(n_1n_2nk)/Norm(n_1n_2nk)
\label{5.5}
\end{eqnarray}
satisfy the commutation relations in the form
(\ref{2.4}) or (\ref{3.8}-\ref{3.10}). In this expression
the sum is taken over all possible values of the
quantum numbers in the modular case.

As noted in Sect. \ref{S3.4}, the solution
satisfying the required commutation relations can be
taken not only in the form (\ref{3.41}) but also
as in Eq. (\ref{3.42}). As noted in Sect. \ref{S3.4},
in the standard theory the requirement selecting
Eq. (\ref{3.41}) is rather artificial.

In the modular case the solution also can be taken
not only as in Eq. (\ref{5.5}) but also as
\begin{eqnarray}
&M^{ab}=\mp\sum M^{ab}(n_1'n_2'n'k';n_1n_2nk)\nonumber\\
&a(n_1n_2nk)a(n_1'n_2'n'k')^*/Norm(n_1n_2nk)
\label{5.6}
\end{eqnarray}
for the cases of anticommutators and commutators,
respectively.
However, as follows from Eqs. (\ref{3.36}), 
(\ref{3.37}) and (\ref{5.3}), the solutions 
(\ref{5.5}) and
(\ref{5.6}) are the same. Therefore in the
modular case there is no need to impose an
additional requirement that all operators
should be written in the normal form.
Moreover, we can define the following

{\it $^*$ symmetry condition: the free
theory should be invariant under the transformation
replacing each operator 
$$\sum A(n_1'n_2'n'k';n_1n_2nk)a(n_1'n_2'n'k')^*
a(n_1n_2nk)$$
by 
$$\mp\sum A(n_1'n_2'n'k';n_1n_2nk)
a(n_1n_2nk)a(n_1'n_2'n'k')^*$$ }
At the same time, since
the meaning of the operators $(a,a^*)$ is not
clear yet, the problem arises whether the vacuum
is the eigenvalue of all the quantized representation
operators with all the eigenvalues equal to zero.
This problem will be discussed in Sect. \ref{S6.1}.

The problem with the treatment of the $(a,a^*)$
operators is as follows.  When the values of
$(n_1n_2n)$ are much less than $p$, the modular IR
corresponds to the standard one (see the preceding
section) and therefore the $(a,a^*)$ operator can
be treated as those describing the particle
annihilation and creation, respectively. However,
when the AdS energy is negative, the operators
$a(n_1n_2nk)$ and $a(n_1n_2nk)^*$ become unphysical
since they describe annihilation and creation,
respectively, in the unphysical region of negative
energies.

Let us recall that at any fixed
values of $n$ and $k$, the quantities $n_1$ and $n_2$
can take only the values described in Eq. (\ref{5.1})
and the eigenvalues of the operators $h_1$ and $h_2$
are given by Eq. (\ref{3.26}), which now should be
considered in $F_p$.
As follows from the results of the preceding section,
the first IR of the sp(2) algebra has the dimension
$N_1(n,k)+1$ and the second IR has the dimension
$N_2(n,k)+1$. If $n_1=N_1(n,k)$
then it follows from Eq. (\ref{3.26}) that the first
eigenvalue is equal to $-Q_1(n,k)$ in $F_p$, and if
$n_2=N_2(n,k)$ then
the second eigenvalue is equal to $-Q_2(n,k)$ in $F_p$.
We use ${\tilde n}_1$ to denote $N_1(n,k)-n_1$ and
${\tilde n}_2$ to denote
$N_2(n,k)-n_2$. Then it follows from Eq.
(\ref{3.26}) that
$e({\tilde n}_1{\tilde n}_2nk)$ is the
eigenvector
of the operator $h_1$ with the eigenvalue $-(Q_1(n,k)+2n_1)$
and the
eigenvector of the operator $h_2$ with the eigenvalue
$-(Q_2(n,k)+2n_2)$.

As noted in Sect. \ref{S2.3}, the standard theory
implicitly involves the idea that creation of the
antiparticle with the positive energy can be treated
as annihilation of the corresponding particle with the
negative energy and annihilation of the
antiparticle with the positive energy can be treated
as creation of the corresponding particle with the
negative energy. In GFQT we can implement this
idea explicitly. Namely, we can define the
operators $b(n_1n_2nk)$ and $b(n_1n_2nk)^*$ in such a
way that they will replace the $(a,a^*)$ operators if
the quantum numbers are unphysical. In addition,
if the values of $(n_1n_2n)$ are much less than $p$,
the operators $b(n_1n_2nk)$ and $b(n_1n_2nk)^*$
should be interpreted as physical operators
describing annihilation and creation of
antiparticles, respectively.

In GFQT the $(b,b^*)$ operators cannot be independent
of the $(a,a^*)$ operators since the latter are defined
for all possible quantum numbers. Therefore the $(b,b^*)$
operators should be expressed in terms of the $(a,a^*)$
ones. We can implement the above idea if the
operator $b(n_1n_2nk)$ is defined in such a way that
it is proportional to $a({\tilde n}_1,{\tilde n}_2,n,k)^*$
and hence $b(n_1n_2nk)^*$ is proportional to
$a({\tilde n}_1,{\tilde n}_2,n,k)$.

Since Eq. (\ref{3.23}) should now be considered in $F_p$,
it follows from the well known Wilson
theorem $(p-1)!=-1$ in $F_p$ (see e.g. \cite{VDW}) that
\begin{equation}
F(n_1n_2nk)F({\tilde n}_1{\tilde n}_2nk) = (-1)^s
\label{5.7}
\end{equation}
We now define the $b$-operators as follows.
\begin{equation}
a(n_1n_2nk)^*=\eta(n_1n_2nk) b({\tilde n}_1{\tilde n}_2nk)/
F({\tilde n}_1{\tilde n}_2nk)
\label{5.8}
\end{equation}
where $\eta(n_1n_2nk)$ is some function.
As a consequence of this definition,
\begin{eqnarray}
&a(n_1n_2nk)=\bar{\eta}(n_1n_2nk) b({\tilde n}_1{\tilde n}_2nk)^*/
F({\tilde n}_1{\tilde n}_2nk)\nonumber\\
&b(n_1n_2nk)^*=a({\tilde n}_1{\tilde n}_2nk)
F(n_1n_2nk)/{\bar \eta}({\tilde n}_1{\tilde n}_2nk)\nonumber\\
&b(n_1n_2nk)=a({\tilde n}_1{\tilde n}_2nk)^*
F(n_1n_2nk)/\eta({\tilde n}_1{\tilde n}_2nk)
\label{5.9}
\end{eqnarray}
Eqs. (\ref{5.8}) and (\ref{5.9}) define a relation
between the sets $(a,a^*)$ and $(b,b^*)$. Although our
motivation was to replace the $(a,a^*)$ operators by the
$(b,b^*)$ ones only for the nonphysical values of the
quantum numbers, we can consider this definition for all
the values of $(n_1n_2nk)$.

We have not discussed yet what exact definition of the
physical and nonphysical quantum numbers should be. This problem
will be discussed in Sect. \ref{S5.5}. However, one might accept

{\it Physical-nonphysical states assumption:
Each set of quantum numbers $(n_1n_2nk)$
is either physical or unphysical. If it is physical then
the set $({\tilde n}_1{\tilde n}_2nk)$ is unphysical
and vice versa.}

With this assumption we can conclude from Eqs. (\ref{5.8})
and (\ref{5.9}) that if some operator $a$ is physical then
the corresponding operator $b^*$ is unphysical and vice
versa while if some operator $a^*$ is physical then
the corresponding operator $b$ is unphysical and vice
versa.

We have no ground to think that the set of the $(a,a^*)$
operators is more fundamental than the set of the $(b,b^*)$
operators and vice versa. Therefore the question arises
whether the $(b,b^*)$ operators satisfy the relations
(compare with Eqs. (\ref{3.36}) and (\ref{3.37}))
\begin{eqnarray}
&\{b(n_1n_2nk),b(n_1'n_2'n'k')^*\}=\nonumber\\
&Norm(n_1n_2nk)
\delta_{n_1n_1'}\delta_{n_2n_2'}\delta_{nn'}\delta_{kk'}
\label{5.10}
\end{eqnarray}
in the case of anticommutators,
\begin{eqnarray}
&[b(n_1n_2nk),b(n_1'n_2'n'k')^*]=\nonumber\\
&Norm(n_1n_2nk)
\delta_{n_1n_1'}\delta_{n_2n_2'}\delta_{nn'}\delta_{kk'}
\label{5.11}
\end{eqnarray}
in the case of commutators, and the operators $M^{ab}$
in terms of the $(b,b^*)$ operators read
\begin{eqnarray}
&M^{ab}=\sum M^{ab}(n_1'n_2'n'k';n_1n_2nk)\nonumber\\
&b(n_1'n_2'n'k')^*b(n_1n_2nk)/Norm(n_1n_2nk)
\label{5.12}
\end{eqnarray}
In contrast with the standard theory, we cannot postulate
these properties independently since the
$(b,b^*)$ operators are not independent of the $(a,a^*)$
operators. It is natural to accept the following

{\it Definition of the AB symmetry: If the $(b,b^*)$
operators satisfy Eq. (\ref{5.10}) in the case of
anticommutators or Eq. (\ref{5.11}) in the case of
commutators, and the operators (\ref{5.5}) in terms of
the $(b,b^*)$ operators have the form (\ref{5.12}) then
it is said that the AB symmetry is satisfied.}

\section{Proof of the AB symmetry}
\label{S5.4}

Our first goal is to investigate whether Eqs. (\ref{5.10})
and (\ref{5.11}) can follow from Eqs. (\ref{3.36}) and
(\ref{3.37}), respectively. As follows from Eqs.
(\ref{5.7}-\ref{5.9}), Eq. (\ref{5.10}) follows from
Eq. (\ref{3.36}) if
\begin{equation}
\eta(n_1n_2nk) {\bar \eta}(n_1,n_2,nk)=(-1)^s
\label{5.13}
\end{equation}
while Eq. (\ref{5.11}) follows from Eq. (\ref{3.37}) if
\begin{equation}
\eta(n_1n_2nk) {\bar \eta}(n_1,n_2,nk)=(-1)^{s+1}
\label{5.14}
\end{equation}
We now represent $\eta(n_1n_2nk)$ in the form
\begin{equation}
\eta(n_1n_2nk)=\alpha f(n_1n_2nk)
\label{5.15}
\end{equation}
where $f(n_1n_2nk)$ should satisfy the condition
\begin{equation}
f(n_1n_2nk) {\bar f}(n_1,n_2,nk)=1
\label{5.16}
\end{equation}
Then $\alpha$ should be such that
\begin{equation}
\alpha {\bar \alpha}=\pm (-1)^s
\label{5.17}
\end{equation}
where the plus sign refers to anticommutators and the minus
sign to commutators, respectively.
If the normal spin-statistics connection is valid,
i.e. we have anticommutators for odd values of $s$ and
commutators for even ones then the r.h.s. of Eq.
(\ref{5.17}) equals -1 while in the opposite case it
equals 1. In Sect. \ref{S6.3} Eq. (\ref{5.17}) is
discussed in detailed and it is shown that it has
solutions in $F_{p^2}$ in both cases.

Our next goal is to prove that all the operators in
Eq. (\ref{5.5}) have the same form in terms of the
$(a,a^*)$ and $(b,b^*)$ operator.
Let us consider Eq. (\ref{5.5}) and
use the fact that in the modular case the trace of the
operators $M^{ab}$ is equal to zero (see Eq. (\ref{5.4})).
Therefore, as follows from Eqs. (\ref{3.36}) and (\ref{3.37}),
one can rewrite Eq. (\ref{5.5}) as
\begin{eqnarray}
&M^{ab}=\mp \sum
M^{ab}(n_1'n_2'n'k';n_1n_2nk)\nonumber\\
&a(n_1n_2nk)a(n_1'n_2'n'k')^*/Norm(n_1n_2nk)
\label{5.19}
\end{eqnarray}
where the minus sign refers to anticommutators
and the plus sign -
to commutators. Then by using Eqs. (\ref{5.8}),
(\ref{5.9}), (\ref{5.16})
and (\ref{5.17}), one gets in the both cases
\begin{eqnarray}
&M^{ab}=-\sum
M^{ab}(n_1'n_2'n'k';n_1n_2nk)
f(n_1'n_2'n'k'){\bar f}(n_1n_2nk)\nonumber\\
&b({\tilde n}_1{\tilde n}_2nk)^*
b({\tilde n}_1'{\tilde n}_2'n'k')/
[F({\tilde n}_1'{\tilde n}_2'n'k')G(k)]=\nonumber\\
&-\sum M^{ab}({\tilde n}_1{\tilde n}_2nk;
{\tilde n}_1'{\tilde n}_2'n'k')
f({\tilde n}_1{\tilde n}_2nk)
{\bar f}({\tilde n}_1'{\tilde n}_2'n'k')\nonumber\\
&b(n_1'n_2'n'k')^*b(n_1n_2nk)/[F(n_1n_2nk)G(n'k')]
\label{5.20}
\end{eqnarray}

We first consider the diagonal operator $h_1$. As follows
from Eq. (\ref{3.26}), the matrix elements of this 
operator are given by
\begin{equation}
h_1(n_1'n_2'n'k';n_1n_2nk)=
(Q_1(n,k)+2n_1)\delta_{n_1n_1'}\delta_{n_2n_2'}
\delta_{nn'}\delta_{kk'}
\label{5.21}
\end{equation}
Therefore, as follows from Eqs. (\ref{5.5}),
(\ref{5.20}) and (\ref{5.21}),
the operator $h_1$ will have the same form in terms
of the $(a,a^*)$ and
$(b,b^*)$ operators if $Q_1(n,k)+2n_1 =-(Q_1(n,k)+2{\tilde n}_1)$.
This relation immediately follows from the
definition of ${\tilde n}_1$.
Analogously we can prove that the quantized
$h_2$ operator has the
same form in terms of $(a,a^*)$ and $(b,b^*)$.

For simplicity we will not investigate general
restrictions imposed
by the AB symmetry on the function $f(n_1n_2nk)$
but instead assume that
\begin{equation}
f(n_1n_2nk)=(-1)^{n_1+n_2+n}
\label{5.22}
\end{equation}
This expression is obviously compatible with Eq. (\ref{5.16}).

Consider now the operator $a_1"$. As follows from Eq. (\ref{3.28}),
its matrix elements are given by
\begin{equation}
a_1"(n_1'n_2'n'k';n_1n_2nk)=\delta_{n_1,n_1'-1}
\delta_{n_2n_2'}\delta_{nn'}\delta_{kk'}
\label{5.23}
\end{equation}
Therefore, as follows from Eqs. (\ref{5.5}),
(\ref{5.20}) and (\ref{5.22}),
the quantized operator $a_1"$ has the same form
in terms of $(a,a^*)$
and $(b,b^*)$. By using the same expressions and
Eq. (\ref{3.28}), one can
also show that the same is valid for the quantized
operator $a_1'$ since
$$n_1[Q_1(n,k)+n_1]=({\tilde n}_1+1)[Q_1(n,k)+{\tilde n}_1]$$
when $n_1'=n_1-1$ and the other quantum numbers
are the same. The proof for
the operators $a_2'$ and $a_2"$ is analogous.

The next step is to investigate whether the remaining
operators $(b',b",L_+,L_-)$ have the same
form in terms of $(a,a^*)$ and $(b,b^*)$. We will discuss only the
operator $b"$ since the proof for the other operators is similar.
As follows from Eq. (\ref{3.31}), the
matrix elements of this operator
are given by
\begin{eqnarray}
&b"(n_1'n_2'n'k';n_1n_2nk)=\{[Q_1(n,k)-1]
[Q_2(n,k)-1]\}^{-1}\nonumber\\
&[k(s+1-k)(q_1-k-1)(q_2+k-2)\delta_{n_1'n_1}\delta_{n_2',n_2+1}
\delta_{n'n}\delta_{k',k-1}+\nonumber\\
&n(m+n-3)(q_1+n-1)(q_2+n-2)\delta_{n_1',n_1+1}\delta_{n_2',n_2+1}
\delta_{n',n-1}\delta_{k'k}+\nonumber\\
&\delta_{n_1'n_1}\delta_{n_2'n_2}\delta_{n',n+1}\delta_{k'k}+
\delta_{n_1',n_1+1}\delta_{n_2'n_2}\delta_{n'n}\delta_{k',k+1}
\label{5.24}
\end{eqnarray}
This expression shows that $n_1'+n_2'+n'=n_1+n_2+n+1$. Therefore,
as follows from Eqs. (\ref{5.5}), (\ref{5.20}) and (\ref{5.22}),
the matrix elements (\ref{5.24}) should satisfy the requirement
\begin{equation}
\frac{1}{G(n'k')}b"({\tilde n}_1{\tilde n}_2nk;
{\tilde n}_1'{\tilde n}_2'n'k')
=\frac{1}{G(nk)}b"(n_1'n_2'n'k';n_1n_2nk)
\label{5.25}
\end{equation}
As follows from the definition of the quantities
$({\tilde n}_1,{\tilde n}_2)$ and Eqs. (\ref{3.23}) and (\ref{5.24}),
this relation is indeed valid.

We conclude that if $\eta(n_1n_2nk)$ is given by
Eq. (\ref{5.15}) and $f(n_1n_2nk)$ is
given by Eq. (\ref{5.22}) then the AB symmetry is valid
regardless of whether the normal spin-statistics
connection is valid or not. The analysis of the proof
for all the representation
operators shows that Eq. (\ref{5.22}) defines
the function $f(n_1n_2nk)$
uniquely up to a constant $c$ such that $c{\bar c}=1$.

\section{Physical and nonphysical states}
\label{S5.5}

Although we have called the sets $(a,a^*)$ and
$(b,b^*)$ annihilation and creation operators for
particles and antiparticles, respectively, it is not
clear yet whether these operators indeed can be treated
in such a way. The operator $a(n_1n_2nk)$ can be called
the annihilation one only if it annihilates the vacuum
vector $\Phi_0$. Then if the operators $a(n_1n_2nk)$ and
$a(n_1n_2nk)^*$ satisfy the relations (\ref{3.36}) or
(\ref{3.37}), the vector $a(n_1n_2nk)^* \Phi_0$ has the
meaning of the one-particle state. The same can be said
about the operators $b(n_1n_2nk)$ and $b(n_1n_2nk)^*$.
For these reasons in the standard theory it is required
that the vacuum vector should satisfy the conditions
\begin{equation}
a(n_1n_2nk)\Phi_0=b(n_1n_2nk)\Phi_0=0\quad
\forall\,\, n_1,n_2,n,k
\label{5.28}
\end{equation}
Then the elements
\begin{equation}
\Phi_+(n_1n_2nk)=a(n_1n_2nk)^*\Phi_0\quad
\Phi_-(n_1n_2nk)=b(n_1n_2nk)^*\Phi_0
\label{5.29}
\end{equation}
have the meaning of one-particle states for particles
and antiparticles, respectively.

However, if one requires the condition (\ref{5.28})
in GFQT then it is obvious from Eqs. (\ref{5.8})
and Eq. (\ref{5.9}), that the elements defined by
Eq. (\ref{5.29}) are null vectors. Note that in the
standard approach the AdS energy is always greater
than $m$ while in GFQT the AdS energy is not
positive definite. We could therefore modify
Eq. (\ref{5.28}) as follows. Suppose that
'Physical-nonphysical states assumption' (see Sect.
\ref{S5.3}) can be substantiated. Then we can break
the set of elements $(n_1n_2nk)$ into two equal
nonintersecting parts, $S_+$ and $S_-$, such that if
$(n_1n_2nk)\in S_+$ then $({\tilde n}_1{\tilde n}_2nk)\in S_-$
and vice versa. Then, instead of the condition
(\ref{5.28}) we require
\begin{equation}
a(n_1n_2nk)\Phi_0=b(n_1n_2nk)\Phi_0=0\quad
\forall\,\, (n_1,n_2,n,k)\in S_+
\label{5.30}
\end{equation}
In that case the elements defined by Eq. (\ref{5.29})
will indeed have the meaning of one-particle states
for $(n_1n_2nk)\in S_+$.

It is clear that if we
want to work with the full set of elements
$(n_1n_2nk)$ then, as follows from Eqs. (\ref{5.8})
and (\ref{5.9}), the operators $(b,b^*)$ are redundant
and we should work only with the operators $(a,a^*)$.
This possibility cannot be excluded (see the discussion in
the subsequent chapter) but if one believes that the
separation of states into particle and antiparticle
ones should be valid at any conditions (not only when
the quantum numbers $(n_1n_2n)$ are much less than $p$)
then the operators $(a,a^*,b,b^*)$ can be independent of
each other only for a half of the elements $(n_1n_2nk)$.

\begin{sloppypar}
Regardless of how the sets $S_+$ and $S_-$ are defined,
the 'Physical-nonphysical states assumption' cannot be
consistent if there exist quantum numbers $(n_1n_2nk)$
such that $n_1={\tilde n}_1$ and $n_2={\tilde n}_2$.
Indeed, in that case the sets $(n_1n_2nk)$ and
$({\tilde n}_1{\tilde n}_2nk)$ are the same what contradicts
the assumption that each set $(n_1n_2nk)$ belongs either
to $S_+$ or $S_-$.
\end{sloppypar}

Since the replacements $n_1\rightarrow {\tilde n}_1$ and
$n_2\rightarrow {\tilde n}_2$ change the signs of the
eigenvalues of the $h_1$ and $h_2$ operators (see Sect.
\ref{S5.3}), the condition that that $n_1={\tilde n}_1$
and $n_2={\tilde n}_2$ should be valid, simultaneously
implies that the eigenvalues of the operators $h_1$ and
$h_2$ should be equal to zero simultaneously. Recall that
(see Sect. \ref{S5.1}) if one considers an IR of the sp(2)
algebra and treats the eigenvalues of the diagonal operator
$h$ not as elements of $F_p$ but as integers, then they
take the values of $q_0,q_0+2,...2p-q_0-2,2p-q_0$. Since
these values are in the range $[q_0,2p-q_0]$, it is clear
that the eigenvalue is equal to zero in $F_p$ only if it
is equal to $p$ when considered as an integer. Since the
AdS energy is defined as $E=h_1+h_2$, it is now clear that
when the above situation takes place, the energy
considered as an integer is equal to 2p. It now follows from
Eq. (\ref{3.27}) that the energy can be equal to $2p$ only
if $m$ is even. Since $m=q_1+q_2$ and $s=q_1-q_2$ we conclude
that $m$ can be even if and only if $s$ is even. In that case
we will necessarily have quantum numbers $(n_1n_2nk)$ such
that the sets $(n_1n_2nk)$ and $({\tilde n}_1{\tilde n}_2nk)$
are the same and therefore the 'Physical-nonphysical states
assumption' is not valid. On the other hand, if $s$ is odd
(i.e. half-integer in the usual units) then there are no
quantum numbers $(n_1n_2nk)$ such that the sets $(n_1n_2nk)$
and $({\tilde n}_1{\tilde n}_2nk)$ are the same.

In view of these observations it seems natural to
implement the 'Physical-nonphysical states assumption'
as follows.

{\it If the quantum numbers $(n_1n_2nk)$ are such that
$E(n)=m+2(n_1+n_2+n) < 2p$ then the corresponding state is
physical and belongs to $S_+$, otherwise the state is
unphysical and belongs to $S_-$.}

This definition has a chance to be consistent only if $s$
is odd and we will assume that this is the case. The
definition also agrees with our expectation
(see Sect. \ref{S5.2}) that real particles
are described by quantum numbers $(n_1n_2n)$ such that
$E(n)\ll p$ while nonphysical states related to
antiparticles are such that $E(n)$ is close to $4p$.
As follows from the definition of the numbers ${\tilde n}_1$
and ${\tilde n}_2$ in Sect. \ref{S5.3}, the transformation
$n_j\rightarrow {\tilde n}_j$ ($j=1,2$) indeed transforms
any physical state to unphysical one and vice versa.
Therefore for proving that all the requirements of
'Physical-nonphysical states assumption' are satisfied,
it is sufficient to prove that the numbers of states in
the sets $S_+$ and $S_-$ are equal.

Let $n$ and $k$ be fixed. Then, as follows from Eq.
(\ref{5.1}), the total number of states $(n_1n_2nk)$ is
$(p-q_1-n+k+1)(p-q_2-n-k+1)$.
Since $q_1=q_2+s$, it is obvious that when $s$ is odd then
one of these multipliers is necessarily odd and the other is
even. Therefore the total number of states is even as it
should be. To complete the prove one has to show that the
numbers of states with the energies $E=m+2n+2N < 2p$ and
$4p-E$ are the same. Suppose, for example, that $2k <s$.
Then $N_1(n,k)<N_2(n,k)$ (see Eq. (\ref{5.1})). If
$N \leq N_1(n,k)$ then there exists $N+1$ combinations
$(n_1,n_2)$ when $E<2p$ and the same number of combinations
when the energy equals $4p-E$, while if $N > N_1(n,k)$ but
$E < 2p$ then there exists $N_1(n,k)+1$ such combinations.
Analogously we can consider the case $2k>s$ (since $s$ is
odd, the case $2k=s$ is impossible).

\vfill\eject

\chapter{Properties of GFQT}
\label{C6}

\section{Symmetries in GFQT}
\label{S6.1}

As argued in Sect. \ref{3.4}, in the AdS version of
the standard theory the CPT invariance is not so
fundamental as in the Poincare invariant version.
The results of the preceding chapter indicate that
in GFQT the AB symmetry plays a fundamental role.
Let us first discuss some features of this symmetry.
In the standard theory the CPT-transformation in
Schwinger's formulation transforms
$a^*$ to $b^T$ \cite{AB,BLP}. For this reason one might
think that Eq. (\ref{5.8}) is a modular analog of the
CPT transformation. However, in the standard theory
the operators $(a,a^*)$ on one hand and $(b,b^*)$
on the other refer to objects described by
different IRs while in our approach they refer to the
same object. While in the standard theory the CPT
transformation is a true transformation relating two
sets of physical operators, Eq. (\ref{5.8})
is not a transformation but {\it a definition} of the
$(b,b^*)$ operators in terms of $(a,a^*)$ ones.

Another possible interpretation of the AB symmetry
is as follows. We can quantize the same physical
system by using either the set of the $(a,a^*)$
operators or the $(b,b^*)$ ones. In terms of the
fashion language, we have to choose the 'gauge',
and the AB symmetry shows that the physical results do
not depend on the choice of the 'gauge'.

It has been noted in Sect. \ref{S3.4} that in the
standard theory the C transformation is defined as
$a\rightarrow \eta_C b$ and for this reason one might
think that the AB symmetry is simply the C invariance.
This is not so for the following reason. The quantized
operators (\ref{3.41}) contain both, the $(a,a^*)$ and
$(b,b^*)$ operators and are invariant under the
C transformation. On the other hand, in the modular
case only the $(a,a^*)$ operators are present in
Eq. (\ref{5.5}) and only the $(b,b^*)$ operators are
present in Eq. (\ref{5.12}). It is also worth noting that
although the AB symmetry seems reasonable in view of
the 'Physical-nonphysical states assumption'
(see Sect. \ref{S5.4}), its validity does not depend on
the validity of the assumption.

We now consider a simple consequence of the AB symmetry.
In the standard theory the $P$ transformation of the
$(a,a^*)$ operators is defined by Eq.
(\ref{3.45}) and the same is valid for the
$(b,b^*)$ operators.
In GFQT we can define the $P$ transformation
analogously. Now we
take the operator $a(n_1n_2nk)$, apply the
transformation (\ref{3.45})
and then use Eq. (\ref{5.9}). On the other hand,
we can first use Eq.
(\ref{5.9}) and then apply the transformation
(\ref{3.45}) to the
operator $b(n_1n_2nk)^*$. These operations commute if
\begin{equation}
{\bar\eta}_P (-1)^{n_1+n_2}=\eta_P (-1)^{{\tilde n}_1+{\tilde n}_2}
\label{6.1}
\end{equation}
As follows from the definition of the quantities
$({\tilde n}_1,{\tilde n}_2)$ in Sect. \ref{S5.3},
definition of
$(Q_1(n,k),Q_2(n,k)$ in Sect. \ref{S3.2} and the
fact that $q_1=q_2+s$,
Eq. (\ref{6.1}) is equivalent to
\begin{equation}
{\bar\eta}_P = (-1)^s \eta_P
\label{6.2}
\end{equation}
As argued in Sect. \ref{SG3}, the value of $p$
in GFQT is probably such that $p=3\,\,(mod\,\, 4)$.
Another argument in favor of such a possibility is
given in Sect. \ref{S6.3}. As noted in Sect.
\ref{SG1}, the field $F_{p^2}$ in this case is
simply a complex extension of $F_p$. Therefore the
automorphism in Eq. (\ref{6.2}) can be understood
as the standard complex conjugation. For this
reason Eq. (\ref{6.2}) is the analog of the well
known result of the standard theory
\cite{AB,BLP,Wein-susy} that the spatial parity is
real for particles with the integer spin (in the
usual units) and imaginary for particles
with the half-integer spin. In GFQT this result
follows from the fact that the sets $(a,a^*)$ and
$(b,b^*)$ are not independent and there is no need
to assume that particles are described
by local covariant equations.

Consider now the notion of C invariance in GFQT.
As noted in Sect. \ref{S3.4}, this invariance can
be defined only if we have only physical sets of
$(a,a^*)$ and $(b,b^*)$ operators. Suppose that
$S_+$ is the set of physical states (see Sect.
\ref{S5.5}). Then we have to write the operators
$M^{ab}$ only in terms of the $(a,a^*)$ and
$(b,b^*)$ operators defined on $S_+$. Consider
the contribution to Eqs. (\ref{5.5}) and
(\ref{5.12}) originating from such values
of $(n_1n_2n)$ where all of them are much less
than $p$. As follows from Eqs. (\ref{5.8}) and
(\ref{5.9}), the contribution of such values to
Eq. (\ref{5.12}) originates from the contribution
to Eq. (\ref{5.5}) where the energy considered as
an integer is close to $4p$ (see Sect. \ref{S5.5}).
Therefore if we consider only such values of
$(n_1n_2n)$ where there exists the correspondence
with the standard theory for particles and
antiparticles then, as follows from Eqs.
(\ref{5.5}) and (\ref{5.12}), their contribution
to the representation operators can be written as
\begin{eqnarray}
&M^{ab}=\sum' M^{ab}(n_1'n_2'n'k';n_1n_2nk)
[a(n_1'n_2'n'k')^*a(n_1n_2nk)+\nonumber\\
&b(n_1'n_2'n'k')^*b(n_1n_2nk)]/Norm(n_1n_2nk)
\label{6.3}
\end{eqnarray}
where $\sum'$ means that only the above values
of $(n_1n_2n)$ are taken into account.

Comparing Eqs. (\ref{3.41}) and (\ref{6.3}), we
conclude that for the contribution of particle and
antiparticle states where all the numbers $(n_1n_2n)$
are much less than $p$, there exists the correspondence
between GFQT and the standard theory. Therefore if the
C transformation is defined as in the standard theory
(see Sect. \ref{S3.4}) then the theory considered only
on such states is C invariant. In addition, it is
obvious from Eqs. (\ref{5.30}) and (\ref{6.3}), that
if only such states are taken into account than the
vacuum is the eigenvector of all the representation
operators with the eigenvalue zero. The problem arises
whether these conclusions are valid if all the states
are taken into account.

Consider first the energy operator $M^{05}$. Recall
that the equivalence of Eqs. (\ref{5.5}) and (\ref{5.12})
has been proved by using the fact that the trace of
all the representation operators equals zero (see
Eq. (\ref{5.4})). However, if we wish to replace in
Eq. (\ref{5.12}) only the nonphysical $(a,a^*)$
operators by the physical $(b,b^*)$ ones, we cannot
use this fact. As follows from Eqs. (\ref{3.27}),
(\ref{5.7}-\ref{5.9}) and (\ref{5.17})
\begin{eqnarray}
&M^{ab}=\sum_{S_+} [m+2(n_1+n_2+n)]
[a(n_1n_2nk)^*a(n_1n_2nk)\mp\nonumber\\
& b(n_1n_2nk)b(n_1n_2nk)^*]/Norm(n_1n_2nk)
\label{6.4}
\end{eqnarray}
where the sum is taken only over physical states
and $\mp$ corresponds to the anticommutators and
commutators, respectively. Therefore, as follows
from Eqs. (\ref{5.10}) and (\ref{5.11})
\begin{eqnarray}
&M^{05}=\{\sum_{S_+} [m+2(n_1+n_2+n)]
[a(n_1n_2nk)^*a(n_1n_2nk)+\nonumber\\
&b(n_1n_2nk)^*b(n_1n_2nk)]/Norm(n_1n_2nk)\}+E_{vac}
\label{6.5}
\end{eqnarray}
where
\begin{equation}
E_{vac}=\mp \sum_{S_+}  [m+2(n_1+n_2+n)]
\label{6.6}
\end{equation}
can be called the vacuum energy (this quantity is
discussed in the subsequent section). As follows
from Eq. (\ref{6.5}), the energy operator is
C invariant and the vacuum is the eigenvector of
the energy operator with the eigenvalue zero if
$E_{vac}=0$.

Analogously, for the diagonal operator $L_z=h_1-h_2$
one can show that
\begin{eqnarray}
&L_z=\{\sum_{S_+} [s+2(n_1-n_2-k)]
[a(n_1n_2nk)^*a(n_1n_2nk)+\nonumber\\
&b(n_1n_2nk)^*b(n_1n_2nk)]/Norm(n_1n_2nk)\}+(L_z)_{vac}
\label{6.7}
\end{eqnarray}
where
\begin{equation}
(L_z)_{vac}=\mp \sum_{S_+}  [s+2(n_1-n_2-k)]
\label{6.8}
\end{equation}
Therefore the operator $L_z$ is
C invariant and the vacuum is the eigenvector of
the energy operator with the eigenvalue zero if
$(L_z)_{vac}=0$.

Consider now the operator $a_1'$. As follows from
Eqs. (\ref{3.28}) and (\ref{5.5}), its quantized form
can be written as
\begin{eqnarray}
&a_1'=\sum n_1[Q_1(n,k)+n_1-1]
a(n_1-1,n_2nk)^*\nonumber\\
&a(n_1n_2nk)/Norm(n_1n_2nk)
\label{6.9}
\end{eqnarray}
where the sum is taken over all possible
quantum numbers. For simplicity, consider the case
when $s$ is odd and the separation of states into
physical and nonphysical ones is accomplished as in Sect.
\ref{S5.5}. Let us consider the contribution to Eq.
(\ref{6.7}) of such quantum numbers that
$m+2(n_1+n_2+n)=2p+1$. Then the operator $a(n_1-1,n_2nk)$
in Eq. (\ref{6.7}) is physical and the operator
$a(n_1n_2nk)$ is unphysical. As follows from Eqs.
(\ref{5.7}-\ref{5.9}), Eq. (\ref{6.9}) can be
represented in the form
\begin{eqnarray}
&a_1'=\{\sum_{S_+} n_1[Q_1(n,k)+n_1-1]
[a(n_1-1,n_2nk)^*a(n_1n_2nk)+\nonumber\\
&b(n_1-1,n_2nk)^*b(n_1n_2nk)]/Norm(n_1n_2nk)\}+\nonumber\\
&\sum" (-1)^s{\bar \eta}(n_1n_2nk)n_1[Q_1(n,k)+n_1-1]\nonumber\\
&a(n_1-1,n_2nk)^*b({\tilde n}_1{\tilde n}_2nk)^*/G(nk)
\label{6.10}
\end{eqnarray}
where $\sum"$ means that the sum is taken over such values
of $(n_1n_2nk)$ that $m+2(n_1+n_2+n)=2p+1$.
Analogously one can consider the other nondiagonal operators
and conclude that
\begin{itemize}
\item In GFQT C invariance is not an exact symmetry even in the
free massive case.
\item In GFQT the vacuum cannot be the eigenvector of all the
representation operators with zero eigenvalue. If the constants
$E_{vac}$ and $(L_z)_{vac}$ are equal to zero then the vacuum
vector $\Phi_0$ satisfies the condition
\begin{equation}
(\Phi_0,M^{ab}\Phi_0)=0
\label{6.11}
\end{equation}
for any representation operator.
\end{itemize}
In other words, even if the quantities $E_{vac}$ and $(L_z)_{vac}$
are equal to zero then all the representation operators
annihilate the vacuum only in the weak sense.

\section{Dirac vacuum energy problem}
\label{S6.2}

The Dirac vacuum energy problem is discussed in practically
every textbook on LQFT. In its simplified form it can be
described as follows. Suppose for simplicity that the
energy spectrum is discrete and $n$ is the quantum number
enumerating the states. Let $E(n)$ be the energy in the state
$n$. Consider for simplicity the electron-positron field.
As a result of quantization one gets for the energy operator
\begin{equation}
E = \sum_n E(n)[a(n)^*a(n)-b(n)b(n)^*]
\label{6.12}
\end{equation}
where $a(n)$ is the operator of electron annihilation in the
state $n$, $a(n)^*$ is the operator of electron creation in
the state $n$, $b(n)$ is the operator of positron 
annihilation in the
state $n$ and $b(n)^*$ is the operator of positron creation in
the state $n$. It follows from this expression that only
anticommutation relations are possible since otherwise
the energy of positrons will be negative. However, if
anticommutation relations are assumed, it follows from
Eq. (\ref{6.12}) that
\begin{equation}
E = \{\sum_n E(n)[a(n)^*a(n)+b(n)^*b(n)]\}+E_0
\label{6.13}
\end{equation}
where $E_0$ is some infinite negative constant. Its presence
was a motivation for developing Dirac's hole
theory (see e.g. Dirac's Nobel lecture \cite{DirNobel}). In
the modern approach it is usually required that the
vacuum energy should
be zero. This can be obtained by assuming that all
operators should be written in the normal form. However,
this requirement is not quite consistent since the result of
quantization is Eq. (\ref{6.12}) where the positron operators
are not written in the that form (see also the discussion
in Sect. \ref{S3.4}).

We have seen in the preceding section that in GFQT one can
obtain Eq. (\ref{6.4}) (which is an analog of Eq. (\ref{6.12}))
without assuming that particles and antiparticles are described
by a local covariant equation. In turn this expression can be
transformed to Eq. (\ref{6.5}) (which is an analog of
Eq. (\ref{6.13})) and the vacuum energy is given by Eq.
(\ref{6.6}). The goal of this section is to explicitly
calculate the quantities $E_{vac}$ and $(L_z)_{vac}$ defined
in Eqs. (\ref{6.6}) and (\ref{6.8}).

Consider first the sum in Eq. (\ref{6.6}) when the values of
$n$ and $k$ are fixed. It is convenient to distinguish the
cases $s > 2k$ and $s<2k$ (recall that $s$ is assumed to be
odd). If $s > 2k$ then, as follows from Eq. (\ref{5.1}),
the maximum value of $n_1$ is such that $m+2(n+n_1)$ is always
less than $2p$. For this reason all the values of $n_1$
contribute to the sum, which can be written as
\begin{eqnarray}
&S_1(n,k) =-\sum_{n_1=0}^{p-q_1-n+k}[(m+2n+2n_1)+\nonumber\\
&(m+2n+2n_1+2)+...+(2p-1)]
\label{6.14}
\end{eqnarray}
Since the result should be considered in $F_p$, one gets
\begin{equation}
S_1(n,k)=\sum_{n_1=0}^{p-q_1-n+k}(n+\frac{m-1}{2}+n_1)^2
\label{6.15}
\end{equation}
In turn, this expression obviously can be represented as
\begin{equation}
S_1(n,k)=\sum_{n_1=1}^{p-1}n_1^2-\sum_{n_1=1}^{n+(m-3)/2}n_1^2-
\sum_{n_1=1}^{(s-1)/2-k}n_1^2
\label{6.16}
\end{equation}
where the last sum should be taken into account only
if $(s-1)/2-k\geq 1$.

The first sum in this expression equals $(p-1)p(2p-1)/6$
and, since we assume that $p\neq 2$ and $p\neq 3$, this
quantity is zero in $F_p$. As a result, $S_1(n,k)$ is
represented as a sum of two terms such that the first one
depends only on $n$ and the second --- only on $k$. Note
also that the second term is absent if $s=1$, i.e. for
particles with the spin 1/2 in the usual units.

Analogously, if $s < 2k$ we have to calculate the sum
\begin{eqnarray}
&S_2(n,k) =-\sum_{n_2=0}^{p-q_2-n-k}[(m+2n+2n_2)+\nonumber\\
&(m+2n+2n_2+2)+...+(2p-1)]
\label{6.17}
\end{eqnarray}
and the result is
\begin{equation}
S_2(n,k)=-\sum_{n_2=1}^{n+(m-3)/2}n_2^2-\sum_{n_2=1}^{k-(s+1)/2}n_2^2
\label{6.18}
\end{equation}
where the second term should be taken into account only
if $k-(s+1)/2\geq 1$.

In the massive case we now should calculate the sum
\begin{equation}
S(n)=\sum_{k=0}^{(s-1)/2}S_1(n,k) +\sum_{k=(s+1)/2}^s S_2(n,k)
\label{6.19}
\end{equation}
and the result is
\begin{eqnarray}
&S(n)=-(s+1)(n+\frac{m-1}{2})[2(n+\frac{m-1}{2})^2-\nonumber\\
&3(n+\frac{m-1}{2})+1]/6-(s-1)(s+1)^2(s+3)/96
\label{6.20}
\end{eqnarray}
Since the value of $n$ is in the range $[0,n_{max}]$
where $n_{max}=p+2-m$
(see Sect. \ref{S5.1}), we have
\begin{equation}
E_{vac}=\sum_{n=0}^{n_{max}}S(n)
\label{6.21}
\end{equation}
In the massless case with $s=1$ (recall that we consider only
odd values of $s$) the result is the same but $n_{max}=p-2$
(see Sect. \ref{S5.1}). In the massless case with $s\geq 3$
the result is
\begin{equation}
E_{vac}=S(0)+\sum_{n=1}^{n_{max}}[S_1(n,0)+S_2(n,s)]
\label{6.22}
\end{equation}
and $n_{max}=p-1-s$.

Consider first Eq. (\ref{6.21}) in the massive case.
Notice that if $l\neq 0$ then
\begin{eqnarray}
&\sum_{n=0}^{p+2-m}[n+(m-1)/2]^l=\sum_{n=1}^{p-1}n^l-\nonumber\\
&\sum_{n=1}^{(m-3)/2}n^l -(-1)^l\sum_{n=1}^{(m-5)/2}n^l
\label{6.23}
\end{eqnarray}
The first term in the r.h.s. equals zero in $F_p$ since we assume
that $p\neq 2,3$. Therefore, as follows from Eqs.
(\ref{6.20}), (\ref{6.21}) and (\ref{6.23}), the final result for
$E_{vac}$ in the massive case is
\begin{equation}
E_{vac}= (m-3)(s-1)(s+1)^2(s+3)/96
\label{6.24}
\end{equation}
One should not be confused by the observation that this value looks
as positive while one might expect from Eq. (\ref{6.4}) that this
value should be negative as in the standard theory. The reason is
obvious since Eq. (\ref{6.24}) has been derived in $F_p$.

In the massless case with $s=1$ the result is $E_{vac}=0$ since
the contribution of $n=p-1$ to the sum in Eq. (\ref{6.21}) is zero
and the result (\ref{6.24}) applies. The result in the massless case
with $s\geq 3$ is
\begin{equation}
E_{vac}= -(s-1)(s+1)(s^2+12s+3)/96
\label{6.25}
\end{equation}

Consider now the quantity $(L_z)_{vac}$ defined by Eq. (\ref{6.8}).
We again consider first the case when the quantities $n$ and $k$
are fixed. If $s>2k$ then, it is clear from Eq. (\ref{6.14}) that
$n_1$ is in the range $[0,p-q_1-n+k]$ and $n_2$ is in the range
$[0,(2p-1-m-2n-2n_1)/2]$. Therefore we have the contribution of
\begin{equation}
S_1(n,k)=\sum_{n_1=0}^{p-q_1-n+k}\quad\sum_{n_2=0}^{(2p-1-m-2n-2n_1)/2}
[s+2(n_1-n_2-k)]
\label{6.26}
\end{equation}
Analogously if $s < 2k$, we have the contribution of
\begin{equation}
S_2(n,k)=\sum_{n_2=0}^{p-q_2-n-k}\quad\sum_{n_2=0}^{(2p-1-m-2n-2n_2)/2}
[s+2(n_1-n_2-k)]
\label{6.27}
\end{equation}
If $k$ is such that $s>2k$ then $s<2(s-k)$. We use this observation and
notice that
\begin{equation}
S_1(n,k)+S_2(n,s-k)=0
\label{6.28}
\end{equation}
As a result, all the terms in the sum (\ref{6.8}) cancel out and
$(L_z)_{vac}=0$.
Our final conclusion in this section is as follows.
\begin{itemize}
\item If $s$ is odd and the separation of states into physical
and nonphysical ones is accomplished as in Sect. \ref{S5.5} then
$(L_z)_{vac}=0$ while $E_{vac}=0$ only if $s=1$ (i.e. $s=1/2$
in the usual units).
\end{itemize}

\section{Neutral particles and spin-statistics theorem}
\label{S6.3}

The nonexistence of neutral elementary particles in
GFQT is one of the most striking differences
between GFQT and the standard theory. For this reason
we discuss this problem in detail.
One could give the following definition of neutral
particle:
\begin{itemize}
\item i) it is a particle coinciding with its
antiparticle
\item ii) it is a particle which does not coincide
with its antiparticle but they have the same properties
\end{itemize}
In the standard theory only i) is meaningful since
neutral particles are described by real (not complex)
fields and this condition is required by Hermiticity.
One might think that the definition ii) is only academic
since if a particle and its antiparticle have the same
properties then they are indistinguishable and can be
treated as the same. However, the cases i) and ii)
are essentially different from the operator point of
view. In the case i) only the $(a,a^*)$ operators
are sufficient for  describing the operators (\ref{3.41}).
This is the reflection of the fact that the real
field has the number of degrees of freedom twice
as less as the complex field. On the other hand,
in the case ii) both $(a,a^*)$ and $(b,b^*)$
operators are required, i.e. in the standard theory
such a situation is described by a complex field.
Nevertheless, the case ii) seems to be rather odd:
it implies that there exists a quantum number
distinguishing a particle from its antiparticle
but this number is not manifested experimentally.
We now consider whether the conditions i) or ii) can
be implemented in GFQT.

Since each operator $a$
is proportional to some operator $b^*$ and vice versa
(see Eqs. (\ref{5.8}) and (\ref{5.9})), it is
clear that if the particles described by the
operators $(a,a^*)$ have some nonzero charge then
the particles described by the operators $(b,b^*)$
have the opposite charge and the number of operators
cannot be reduced. However, if all possible charges are
zero, one could try to implement i) by requiring that
each $b(n_1n_2nk)$ should be proportional to
$a(n_1n_2nk)$ and then $a(n_1n_2nk)$ will be
proportional to $a({\tilde n}_1,{\tilde n}_2,nk)$.
In this case the operators $(b,b^*)$ will not be
needed at all.

In GFQT such a possibility is unacceptable for several
reasons. Indeed, if one requires that $a(n_1n_2nk)$ is
proportional to $a({\tilde n}_1,{\tilde n}_2,nk)$ then
the same considerations as in the derivation of Eq.
(\ref{6.3}) show that instead of that equation
we will have
\begin{eqnarray}
&M^{ab}=2\sum' M^{ab}(n_1'n_2'n'k';n_1n_2nk)\nonumber\\
&a(n_1'n_2'n'k')^*a(n_1n_2nk)/Norm(n_1n_2nk)
\label{6.29}
\end{eqnarray}
Note that in Eq. (\ref{6.3}) the parts of the $M^{ab}$
operators described by the $(a,a^*)$ and $(b,b^*)$
operators satisfy the correct commutation relations
of the AdS algebra separately. At the same time, if
the $(b,b^*)$ operators are thrown away and the result
is multiplied by two then the commutation relations will
be broken since the l.h.s. of Eq. (\ref{2.4}) will
be multiplied by four while the r.h.s. --- by two.

The fact that the commutation relations will be broken
is also clear from the following observation. Suppose
that the operators $(a,a^*)$ satisfy the commutation
relations (\ref{3.37}). In that case
the operators $a(n_1n_2nk)$ and $a(n_1'n_2'n'k')$
should commute if the sets $(n_1n_2nk)$ and
$(n_1'n_2'n'k')$ are not the same.
In particular, one should have $[a(n_1n_2nk),
a({\tilde n}_1{\tilde n}_2nk)]=0$ if either
$n_1\neq {\tilde n}_1$ or $n_2\neq {\tilde n}_2$.
On the other hand, if $a({\tilde n}_1{\tilde n}_2nk)$
is proportional to $a(n_1n_2nk)^*$, it follows from
Eq. (\ref{3.37}) that the commutator cannot be zero.
Analogously one can consider the case of anticommutators.

An especially striking contradiction arises in the case
of bosons with the integer spin (in the usual units).
As noted in Sect. \ref{S5.5}, in that case there
exist such values of $(n_1,n_2)$ that
$n_1={\tilde n}_1$ and $n_2={\tilde n}_2$. Then
$a(n_1n_2nk)$ is proportional to $a(n_1n_2nk)^*$
and their commutator equals zero. On the other hand,
this contradicts Eq. (\ref{3.37}).

The fact that the number of operators cannot be
reduced is also clear from the observation that the
$(a,a^*)$ or $(b,b^*)$ operators describe an
irreducible representation in which the number of
states (by definition) cannot be reduced. In other
word, since in GFQT one IR necessarily describes a
particle and its antiparticle simultaneously,
there is no analog of real fields. Our conclusion
is that in GFQT the definition of neutral particle
according to i) is fully unacceptable.

Consider now whether it is possible to implement
the definition ii) in GFQT. Recall that we started from
the operators $(a,a^*)$ and defined the operators
$(b,b^*)$ by means of Eq. (\ref{5.8}). Then the
latter satisfy the same commutation or
anticommutation relations as the former and the AB
symmetry is valid. Does it mean that the particles described
by the operators $(b,b^*)$ are the same as the ones
described by the operators $(a,a^*)$? If one starts
from the operators $(b,b^*)$ then, by analogy with Eq.
(\ref{5.8}), the operators $(a,a^*)$ can be defined as
\begin{equation}
b(n_1n_2nk)^*=\eta'(n_1n_2nk) a({\tilde n}_1{\tilde n}_2nk)/
F({\tilde n}_1{\tilde n}_2nk)
\label{6.30}
\end{equation}
where $\eta'(n_1n_2nk)$ is some function. By analogy
with the consideration in Sect. \ref{S5.4} one
can show that
\begin{equation}
\eta'(n_1n_2nk)=\beta (-1)^{n_1+n_2+n}
\label{6.31}
\end{equation}
and $\beta$, as well as $\alpha$, should be such that
\begin{equation}
\beta {\bar \beta}=\mp 1
\label{6.32}
\end{equation}
where the minus sign refers to the normal
spin-statistics connection and the plus sign ---
to the broken one.

As follows from Eqs. (\ref{5.7}, (\ref{5.13}),
(\ref{5.15}), (\ref{5.22}), (\ref{6.30}), (\ref{6.31})
and the definition of the quantities ${\tilde n}_1$
and ${\tilde n}_2$ in Sect. \ref{S5.3}, the relation
between the quantities $\alpha$ and $\beta$ is
\begin{equation}
\alpha {\bar \beta}=1
\label{6.33}
\end{equation}
Therefore, as follows from Eq. (\ref{6.32}), there
exist only two possibilities
\begin{equation}
\beta = \mp \alpha
\label{6.34}
\end{equation}
depending on whether the normal spin-statistics
connection is valid or not.
We conclude that the broken spin-statistics connection
implies that $\alpha{\bar \alpha}=\beta{\bar\beta}=1$
and $\beta=\alpha$ while the normal spin-statistics
connection implies that
$\alpha{\bar \alpha}=\beta{\bar\beta}=-1$
and $\beta=-\alpha$.

In the first case we have a
situation when particles and antiparticles have the
same properties, i.e. they are indistinguishable. In
this case solutions for $\alpha$ and $\beta$ obviously
exist and the particle and its antiparticle can be
treated as neutral in the sense of the definition ii).
Since such a situation is clearly unphysical, one can
treat the spin-statistics theorem as a requirement
excluding neutral particles in the sense ii).

In the case of normal spin-statistics connection a
particle and its antiparticle necessarily have different
properties since $\beta \neq \alpha$. However, one
might wonder whether in this case it is possible to
find a solution of the relation
\begin{equation}
\alpha {\bar \alpha}=-1
\label{6.35}
\end{equation}
In the standard theory this relation is
impossible, but when $\alpha\in F_{p^2}$,
a solution always exists. Indeed,
in Sect. \ref{SG2} it has been noted that any
element of $F_{p^2}$ can be represented as a power
of the primitive root $r$ and $F_{p^2}$ has only one
nontrivial automorphism which is defined as
$\alpha\rightarrow {\bar \alpha}=\alpha^p$.
Therefore if $\alpha =r^k$ then $\alpha{\bar \alpha}=
r^{(p+1)k}$. On the other hand, since $r^{(p^2-1)}=1$, we
conclude that $r^{(p^2-1)/2}=-1$. Therefore there exists at
least a solution with $k=(p-1)/2$.

There exist, however, another more interesting
treatment of the spin-statistics theorem.
In Sect. \ref{SG2} it has been shown that there exists
the correspondence between the standard theory and GFQT
if the latter is based on the field $F_{p^2}$. At the
same time, it has been argued in Sect. \ref{SG3} that,
since GFQT is treated as a more general theory than
the standard one, the field should not be chosen from
the requirement that there should exist a
correspondence with the standard theory. Although in
the consideration in the preceding and this chapters
it has been implicitly assumed that the states are
described by elements in a space over $F_{p^2}$,
this fact has not been used and all the results are
valid for any extension of $F_p$. Therefore the
problem arises whether there are arguments in favor
of the choice $F_{p^2}$.

In Sect. \ref{SG3} it has been argued that
$p=3\,\, (mod\, 4)$ rather than $p=1\,\, (mod\, 4)$,
and a possible motivation of the choice of $F_{p^2}$
is that this is a minimal extension of $F_p$ such that
the representation operators are fully decomposable.
We now propose another motivation: {\it the spin-statistics
theorem is precisely the requirement that GFQT should
be based on $F_{p^2}$}. Indeed, if $\alpha{\bar\alpha}=1$,
the solution for $\alpha$ can be obviously found without
extending $F_p$. Let us now discuss whether one can
find solutions of Eq. (\ref{6.35}) without extending
$F_p$. This is possible if
$-1$ can be represented as a square of an element from
$F_p$ or, in the terminology of number theory, if $-1$
is a quadratic residue in $F_p$. As noted in
Sect. \ref{SG2}, a well known fact in number theory
is that $-1$ is a quadratic residue in $F_p$ if
$p = 1\,\, (mod\, 4)$, and a quadratic nonresidue if
$p = 3\,\, (mod\, 4)$. For example, if $p=5$
then $-1$ is a square in $F_5$
since $2\times 2=4=-1 \,\, (mod\, 5)$, but if $p=7$
then $-1$ cannot be represented as a square of an
element from $F_7$. We conclude that if
$p = 1\,\, (mod\, 4)$ then there exist solutions of
Eq. (\ref{6.35}) in $F_p$ but if $p = 3\,\, (mod\, 4)$
then the field should be necessarily extended and the
minimal extension is $F_{p^2}$.

Let us now discuss a somewhat different approach to
the AB symmetry. Note first that this symmetry has 
been formulated as the
condition that the representation operators have 
the same form in terms of $(a,a^*)$ and $(b,b^*)$
(see Sect. \ref{S5.3}). In that case the operators
$(b,b^*)$ are defined in terms of $(a,a^*)$ by Eqs. 
(\ref{5.8}) and (\ref{5.9}). A desire to have 
operators which can be interpreted as those relating 
separately to particles and antiparticles is natural 
in view of our
experience in the standard approach. However, one
might think that in the spirit of GFQT there is no 
need to have separate operators for
particles and antiparticles since they are different states
of the same object. For this reason the operators $(b,b^*)$
are strictly speaking redundant. We can therefore 
reformulate the AB symmetry as follows. Instead of 
Eqs. (\ref{5.8}) and (\ref{5.9}), we consider a 
{\it transformation} defined as 
\begin{eqnarray}
&a(n_1n_2nk)^*\rightarrow \eta(n_1n_2nk) 
a({\tilde n}_1{\tilde n}_2nk)/
F({\tilde n}_1{\tilde n}_2nk)\nonumber\\ 
&a(n_1n_2nk)\rightarrow \bar{\eta}(n_1n_2nk) 
a({\tilde n}_1{\tilde n}_2nk)^*/
F({\tilde n}_1{\tilde n}_2nk)
\label{AB}
\end{eqnarray}
Then the AB symmetry can be formulated as a 
requirement that physical results should be
invariant under this transformation.
Strictly speaking, the name 'AB symmetry' is 
not appropriate anymore but we retain it for 
'backward compatibility'. 

\begin{sloppypar}
Let us now apply the AB transformation twice. 
Then, by analogy with the derivation of Eq. (\ref{5.17}),
we get 
\begin{equation}
a(n_1n_2nk)^*\rightarrow \mp a(n_1n_2nk)^*\quad
a(n_1n_2nk)\rightarrow \mp a(n_1n_2nk)
\label{AB1}
\end{equation} 
for the normal and broken spin-statistic connections,
respectively. Therefore, as a consequence of the 
spin-statistics theorem, any particle
(with the integer or half-integer spin) has the
AB$^2$ parity equal to $-1$. Therefore in GFQT any 
interaction can involve only an even number of
creation and annihilation operators. In particular,
this is additional demonstration of the fact that in
GFQT the existence of neutral elementary particles
is incompatible with the spin-statistics theorem.
\end{sloppypar}

\section{Charge operator}
\label{S6.4}

In Sect. \ref{S6.2} we mentioned
the Dirac vacuum energy problem in the standard
quantum theory. The vacuum charge problem is
analogous and with the same notations it can be
described as follows. Consider for simplicity the
electron-positron field and let $e$ be the
electron charge. Then the quantization of the total
charge results in the operator
\begin{equation}
Q=e\sum_n [a(n)^*a(n)+b(n)b(n)^*]
\label{6.36}
\end{equation}
If the $(b,b^*)$ operators anticommute then
\begin{equation}
Q=e\{\sum_n [a(n)^*a(n)-b(n)^*b(n)]\}+Q_0
\label{6.37}
\end{equation}
This result shows that a particle and its
antiparticle have opposite charges as it should be.
At the same time, the presence of the infinite
constant $Q_0$ (vacuum charge) is clearly undesirable.

A possible prescription for eliminating $Q_0$ is to
again require that the operators
in Eq. (\ref{6.36}) should be written in the normal
form. However, for the same reason as in the Dirac
vacuum energy problem (see Sect. \ref{S6.2}), such
a prescription is not quite consistent. Another
well known prescription \cite{AB,BLP,Wein} is to
require that {\it after quantization} the current
density operator should be written as a commutator
$(e/2)[{\bar\psi}(x)\gamma^{\mu},\psi(x)]$ rather
than $e{\bar\psi}(x)\gamma^{\mu}\psi(x)$. Then
the vacuum charge disappears as a good illustration
of Polchinski's joke \cite{Polchinski} that the
standard approach is essentially based on the formula 
'$\infty -\infty = physics$'.

In the standard theory the charge operator should
commute with all the representation operators of
the symmetry algebra and the same requirement
should be imposed in GFQT, i.e.
\begin{equation}
[M^{ab},Q] =0\quad \forall\,\, a,b
\label{charge}
\end{equation}
Then if the representation operators are described 
in terms
of the $(a,a^*)$ operators (see Eq. (\ref{5.5}))
and one seeks the charge operator in the form
$$\sum c(n_1n_2nk)a(n_1n_2nk)^*a(n_1n_2nk),$$
the only operator (up to a multiplicative
constant) which commutes with all the representation
operators is
\begin{equation}
Q=e \sum [a(n_1n_2nk)^*a(n_1n_2nk)/Norm(n_1n_2nk)]
\label{6.38}
\end{equation}
where $e$ is some constant and the sum is
taken over all possible quantum numbers $(n_1n_2nk)$.

As noted in Sect. \ref{S5.3}, in GFQT there is no need
to require that the $a$ operators should always precede
the $a^*$ ones. As follows from the $^*$ symmetry 
condition (see Sect. \ref{S5.3}), the operator
\begin{equation}
Q'=\mp e \sum [a(n_1n_2nk)a(n_1n_2nk)^*/Norm(n_1n_2nk)]
\label{6.39}
\end{equation}
should be the same as $Q$. Let
$Dim(m,s)$ be the total number of possible states
for the particle with the mass $m$ and spin $s$. Then,
as follows from Eqs. (\ref{6.38}) and (\ref{6.39}),
\begin{equation}
Q'=Q\mp eDim(m,s)
\label{6.40}
\end{equation}
The quantity $Dim(m,s)$ has been calculated in Sect.
\ref{S5.1} (see Eqs. (\ref{massive}-\ref{singleton})).
When $Dim(m,s)$ is treated as the dimension of IR, these
expressions should be understood in the usual sense, not
in the sense of $F_p$. However, since the charge operator
in GFQT should be treated as an operator in a space over
$F_{p^2}$, the quantity $Dim(m,s)$ in Eq. (\ref{6.40})
should be treated in the sense of $F_p$. As a result,
$Q=Q'$ only in the massless cases with $s=0$ and
$s=1$.

On the other hand, one can describe the representation
operators only in terms of $(b,b^*)$ operators
(see Eq. (\ref{5.12})). Then the only operator
(up to a multiplicative constant) which commutes with
the representation operators, is bilinear in $(b,b^*)$
and the $b$ operators precede the $b^*$ ones is
\begin{equation}
Q_1=-e \sum [b(n_1n_2nk)^*b(n_1n_2nk)/Norm(n_1n_2nk)]
\label{6.41}
\end{equation}
By analogy with the above discussion, there is no need
to require that the $b$ operators should precede 
the $b^*$ ones if $Dim(m,s)=0$ in $F_p$.

The minus sign in the r.h.s. of Eq. (\ref{6.41})
(compare with Eq. (\ref{6.38})) reflects the fact that 
particles and antiparticles have opposite charges.
The problem arises whether the operators $Q$ and $Q_1$
are the same. If this is the case then the AB parity
of the charge operator is -1, while as shown in Sect.
\ref{5.4}, the operators $M^{ab}$ are invariant under
the AB transformation (i.e. their AB parity is 1).
This is not a problem since the r.h.s. of 
Eq. (\ref{charge}) is zero and therefore this relation
will be invariant under the AB transformation. An
analogous situation exists in the standard theory
where the operators $M^{ab}$ have the C parity 
equal to 1 while $Q$ has the C parity equal to -1.

Another problem is that the vacuum should
be the eigenvector of the charge operator with the
eigenvalue zero. To understand whether this is the case
one has to write the charge operator only in terms of
physical $(a,a^*)$ and $(b,b^*)$ operators (see Sect.
\ref{S5.5}). By analogy with Eq. (\ref{6.37}) one
would expect that the charge operator will then have
the form
\begin{eqnarray}
&Q_2=e \sum_{S+} \{[a(n_1n_2nk)^*a(n_1n_2nk)-\nonumber\\
&b(n_1n_2nk)^*b(n_1n_2nk)]/Norm(n_1n_2nk)\}
\label{6.42}
\end{eqnarray}
Indeed, such an operator annihilates the vacuum in view
of Eq. (\ref{5.30}).

As follows from Eqs. (\ref{5.7}-\ref{5.9}), (\ref{5.13})
and (\ref{5.14}), the operator (\ref{6.38}) can be
written as
\begin{equation}
Q=\pm e \sum [b(n_1n_2nk)b(n_1n_2nk)^*/Norm(n_1n_2nk)]
\label{6.43}
\end{equation}
for the cases of anticommutators and commutators,
respectively. Therefore, as follows from Eq. (\ref{6.41}),
\begin{equation}
Q=Q_1 \pm e Dim(m,s)
\label{6.44}
\end{equation}
One also can replace the $(a,a^*)$ operators
by the $(b,b^*)$ ones only for nonphysical states. Then
by analogy with Eq. (\ref{6.43}) one gets the result
analogous to Eq. (\ref{6.36})
\begin{eqnarray}
&Q= e \sum_{S_+} \{[a(n_1n_2nk)^*a(n_1n_2nk)\pm \nonumber\\
&b(n_1n_2nk)b(n_1n_2nk)^*]/Norm(n_1n_2nk)\}
\label{6.45}
\end{eqnarray}
and therefore, as follows from Eq. (\ref{6.42})
\begin{equation}
Q=Q_2  \pm e Dim(m,s)/2
\label{6.46}
\end{equation}

Our conclusion is as follows. {\it $Q=Q'=Q_1=Q_2$ in the
massles case with
$s=1$ while in the massless case with $s=0$ this relation
is valid if $p=3\,\, (mod\, 4)$. In all the other cases
the above relation is not valid.}

In Sect. \ref{S5.5} it has been noted that when $s$ is even,
the separation of states into physical and nonphysical ones
becomes problematic. Therefore, strictly speaking, the
above conclusion is valid in the massless case with $s=0$
only if one assumes additionally that there exists a
meaningful separation of states in that case.

\begin{sloppypar}
Finally, let us consider the meaning of electric charge in
GFQT. The main ingredient of the standard QED is the 
dimensionless
fine structure constant which is approximately equal to 1/137.
Since in GFQT only integers should be used, one might think
that this theory cannot describe the electromagnetic interaction
since the nearest integer to 1/137 is zero in which case the
interaction does not exist.
However, at least two objections can be made. First, the quantity
$e$ in the above expressions is the bare electric charge, not the
effective one. It is well known that in the standard QED the former
is much greater than the latter. The second (and more important)
objection is as follows. The electric
charge or the fine structure constant can be measured not
directly but only by their contribution to the 
energy, momentum
or angular momentum. For example, in the standard 
nonrelativistic
classical electrostatics, the interaction energy 
of two particles
with the electric charges $e_1$ and $e_2$ is 
$E=e_1e_2/r$ where
$r$ is the distance between them. Therefore
the AdS interaction energy is $E_{AdS}=2Re_1e_2/r$.
If one wishes to write this expression as 
$E_{AdS}=e_{1eff}e_{2eff}/r$
then the effective electric charges will 
not be dimensionless
and their values in the usual units will 
be very large. So the notion of the effective 
electric charge in de Sitter invariant
theories has no fundamental meaning.
\end{sloppypar}

\section{Supersymmetry in GFQT: pros and contras}
\label{S6.5}

Supersymmetry is now one of the 'hottest' problems
in particle physics, and the arguments in favor of
supersymmetry are well known (see e.g. Refs. 
\cite{GSW,Wein-susy}). As noted in Sect. \ref{S3.6},
in the AdS theory there also exists a new strong
argument that all the AdS operators are bilinear in
fermionic ones. This makes it possible to treat
supersymmetry as a square root of the AdS symmetry
and the number of independent operators becomes four
instead of ten.

On the other hand, since supersymmetry is not
discovered yet, one cannot definitely conclude that
the following very naive point of view is incorrect.
What was the reason for nature to create elementary 
particles with both half-integer and integer 
spins if the latter can be built of the former? 
A well known historical analogy is that before the 
discovery of the Dirac equation it was believed 
that nothing could be simpler than the Klein-Gordon 
equation for spinless
particles. However, it has turned out that the spin 1/2
particles are simpler since the covariant equation
for them is of the first order, not the second one as the
Klein-Gordon equation. A very interesting possibility
(which has been probably considered first by
Heisenberg) is that only spin 1/2 particles could be
elementary.

It has been noted in Sect. \ref{S3.6} that in the AdS
case one IR of supersymmetry contains several IRs
of the AdS symmetry. The same result is valid in GFQT
\cite{levsusy}. A full set of annihilation and creation
operators now contains operators 
$(a(n_1n_2nk;j),a(n_1n_2nk;j)^*)$ where $j$ enumerates
the components of the supermultiplet \cite{levsusy}.
While the AdS operators act only within one IR of the
AdS algebra, the fermionic operators are necessarily
nondiagonal with respect to the quantum number $j$.
By analogy with the AdS case, one can define a set
of $(b,b^*)$ operators and prove that the AB symmetry
is valid not only for the AdS operators but for the
fermionic operators as well \cite{levsusy}.

Since one supersymmetry multiplet necessarily
contains AdS IRs with both half-integer and integer
spins (in the usual units), the problem mentioned in
Sect. \ref{S5.5} arises that in the case of integer
spins the separation of states into particle and
antiparticle ones encounters difficulties. The reason
is that there exist states for which ${\tilde n}_1=n_1$
and ${\tilde n}_2=n_2$ and it is not clear whether
such states should be treated as those describing 
the particle or antiparticle. In particular, if the
$(a,a^*)$ operators describe states with the charge
$e$ and the $(b,b^*)$ operators --- states with
the charge $-e$ then should the above states have
the charge $e$ or $-e$? In other words, it is not 
clear whether the vacuum condition for such states 
should be $a\Phi_0=0$ or $b\Phi_0=0$. In particular, 
since 
there necessarily exist sp(2)$\times$sp(2) IRs for
which the total number of states is odd, any choice
of the above possibilities necessarily breaks the
symmetry between particles and antiparticles.

A possible objection is as follows. In GFQT one IR
describes an object and $(a,a^*)$ is a complete set
of operators for this object. Therefore in the spirit of
GFQT there is no need to define new operators $(b,b^*)$.
A desire to have the both sets, $(a,a^*)$ and $(b,b^*)$,
is a reflection of our experience that any state of
the object can be treated either as a particle or 
antiparticle. This is surely true if the energies 
are not asymptotically large but nothing guarantees
that in the region of asymptotically large energies,
where the conditions ${\tilde n}_1=n_1$
and ${\tilde n}_2=n_2$ can be satisfied simultaneously,
the separation of states into particle and antiparticle
ones is still meaningful. A well known analogy is the
situation with the Dirac equation in the external 
field. If the field is weak enough and positive and
negative energy states are separated by a gap, the
notion of particle and antiparticle is meaningful.
However, if the field is so strong that the gap
disappears, a clear separation of states into
particle and antiparticle ones does not exist.  
It is also shown in Sect. \ref{S6.3}, that the
AB symmetry can be formulated without involving
the $(b,b^*)$ operators at all. 

Our conclusion is as follows. In GFQT there exists
the following dilemma which should be investigated:
\begin{itemize}
\item The notion of particle and antiparticle is
valid at all energies. Then only particles with the
half-integer spin (in usual units) can be elementary
and supersymmetry is impossible.
\item The notion of particle and antiparticle is
valid only when the energies are not asymptotically
large and all the quantum numbers $(n_1n_2n)$ are
much less than $p$. Then supersymmetry is possible.
\end{itemize}

\section{Discussion}

In the present paper we have discussed in detail
the description of free elementary particles in a
quantum theory based on a Galois field (GFQT). 
As noted in Chap. \ref{C1}, GFQT does not contain
infinities at all and all operators are 
automatically well defined. In my discussions with
physicists, some of them commented this fact as 
follows. This is the approach where a cutoff
(the characteristic $p$ of the Galois field) is
introduced from the beginning and for this reason
there is nothing strange in the fact that
the theory does not have infinities. It has a 
large number $p$ instead and this number can be
practically treated as infinite. 

However, the difference between Galois fields 
and usual numbers is not only that the former are 
finite and the latter are infinite. If the set of 
usual numbers is visualized as a straight line from 
$-\infty$ to $+\infty$ then the simplest Galois field
can be visualized not as a segment of this line but
as a circle (see Fig. 4.1 in Sect. \ref{SG1}). This 
reflects the fact that in Galois fields the rules of 
arythmetic are different and, as a result, GFQT has 
many unusual features which have no
analog in the standard theory. 

The original motivation for investigating GFQT was
as follows. Let us take the standard QED in
dS or AdS space, write the Hamiltonian and other
operators in angular momentum representation and 
replace standard irreducible representations (IRs)
for the electron, positron and
photon by corresponding modular IRs. Then we will
have a theory with a natural cutoff $p$ and all
renormalizations will be well defined. In other
words, instead of the standard approach, which,
according to Polchinski's joke \cite{Polchinski}, 
is essentially based on the formula 
'$\infty - \infty = physics$', we
will have a well defined scheme. One might treat 
this motivation as an attempt to substantiate
standard momentum regularizations (e.g. the
Pauli-Villars regularization) at momenta $p/R$
(where $R$ is the radius of the Universe).
In other terms this might be treated as introducing
fundamental length of order $R/p$. We now discuss
reasons explaining why this naive attempt fails.

Consider first
the construction of modular IR for the electron. We
start from the state with the minimum energy (where
energy=mass) and gradually construct states with
higher and higher energies. In
such a way we are moving counterclockwise along the 
circle on Fig. 4.1 in Sect. \ref{SG1}. Then sooner or 
later we will arrive at the left half of the
circle, where the energy is negative, and finally 
we will arrive at the point where energy=-mass 
(in fact, as noted in Sect. \ref{S5.2}, almost
four full rotations are needed for that). In other
words, instead of the modular analog of IR describing
only the electron, we obtain an IR describing the 
electron and positron simultaneously. 

In the standard theory a particle and its antiparticle
are described by different IRs but they are combined
together by a local covariant equation (in the given
case this is the Dirac equation). We see that in GFQT
the idea of the Dirac equation is implemented without
assuming locality but already on the level of IRs.
This automatically explains the existence of 
antiparticles, shows that a particle cannot exist 
by itself without its antiparticle and that a particle
and its antiparticle are necessarily {\it different} 
states of the same object. In particular, there are
no elementary particles which in the standard theory 
are called neutral. 

One might immediately conclude that since in GFQT the
photon cannot be elementary, this theory cannot be
realistic and does not deserve attention. 
We believe however, that the nonexistence of 
neutral elementary 
particles in GFQT shows that the photon, the graviton
and other neutral particles should be considered on a
deeper level. For example, several authors considered
a model where the photon is a composite state of Dirac
singletons \cite{FF}. In the present paper the Dirac 
singleton was briefly discussed in Sects. \ref{S3.2}
and \ref{S5.2}. It has several unusual properties and
probably the importance of singleton IRs is not realized
yet. It is interesting to note that Dirac called his
paper \cite{DiracS} 'A remarkable representation of the 
3 + 2 de Sitter group'.

The nonexistence of neutral elementary particles in GFQT
is discussed in detail in Sect. \ref{S6.3}. A possible
elementary explanation is as follows. In the standard
theory, elementary particles having antiparticles are
described by complex fields while neutral elementary
particles --- by real ones. In GFQT there is no 
possibility to choose between real and complex fields.
As noted in Sect. \ref{S6.3}, a possible treatment of 
the spin-statistics theorem is simply that this is a
requirement that quantum theory should be based on
complex numbers, and this requirement excludes the
existence of neutral elementary particles.

The Dirac vacuum energy problem discussed in Sect. 
\ref{S6.3} is a good illustration of the fact that 
replacement of usual numbers by a Galois field 
results in a qualitatively new picture. Indeed, in the
standard theory the vacuum energy is infinite and,
if GFQT is treated simply as a theory with a cutoff
$p$, one would expect that the vacuum energy 
will be of order $p$. However, since the rules
of arithmetic in Galois fields are different from the
standard ones, the result for the vacuum energy is
exactly zero. The consideration of the vacuum energy 
also poses the following very interesting problem.
The result is based on the prescription of Sect. 
\ref{S5.5} for separating physical and nonphysical 
states. With such a prescription the vacuum energy is
zero only for particles with the spin 1/2. Is this an
indication that only such particles can be elementary
or the prescription (although it seems very
natural) should be changed?

The Dirac vacuum energy problem is one of the examples 
demonstrating that the standard quantization procedure 
does not define the order 
of annihilation and creation operators uniquely and 
for this reason one has to require additionally  
that the operators should be written in the normal
form. This requirement does not follow from the
theory and only reflects our desire to have
correct properties of the vacuum (see the discussion in 
Sects. \ref{S3.4} and \ref{S5.3}). As noted in Sects. 
\ref{S5.3},
in GFQT this problem has a natural solution and
such a requirement is not needed.

In the standard theory there also exists a problem
that different definitions of the charge operator
are not compatible with each other. As shown in 
Sect. \ref{S6.4}, the charge operator in
GFQT satisfies all the necessary requirements only
for massless particles with the spins 0 and 1/2.
This is in the spirit of modern ideas
that in the limit of unbroken symmetries all
existing elementary particles are massless. It 
might also be an indication that the mass of any
elementary particle can arise only as a result of
interactions while 'bare' elementary particles
should be massless.

In Sect. \ref{S6.5} the arguments in favor and 
against supersymmetry in GFQT are discussed in 
detail. We argue that in GFQT there necessarily 
exists a dilemma that if the notion of particle
and antiparticle is valid at all energies then
only particles with half-integer spins can be
elementary, while if this notion is valid only
when energies are not asymptotically large then
supersymmetry is possible. 

Since in GFQT one IR necessarily describes a
particle and its antiparticle simultaneously,
the problem arises whether it is possible to
implement this requirement in the standard 
theory. This problem has been discussed in
our recent work \cite{jpa}. It has been shown
that with such a modification of the standard
theory, among the Poincare, AdS and dS groups
only the latter can be the symmetry group and
then only fermions can be elementary. As
discussed above, in GFQT there exist much
stronger limitations but such a conclusion
can be drown only if the above dilemma has
the first solution. In this respect the 
difference between
GFQT and the standard theory is as
follows. If $a ={\bar \eta} b^*$ then necessarily 
$a^* = \eta b$, $\{a,a^*\}=\eta {\bar \eta} \{b,b^*\}$ 
and $[a,a^*]=-\eta {\bar \eta}[b,b^*]$. It is easy to 
satisfy the condition $\eta{\bar \eta} =1$ but in the 
field of complex numbers it is impossible 
to satisfy the condition $\eta{\bar \eta} =-1$.
On the other hand, in GFQT this condition is
possible and moreover, as shown in Sect.
\ref{S6.3}, it is the key condition in the
spin-statistics theorem. 

The above discussion shows that GFQT has very
interesting features which might be very important for
constructing new quantum theory (which, according
to Weinberg \cite{Wein2}, may be 'centuries away').
We believe, however, that not only this 
makes GFQT very attractive.

For centuries, scientists and philosophers 
have been trying to
understand why mathematics is so successful in explaining
physical phenomena (see e.g. Ref. \cite{Wigner1}). 
However, such a branch of mathematics as number 
theory and, in particular, Galois fields, have 
practically no implications in particle physics.
Historically, every new physical theory usually involved
more complicated mathematics. The standard mathematical
tools in modern quantum theory are differential and
integral equations, distributions, analytical functions,
representations of Lie algebras in Hilbert spaces etc.
At the same time, very impressive results of 
number theory about properties of natural numbers 
(e.g. the Wilson theorem) and even the notion of primes 
are not used at all! The reader can easily notice that GFQT 
involves only arithmetic of Galois fields (which are even 
simpler than the set of natural numbers). The very 
possibility that the future quantum theory could be 
formulated in such a way, is of indubitable interest. 
 
{\it Acknowledgements:} The author is grateful to 
V. Karmanov, V. Nechitailo, E. Pace, M. Partensky, 
G. Salme and M. Saniga for useful discussions.

\end{document}